\documentclass{aa}

\usepackage[nobiblatex]{xurl}
\usepackage[colorlinks=true, linkcolor=blue, citecolor=blue, urlcolor=magenta]{hyperref}

\usepackage{amsmath}
\usepackage{graphicx}
\usepackage{graphbox}
\usepackage{tikz}
\usepackage{float}
\usepackage{multirow}
\usepackage{color}
\usepackage{makecell}
\usepackage{wrapfig}
\usepackage{comment}
\usepackage{changepage}
\usepackage{booktabs}
\usepackage{sidecap}

\usepackage{natbib}
\bibpunct{(}{)}{;}{a}{}{,} 

\defcitealias{Lee_2013_clouds}{L13}
\defcitealias{Dyrek_2023}{D23}

\usepackage{lineno}
\linenumbers

\begin{document}

\title{ Cloud and Haze Parameterization in Atmospheric Retrievals: Insights from Titan's Cassini Data and JWST Observations of Hot Jupiters}

\author{Q. Changeat \inst{1, 2}, 
        D. Bardet \inst{3}, 
        K. Chubb \inst{4},
        A. Dyrek \inst{5, 6}, 
        B. Edwards \inst{7, 2},
        K. Ohno \inst{8},
        O. Venot \inst{3}}   

\institute{
            \inst{1}Kapteyn Institute, University of Groningen, 9747 AD Groningen, NL.\\
            \inst{2}Department of Physics and Astronomy, University College London, WC1E 6BT London, UK.\\
            \inst{3}Université Paris Cité and Univ Paris Est Creteil, CNRS, LISA, F-75013 Paris, France. \\
            \inst{4}University of Bristol, School of Physics, HH Wills Physics Laboratory, Tyndall Avenue, Bristol BS8 1TL, UK. \\
            \inst{5}Universite Paris Cite, Universite Paris-Saclay, CEA, CNRS, AIM, F-91191 Gif-sur-Yvette, France. \\
            \inst{6}Space Telescope Science Institute, 3700 San Martin Drive, Baltimore, MD 21218, USA. \\
            \inst{7}SRON, Netherlands Institute for Space Research, Niels Bohrweg 4, NL-2333 CA, Leiden, NL. \\
            \inst{8}Division of Science, National Astronomical Observatory of Japan, 2-21-1 Osawa, Mitaka-shi, Tokyo, Japan. \\ \\
            \email{q.changeat@rug.nl}}

\titlerunning{Cloud and Haze Parameterization in Atmospheric Retrievals}
\authorrunning{Changeat et al. 2025}

\date{Received November 26, 2024; accepted May 19, 2025}


\renewcommand{\floatpagefraction}{.95}
\renewcommand{\thefootnote}{\arabic{footnote}}

\abstract{Before JWST, telescope observations were not sensitive enough to constrain the nature of clouds in exo-atmospheres. Recent observations, however, have inferred cloud signatures as well as haze-enhanced scattering slopes motivating the need for modern inversion techniques and a deeper understanding of the JWST information content.}
{We aim to investigate the information content of JWST exoplanet spectra. We particularly focus on designing an inversion technique able to handle a wide range of cloud and hazes.}
{We build a flexible aerosol parameterization within the \textsc{TauREx} framework, enabling us to conduct atmospheric retrievals of planetary atmospheres. The method is evaluated on available {\it Cassini} occultations of Titan. We then use the model to interpret the recent JWST data for the prototypical hot Jupiters HAT-P-18\,b, WASP-39\,b, WASP-96\,b, and WASP-107\,b. In parallel, we perform complementary simulations on controlled scenarios to further understand the information content of JWST data and provide parameterization guidelines.}
{Our results use free and kinetic chemistry retrievals to extract the main atmospheric properties of key JWST exoplanets, including their main molecular abundances (and elemental ratios), thermal structures, and aerosol properties. In our investigations, we show the need for a wide wavelength coverage to robustly characterize clouds and hazes---which is necessary to mitigate biases arising from our lack of priors on their composition---and break degeneracies with atmospheric chemical composition. With JWST, the characterization of clouds and hazes might be difficult due to the lack of simultaneous wavelength coverage from visible to mid-infrared by a single instruments and the likely presence of temporal variability between visits (from e.g., observing conditions, instrument systematics, stellar host variability, or planetary weather).}{}


\maketitle

\section{Introduction}

Population surveys by the Hubble Space Telescope (HST) have long revealed the ubiquitous presence of aerosols (i.e., suspended solid or liquid particles, including clouds and hazes) in exoplanet atmospheres \citep{sing_pop, tsiaras_30planets, Min_2020, Estrela_2022, edwards_pop, Fairman_2024}, importantly affecting their chemical and energy balance. Recently, observations have provided clues about their composition \citep{Miles_2023, Dyrek_2023, Grant_2023, Voyer_2024}, their complex spatial distribution \citep{Miles_2023}, and their time variability \citep{Changeat_2024}. With the increase in data quality offered by the recently launched James Webb Space Telescope (JWST), atmospheric retrievals---the statistical inversion technique used to extract the information from spectroscopic exoplanet data---need to incorporate complex representations of clouds and hazes. Simplified assumptions from the HST-era (e.g., homogeneous grey clouds, or heuristic power laws) are no longer expected to hold with JWST and should be challenged in the context of those novel observations. Given the expected diversity of aerosol signatures \citep{Gao_2021} and the various observing options provided by JWST, it is difficult to know a priori the level of sophistication required to interpret a given dataset. In this work, we use a flexible parameterization for clouds and hazes incorporated in the \textsc{TauREx} retrieval framework \citep{2019_al-refaie_taurex3, al-refaie_2021_taurex3.1} to conduct such an exploration. The overarching goal of this paper is twofold: 1) to present a flexible parameterization for aerosols compatible with the \textsc{TauREx} suite, 2) to explore {\it retrievable} aerosol properties from JWST observations. Our software and methodology are described in Section \ref{sec:meth}. In Section \ref{sec:titan}, we evaluate our retrieval strategy on the solar-system moon Titan, as observed by the {\it Cassini} spacecraft. Section \ref{sec:jwst} explores a few JWST cases with the JWST/NIRISS instrument (i.e., HAT-P-18\,b, WASP-39\,b, and WASP-96\,b), and provides a comprehensive dive into the full WASP-107\,b dataset (HST/WFC3, JWST/NIRSpec, JWST/NIRCam, and JWST/MIRI data), focusing on characterization of aerosols. Finally, in Section \ref{sec:disc}, we discuss the more general implications of our experiments for retrievals of clouds and hazes in the JWST era, and provide complementary simulated cases.  

\section{Methodology}\label{sec:meth}

We employ the \textsc{TauREx3} retrieval framework \citep{2019_al-refaie_taurex3}. \textsc{TauREx3} relies on a modern code design with a convenient plugin system for rapid prototyping of new functions \citep{al-refaie_2021_taurex3.1}. To conduct our exploration, we utilize the plugin feature to introduce three open-source plugins: \textsc{TauREx-PyMieScatt}\footnote{\textsc{TauREx-PyMieScatt}: \url{https://github.com/groningen-exoatmospheres/taurex-pymiescatt}}, \textsc{TauREx-MultiModel}\footnote{\textsc{TauREx-MultiModel}: \url{https://github.com/groningen-exoatmospheres/taurex-multimodel}}, and \textsc{TauREx-InstrumentSystematics}\footnote{\textsc{TauREx-InstrumentSystematics}: \url{https://github.com/groningen-exoatmospheres/taurex-instrumentsystematics}}. Central to this paper, \textsc{TauREx-PyMieScatt} provides flexible modeling and retrieval of parameterized aerosols using the open-source python library \textsc{PyMieScatt} \citep{Sumlin_2018} and the phenomenological cloud model from \cite{Lee_2013_clouds}, hereby \citetalias{Lee_2013_clouds}. It is described in more details in the Appendix. \textsc{TauREx-MultiModel} allows to simulate in-homogeneous atmospheres by combining multiple \textsc{TauREx} forward models---with independent or coupled parameters---and by weighting their contribution to produce the final observed signal (i.e., eclipse flux ratio or transit depth). \textsc{TauREx-InstrumentSystematics} allows to treat instrumental systematics when combining observations from different instruments and/or epochs \cite[see also][]{Yip_2021_W96}.

In \textsc{TauREx-PyMieScatt}, cloud and haze layers are defined by their extinction coefficient ($Q_\mathrm{ext}$), the particle size distribution (i.e., modeled as a log-normal, a modified gamma, or a one-parameter gamma distribution), the mean pressure of the layer ($P$) and its range ($\Delta P$), and the particle number density ($\chi$). Multiple particle size distributions are available but for this work we primarily employ the one-parameter gamma distribution introduced in \cite{Budaj_2015}, which is uniquely defined by the critical radius of the aerosol particles ($\mu_r$). \textsc{TauREx-PyMieScatt} can also model porous particles and aggregates, but in this work, we concentrate on spherical particles since such level of characterization is difficult with JWST spectra (see Section \ref{sec:disc}). Additionally, recent studies have highlighted the likely presence and signature of spatially scattered (or partial) clouds, which could be induced by atmospheric circulation. Partial cloud coverage can be modeled by combining a clear and a cloudy forward model with the \textsc{TauREx-MultiModel} plugin. To illustrate the developments made in this work, we show forward model spectra for various possible cloud scenarios in Figure \ref{fig:cloud_W107_exploration}, inspired by the recent results of the prototypical hot Jupiter WASP-107\,b \citep{Dyrek_2023, Welbanks_2024, Sing_2024}. Relevant literature references for the opacity sources we used in this work are available in Table \ref{tab:opacityRef} with the data being available online. \textsc{TauREx} and its plugins are used to perform atmospheric retrievals of {\it Cassini} and JWST data for which the setups are briefly described below.

\begin{table*}
\centering
\caption{\protect\rule{0ex}{1cm}Reference list for the opacity sources used in this work.}
\begin{tabular}{ |c|c|c|c| } 
 \hline
 Gas & References & Gas & References \\ \hline
 H$_2$O & \cite{polyansky_h2o} & CO & \cite{li_co_2015} \\
 CO$_2$ & \cite{Yurchenko_2020} & CH$_4$ & \cite{ExoMol_CH4_new} \\
 C$_2$H$_2$ & \cite{Chubb_2020_c2h2} & C$_2$H$_4$ & \cite{2018_Mant_C2H4} \\
 C$_2$H$_6$ & \cite{Harrison_2010_C2H6} & C$_3$H$_8$ & \cite{Harrison_2010_C3H8} \\
 NH$_3$ & \cite{al-derzi_2015_nh3, Coles_2019_nh3} & SO$_2$ & \cite{Underwood_2016} \\
 H$_2$S & \cite{Azzam_2016_h2s, Chubb_2021_exomol} & HCN & \cite{Barber_2013_HCN, Chubb_2021_exomol} \\
 Na & \cite{Allard_2019_Na} & K & \cite{Allard_2016_K} \\ \hline\hline
 Aerosols & References & Aerosols & References \\ \hline
 Tholins & \cite{Khare_1984, Rannou_2010} & CH$_{4\mathrm{(l)}}$ & \cite{Martonchik_1994} \\
 KCl$_\mathrm{(c)}$ & \cite{Querry_1987} & Na$_2$S$_\mathrm{(c)}$ & \cite{Khachai_2009} \\
 ZnS$_\mathrm{(c)}$ & \cite{Querry_1987} & MgSiO$_{3\mathrm{(a)}}$ & \cite{Scott_1996} \\
 SiO$_\mathrm{(a)}$ & \cite{Wetzel_2013} & SiO$_{2_\mathrm{(a)}}$ & \cite{Palik_1991, Kitzmann_2018} \\
 \hline\hline
 Others & References & Others & References \\ \hline
 H$_2$-H$_2$ & \cite{abel_h2-h2, fletcher_h2-h2} & H$_2$-He & \cite{abel_h2-he} \\
 N$_2$-N$_2$ & \cite{Borysow_1987} & N$_2$-CH$_4$ & \cite{Borysow_1993} \\
 CH$_4$-CH$_4$ & \cite{Chistikov_2019} & Rayleigh & \cite{cox_allen_rayleigh} \\ \hline
\end{tabular}%
\tablefoot{The ExoMol cross-sections for H$_2$O, CO, CO$_2$, CH$_4$, NH$_3$, SO$_2$, H$_2$S, and HCN are available at resolution $\mathcal{R}=50,000$ \citep{Chubb_2021_exomol}. Other cross-sections employed here are computed at $\mathcal{R}=15,000$. (l): liquid, (c): crystaline, (a): amorphous. Compatible aerosol species can also be found at \cite{natasha_batalha_2020_5179187}. The \textsc{TauREx} compatible data used for this work is available online at Zenodo \citep{changeat_2025_15495830}.}\label{tab:opacityRef}
\end{table*}

\paragraph{Cassini retrievals:} From 2006 to 2011, the {\it Cassini} spacecraft obtained 10 occultations of Titan using the Visual and Infrared Mapping Spectrometer (VIMS), four of which are consolidated in \cite{Robinson_2014}. The other six could not be recovered due to technical issues. The {\it transit-like} spectra in \cite{Robinson_2014} cover the wavelengths $\lambda \in [0.88,5]\,\mu$m and are ideal to benchmark exoplanet models because of our extensive knowledge of Titan's atmosphere and in-situ measurements. However, our exploration of this Titan data does not aim to replicate the breadth of knowledge accumulated over half a century of observations of this atmosphere. Instead, we seek to illustrate the retrievable atmospheric properties that could be deduced should a similar exoplanet atmosphere be observed. We perform \textsc{TauREx} retrievals on the data of \cite{Robinson_2014} focusing on the recovery of the chemistry and aerosol properties. The retrievals include Rayleigh Scattering, Collision Induced Absorption (CIA for N$_2$-N$_2$, N$_2$-CH$_4$, and CH$_4$-CH$_4$), molecular absorption (CH$_4$, CO, C$_2$H$_2$, C$_2$H$_4$, C$_2$H$_6$, and C$_2$H$_8$), and aerosols. Here, we consider tholins hazes, and condensate CH$_{4\mathrm{(l)}}$ clouds. Hazes are thought to produce the blue-ward scattering slope seen in the Titan data, as well as contributing to enhanced absorption around $\lambda = 3.4\,\mu$m \citep{Bellucci_2009, Cours_2020} from C--H stretching. We combine the most recent optical data from \cite{Rannou_2010} with those from \cite{Khare_1984} to construct an updated, broad wavelength ($\lambda \in [0.4, 5.5]\,\mu$m), set of refractive indexes for tholins (see Table \ref{tab:hazesRef}). We perform two sets of retrievals: 1) retrievals with the exoplanet optimized cross-sections from the ExoMol project \citep{Tennyson_exomol, Chubb_2021_exomol, Tennyson_2024}, and 2) retrievals with HITRAN generated cross-sections\footnote{An Earth N2/O2 background mixture is assumed since the broadening coefficients for Titan's composition are not available.} using the HAPI tool \citep{Kochanov_2016}. For each set of cross-sections, we test a haze only case and a haze+CH$_4\mathrm{(l)}$ case, using independent aerosol layers of constant particle number density (this hypothesis is discussed more in Section \ref{sec:disc}). Here, the CH$_4\mathrm{(l)}$ condensate cloud layer acts as a simple proxy for the complex methane/ethane aerosol cycle happening in the troposphere of Titan. However, note that this is not intended to fully model the region. In the retrievals presented in the main text, the atmosphere is mainly composed of N$_2$ and CH$_4$, with $X_\mathrm{CH_4}$ = 0.0148\footnote{This is the averaged stratospheric abundance. Note that the surface abundance was determined to be around $X_\mathrm{CH_4}$ = 0.0565, and that other works have suggested atmospheric temporal variations of up to 40\% \cite[e.g.,][]{Adamkovics_2016}.} \citep{Niemann_2010} and we fix the thermal profile to the one presented in \cite{Fulchignoni_2005}. In Figure \ref{fig:titan_freeTP} of the appendix, we remove these two assumptions (i.e., we retrieve the thermal profile and abundance of CH$_4$), highlighting retrieval degeneracies commonly faced in exoplanet transit studies \citep{Rocchetto_2016, Changeat_2020_mass, DiMaio2023, Schleich_2024}. We retrieve the abundances of trace molecules using log Gaussian priors \cite[$\mu_\textsc{S}$ from][$\sigma_\textsc{S}$ = 0.5]{Coustenis_2016} and the cloud parameters using uninformative uniform priors.

\paragraph{JWST retrievals:} Recently JWST has obtained exoplanet spectra of similar signal-to-noise and resolution to the {\it Cassini} occultation data. We here explore the information contained at short wavelength using transit observations of HAT-P-18\,b, WASP-39\,b, and WASP-96\,b by the Near Infrared Imager and Slitless Spectrograph (NIRISS). We then perform a more complete exploration of WASP-107\,b, which has been extensively observed. WASP-107\,b has observations at short wavelengths---from the Hubble Wide Field Camera 3 G102 and G141 Grisms---and recent data from the Near Infrared Spectrograph G395H (NIRSpec-G395H), the Near Infrared Camera with F322W2 and F444W filters (NIRCam-F322W2 and NIRCam-F444W), as well as the Mid-Infrared Instrument Low Resolution Spectrograph (MIRI-LRS). We aim to highlight the advantage of MIRI in constraining the silicate cloud feature around 10$\mu$m, which appears to be present in many hot atmospheres.

The JWST/NIRISS data was obtained from two sources: \cite{Holmberg_2023} for WASP-39\,b and WASP-96\,b, and \cite{Fu_2022} for HAT-P-18\,b. To investigate those observations, we use a similar retrieval strategy to the Titan case. We include absorption from Rayleigh Scattering, CIA, the relevant molecules (i.e., H$_2$O, CO, CO$_2$, CH$_4$, Na, and K), and aerosols. Note that including aerosols is needed to fit the JWST data, except for the WASP-96\,b case as we will show in the result section. For the aerosols, we initially considered layers of KCl, Na$_2$S, and ZnS clouds, which are appropriate for the temperature regimes of those objects ($T_{eq} \in [900, 1400]$\,K). However, comparing the Bayesian evidences, those cloud species did not provide an advantage over the more heuristic cloud model from \citetalias{Lee_2013_clouds}, so we concentrate our main discussion on the cases using the \citetalias{Lee_2013_clouds} model. Using \textsc{TauREx-MultiModel}, we consider spatially scattered aerosols, retrieving the fraction of aerosol coverage (F$_a$). Note that \textsc{TauREx-MultiModel} can perform more complex separations of the terminator (i.e., using two completely separate atmospheres to represent the west-east limbs) as suggested in \cite{Arfaux_2024}. We explored this option but this did not seem to improve our fits, so we only present the scattered cloud results. As opposed to Titan, the radius and thermal structure of those objects are unknown, so they must be retrieved directly. The radius is retrieved at a pressure of $p = 10$\,bar using uninformative priors. The thermal structure is modeled by seven freely moving $T-p$ nodes \citep[at fixed pressure, see e.g., ][]{changeat_2021_phasecurve2, Schleich_2024}. For each planet, two retrievals are performed: 1) a free chemistry retrieval using a constant-with-altitude chemical profile for each species (referred as {\it free}), and 2) a kinetic chemistry retrieval using \textsc{FRECKLL} \citep{Al-Refaie_2022_FRECKLL} to compute the altitude dependent abundances of H/He/C/N/O species\footnote{Here, the reduced chemical scheme---without photochemistry---from \cite{Venot_2020_new} is used. The relevant free parameters are metallicity (Z), C/O ratio, and vertical mixing coefficient (Kzz).} and using free abundances for Na and K (referred as \textsc{FRECKLL}). All the free parameters in those retrievals are explored using uninformative priors. 

\begin{figure*}
\centering
    \includegraphics[width = 0.98\textwidth]{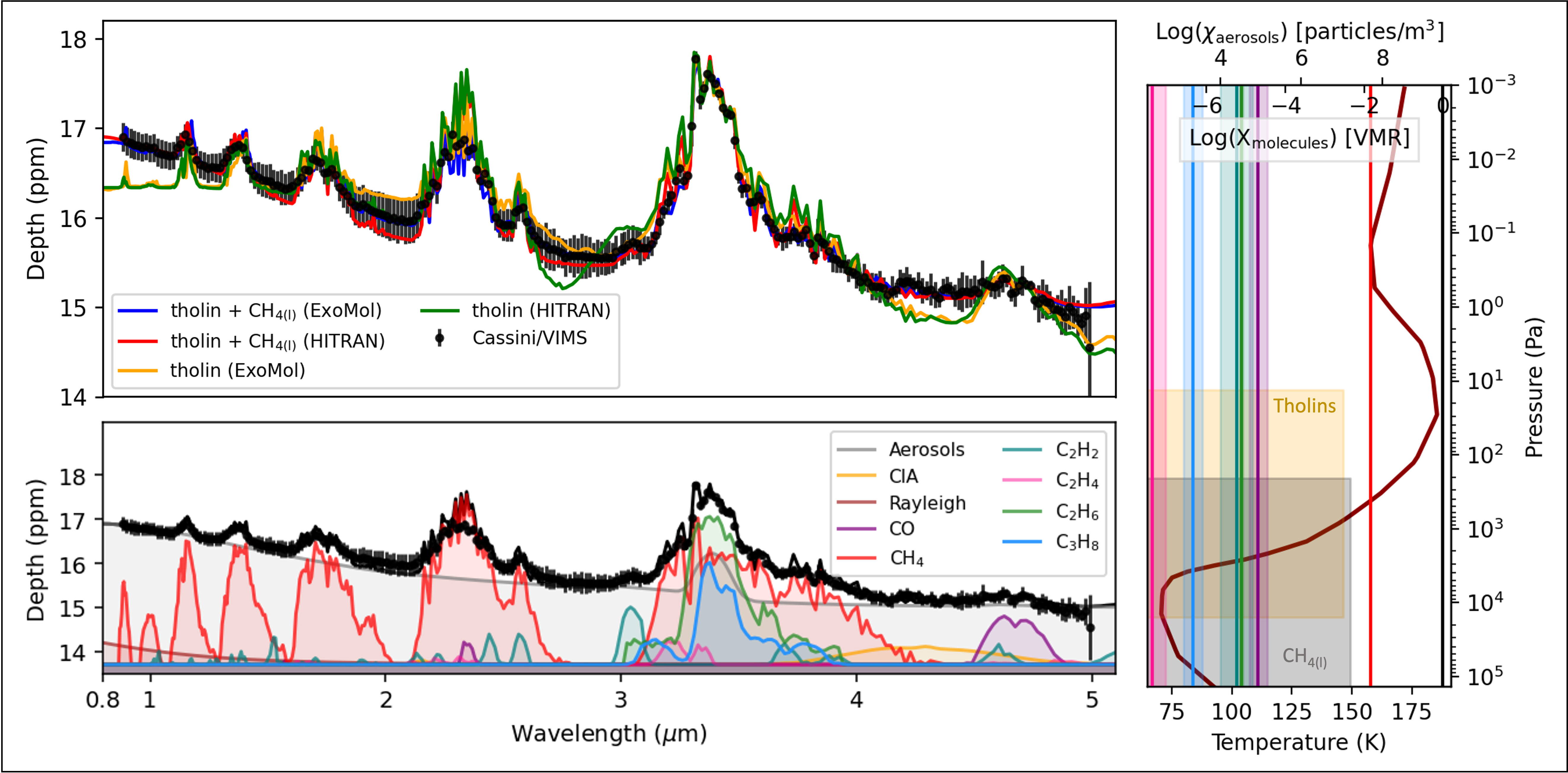}
    \caption{Retrievals of {\it Cassini}/VIMS occultation of Titan. Top left: observations and best-fit \textsc{TauREx} retrievals. Bottom left: breakdown of the extinction contributions in the tholin + CH$_{4\mathrm{(l)}}$ (HITRAN) case. Right: Atmospheric properties inferred for the  tholin + CH$_{4\mathrm{(l)}}$ (HITRAN) case. The source for the opacity data (i.e, HITRAN or ExoMol) is important and can slightly change the best-fit spectrum and retrieval interpretation. Overall, the {\it Cassini} data is dominated by molecular absorption from CH$_4$, CO, C$_2$H$_6$, C$_3$H$_8$ and extinction by hydrocarbon aerosols (tholins + CH$_{4\mathrm{(l)}}$). The shape of the 3.4\,$\mu$m absorption feature is likely best explained by the contribution of multiple species including some contribution from tholins. Overall, retrieval interpretations of Titan's atmosphere are compatible with stratospheric inferences from previous works (see Figure \ref{fig:titan_corner}). }\label{fig:titan_spec}
\end{figure*}

For WASP-107\,b, we considered the data from \cite{Welbanks_2024} for HST/WFC3, JWST/NIRCam, and JWST/MIRI and the data from \cite{Sing_2024} for JWST/NIRSpec. Note that many different reduction of the data exist \cite[see for instance MIRI: ][]{Dyrek_2023, Welbanks_2024}, but we here concentrate on the main \textsc{Eureka!} reduction labeled in these articles. A similar setup to the NIRISS retrievals was used for this data. However, we caution our results, highlighting that the combination of multiple observations---from HST and various JWST instruments---could potentially produce biases from reduction or time-variable incompatibilities \cite[see e.g., ][]{Yip_2021_W96, Changeat_2024, Edwards_2024_K11}. For instance, comparison of the \textsc{petitRADTRANS} retrievals of WASP-17\,b in \cite{Grant_2023} show such behaviour, with the $T$ and cloud solutions varying significantly whether the HST data is included or not. 
We retrieve the planetary radius, a temperature profile composed of seven $T-p$ nodes, the relevant molecular abundances (i.e., H$_2$O, CO, CO$_2$, SO$_2$, H$_2$S, NH$_3$), and clouds (SiO, SiO$_{2}$, MgSiO$_{3}$, and Mg$_2$SiO$_{4}$). Heuristic aerosols like those in \citetalias{Lee_2013_clouds} cannot model the 9--10\,$\mu$m Si--O stretch spectral feature from silicate particles, requiring the use of Mie theory. The aerosols are included as independent layers. The retrievals also include vertical offsets for the HST and the MIRI data using the \textsc{TauREx-InstrumentSystematics} plugin. 

\section{Titan retrievals of Cassini occultations} \label{sec:titan}

The consolidated {\it Cassini} occulations of Titan from \cite{Robinson_2014} and our best-fit retrieval models are shown in Figure \ref{fig:titan_spec}. The haze only retrievals do not fit the observation well and the continuum shape of the spectrum requires the addition of a second aerosol component (i.e., a single haze layer does not seem to have enough flexibility). When hydrocarbon clouds, here assumed to be from condensed methane (CH$_{4\mathrm{(l)}}$)---but which in reality are more complex and also made from many other organics like ethane---are added, both ExoMol and HITRAN retrievals can reproduce the {\it Cassini} observations, albeit with slightly different retrieved parameters for the hazes (see Figure \ref{fig:titan_corner}). The VIMS spectrum has clear features of aliphatic hydrocarbons, in particular around $\lambda = 3.4\,\mu$m as highlighted in \cite{Rannou_2022}, where CH$_4$, C$_2$H$_6$, and C$_3$H$_8$ are detected. C$_2$H$_2$ and C$_2$H$_4$ are also likely present, but they contribute less to the atmospheric extinction (see contribution function). Enhanced absorption from CO is also visible at $\lambda = 4.5\,\mu$m, and correctly identified by the retrievals. In Figure \ref{fig:titan_corner} we show the posteriors distributions of the haze+CH$_{4\mathrm{(l)}}$ retrievals, highlighting the broad agreement between our retrieved Titan chemistry and the more precise results from resolved observations in the literature \cite[i.e.,][]{Horst_2017, Sylvestre_2018, Rannou_2022}. Note however that these studies employed temporally and spatially resolved observations---showing the highly variable nature of Titan---whereas our consolidated dataset from \cite{Robinson_2014} combines multiple observations: a direct comparison could therefore be misleading. Nevertheless, our retrievals probe aerosol properties (likely in the stratosphere) favoring small haze particles (i.e, $\mu_{\mathrm{tholins}} \sim 0.2\,\mu$m). This is consistent---in location and properties---with the well-known main haze layer at pressures up to $p = 1$\,Pa. Previous micro-physical aerosol models of the main haze layer \citep{McKay_2001} roughly predict similar aerosol characteristics: sub-micrometer size particles (i.e, $\mu$ $\sim0.2\,\mu$m) with a particle number density of log($\chi$)\,$\sim\,[6, 9]$\,particles/m$^{3}$. For instance, the Huygens observations \citep{Tomasko_2008} used in \cite{Lavvas_2010} suggest a slighly larger volume average haze radii of $\mu$\,$\in\,[0.7, 2]\,\mu$m with similar particle densities of $\chi\,\sim\,5. 10^6$\,particles/m$^{3}$. Differences are to be expected since our retrievals do not consider aggregates, which are known to be crucial for Titan. In particular, \cite{Rannou_2022} performed retrievals on the individual VIMS data (the original source of our spectrum) taken from \cite{Maltagliati_2015}. Overall they show similar results to us (see e.g., their Figure 2) but also highlighting the complex morphology of Titan's hazes (aggregate of $\sim$3000 monomers and fractal dimension D$_f$ = 2.3), which is not considered in our work. Their methodology and aerosol modeling is different, selecting the continuum wavelength windows ($\lambda \in [0.5, 2.5]$) to constrain fractal hazes of constant refractive indexes, before focusing on other molecules. We also find small differences in haze retrieved parameters induced by our choice of opacity source (HITRAN vs ExoMol). For instance, $\mu$ varies from $\sim 0.15\,\mu$m to $\sim 0.3\mu$m just from our choice of opacities. This is likely due to differences in broadening and line completion at low temperatures---note that ExoMol line-lists are broadened by a H$_2$/He mixture, while HITRAN line-lists are broadened by an Earth mixture---clearly highlighting opacity sources as an additional difficulty in the study of cold secondary exo-atmospheres \citep{Anisman_2022, Niraula_2022, Garib_2024, Chubb_2024}. These differences are also clearly visible in the investigations with free thermal profiles and methane abundances (see Figure \ref{fig:titan_freeTP}). In this case, differences in the cross-sections affect the retrieved $T-p$ profiles, highlighting the difficulties of extracting robust temperature information in this case. In the retrievals with fixed thermal structure and CH$_4$ abundance, a deep condensate cloud layer is needed to fit the data. It is made of larger particles ($\mu_{\mathrm{clouds}}\,\sim 1\,\mu$m), and possibly extends up to $p\,\sim\,10$ Pa. We model this using methane condensates---which should form on Titan via tropospheric convection---but other types of stratified organic ice clouds are also expected \citep{Barth_2017, anderson_2018_titan} and could extent to the stratosphere. Titan's condensate cloud particles are known to be typically larger than those we retrieve (with $\mu_{\mathrm{clouds}}\,\sim\,10\,\mu$m for CH$_4$ condensates, and $\mu_{\mathrm{clouds}}\,\sim\,1\,\mu$m for C$_2$H$_6$ condensates).\\

Employing a general purpose retrieval code designed for exoplanet applications, and using only the averaged occultation data, we do not expect to extract the exact structure of Titan's aerosols, which is not the objective of this study. As previously said, other in situ and spatially resolved spectroscopic measurements from {\it Cassini} have provided a more detailed and complex picture of Titan's atmosphere. However, we approached this occultation data similarly to how one would approach exoplanet data, showing that these data are sensitive to the broad properties of the atmosphere, including aerosols. This experiment offers guidance on which atmospheric properties can be reliably retrieved and could serve as a controlled benchmark for future modeling strategies. The retrievals performed here have key limitations: 1) cloud and haze particles are assumed to be compact spheres, 2) simple particle size distributions are employed, 3) vertical number density is modeled using simple laws, and 4) we employ a combined (i.e., not spatially and temporally resolved) dataset. For Titan, haze particles in the upper atmosphere are known to be fractal aggregates rather than spheres \cite[e.g.,][]{West&Smith91,Cabane+93,Rannou+95, Rannou_2022, Perrin_2025}. In Section \ref{sec:disc}, we discuss the relevance of some of those assumptions in the context of recent JWST data. Overall, the inferred atmospheric properties from our Titan retrievals of the {\it Cassini}/VIMS data, shown in Figure \ref{fig:titan_spec}, provide a bulk picture consistent with the actual structure on Titan. Our exploration of the Titan data suggests that parameterized approaches can be used to explore the information content of exoplanet spectra similar to those of Titan.

\begin{figure*}[!h]
\centering
    \includegraphics[width = 0.95\textwidth]{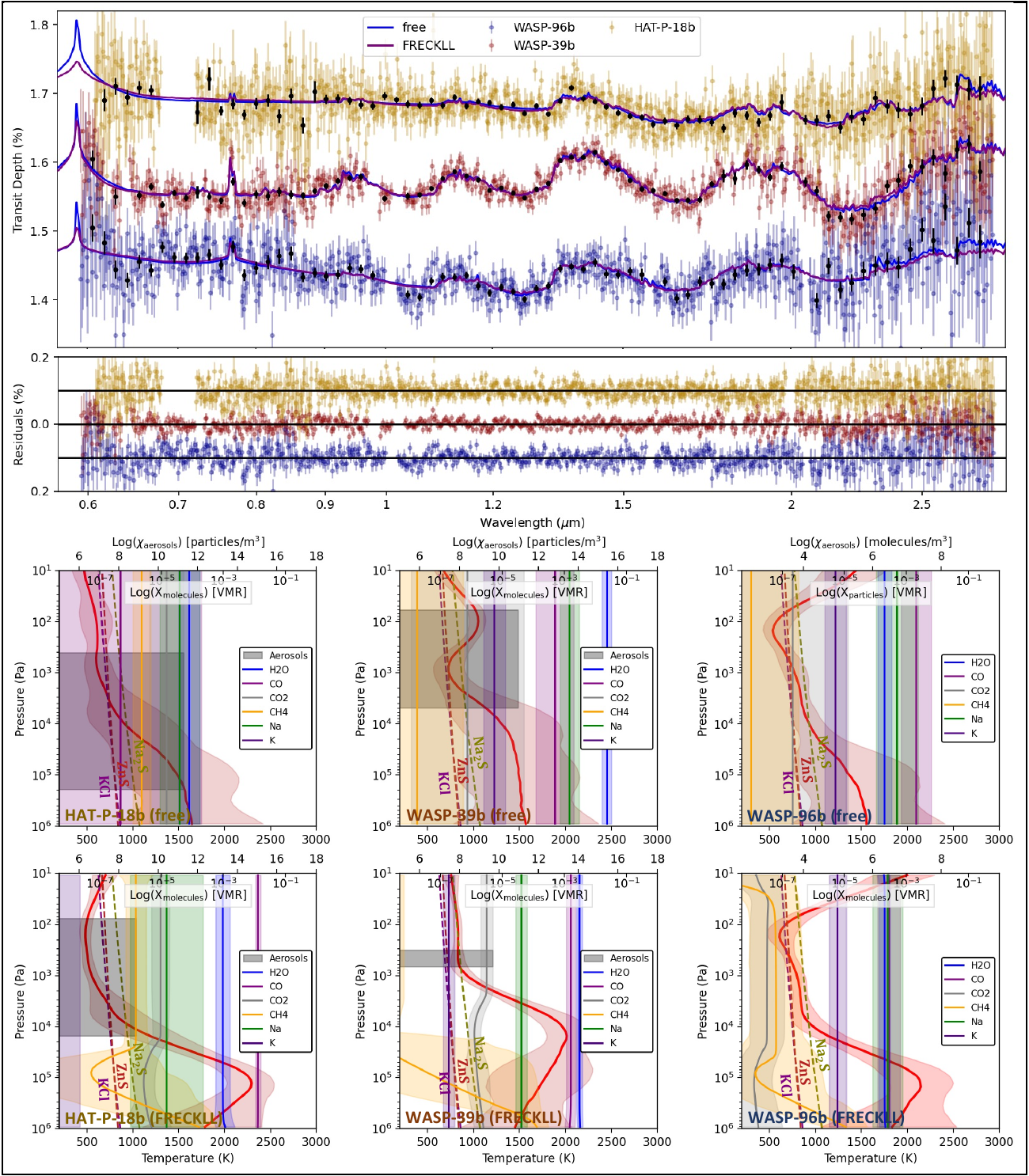}

    \caption{Results of the atmospheric retrievals using the \cite{Lee_2013_clouds} cloud model for the NIRISS observations. The observed spectra, best-fit models (free chemistry in blue and \textsc{FRECKLL} chemistry in purple) and residuals for the \textsc{FRECKLL} retrieval are shown in the top panel. The lower panels show retrieved $T-p$ and chemistry---free chemistry in middle row, and \textsc{FRECKLL} chemistry in bottom row---with 1\,$\sigma$ confidence intervals, as well as best-fit aerosol solutions. We also indicate with dashed lines the relevant cloud condensation curves from \cite{Wetzel_2013, Gao_2021} for context. Retrieved atmospheric properties are not always fully consistent despite similar Bayesian evidences.
    The accompanying posterior distributions of these retrievals are available in Fig \ref{fig:niriss_corner}.}
    \label{fig:niriss_spec}
\end{figure*}

\section{Applications to JWST exoplanets}\label{sec:jwst}

For HAT-P-18\,b, WASP-39\,b and WASP-96\,b, we consider the data obtained by the JWST/NIRISS instrument. For WASP-107\,b, we consider the data obtained by HST/WFC3, JWST/NIRSpec-G395H, JWST/NIRCam-F322W2, JWST/NIRCam-F444W and JWST/MIRI. We detail our retrieval results in the next sections.

\subsection{Aerosol scattering slopes: example from HAT-P-18\,b, WASP-39\,b, and WASP-96\,b JWST/NIRISS observations}

For HAT-P-18\,b, WASP-39\,b, and WASP-96\,b, results of our retrievals using the \citetalias{Lee_2013_clouds} aerosol model are shown in Figure \ref{fig:niriss_spec}. We here focus on the \citetalias{Lee_2013_clouds} retrievals since the posterior distributions for \textsc{PyMieScatt} retrievals (see HAT-P-18\,b case in Figure \ref{fig:corner_hp18}) are independent from the assumed cloud composition when using JWST/NIRISS data alone. Similarly, complex terminator modeling (i.e., two distinct regions) is not presented as they did not lead to improved Bayesian Evidence over the simpler retrievals. Figure \ref{fig:niriss_contrib} shows a breakdown of the contributions for the \textsc{FRECKLL} retrievals. Figure \ref{fig:niriss_corner} shows the corner plots and computed elemental ratios from both free and \textsc{FRECKLL} runs. The retrieved quantiles of our retrievals are available in Table \ref{tab:niriss_retrievals}. The three planets have a roughly similar equilibrium temperature and consistent retrieved $T-p$ structures that are crossing the same condensate curves. However Figure \ref{fig:niriss_spec} clearly shows spectra with different shape, suggesting disparate chemistries and cloud structures. We here describe our retrieval findings. \\

In HAT-P-18\,b, we detect the spectral signatures of H$_2$O, CH$_4$, CO$_2$, and Na. The Na abundance is consistent between our free and \textsc{FRECKLL} retrievals with log(Na) $\sim -4.5$ (about 10x solar). However, the metallicity of the atmosphere depends on the chemical model, with for instance a recovered log(H$_2$O)\,$\in [-4.45,-3.76]$\,dex corresponding to a depleted (Z $\sim$ 0.17x solar) atmosphere for the free retrieval, while the \textsc{FRECKLL} retrieval recovers Z $\sim$ 10x solar metallicity. The metallicity inferred in the free run, from C and O species as well as the refractories (including Na and K), might not be very accurate due to missing species in this estimate. The C/O from the \textsc{FRECKLL} retrieval is estimated to C/O $\sim 0.9$, while the free retrieval estimate for C/O is much lower  and difficult to interpret since major reservoirs of C (e.g., CH$_4$, and CO) are not detected. In general, we caution against over-interpreting C/O computed from the free abundance retrieval, as many C-bearing species are not constrained by the data and because O can easily be sequestered in e.g., other species or aerosols \citep{Fonte_2023}. Note that the \textsc{FRECKLL} retrievals used the reduced scheme of \cite{Venot_2020_new}, most certainly missing processes (i.e., photochemistry) that are efficiently removing CH$_4$ from the upper atmosphere and could bias the results if mixing is important and the deeper atmosphere is affected: transit data in the near-IR typically probe the 10$^5$-100\,Pa deeper regions \cite[][]{Caldas_2019, Changeat_2019_2l, Rustamkulov_2023, Fenstein_2023}. The detection of Na---which originates from an apparent enhanced absorption at the blue end of the spectrum, in the red wing of the 0.6\,$\mu$m Na line---is a difficult feature to reproduce with the aerosol models used here but we note the precarious nature of this detection (and abundance constraints), which relies on very few datapoints and could also be impacted by instrumental systematics. For the cloud solutions, while the retrieved altitudes for the layers are slightly different between the two runs, they are consistent with a think layer of sub-micrometer sized particles extending above 1\,mbar and extending across the full terminator limb (i.e., F$_\mathrm{lee}$ consistent with 1). Salt clouds (i.e., KCl, Na$_2$S, and ZnS are likely contenders as shown in Figure \ref{fig:niriss_spec}: the retrieved $T-p$ profiles cross the condensation lines for these species, but we are unable to distinguish a most likely candidate from the data alone. This is shown in Figure \ref{fig:corner_hp18}, where the solutions of three retrievals with the \textsc{TauREx-PyMieScatt} parameterization, and using respectively KCl, Na$_2$S, and ZnS refractive indexes, lead to the exact same posteriors and evidence. The figure also demonstrates the stability of our solution to the cloud type assumptions and justifies the use of a simpler model \cite[i.e., ][]{Lee_2013_clouds} for our main conclusions with NIRISS. The formation of KCl or Na$_2$S clouds should have a significant impact on the gaseous abundances of K or Na respectively. In particular, gaseous K is constrained to low abundances by this HAT-P-18\,b data with log(K) $< -9$, which could suggest sequestration of K in KCl clouds. However, since gases and condensates are included separately in our inversions, we cannot verify this statement and complementary, self-consistent approaches connecting gas and condensed phases, such as done in \cite{Ormel&Min19, Ma_2023, Arfaux_2024}, are needed to properly assess this hypothesis---and more generally to invert cloud types from the NIRISS data. Similar results were found for WASP-39\,b and WASP-96\,b. The abundance constraints of our free run are similar to those from \cite{Fournier_2024} despite not accounting for potential light-source effects. CH$_4$, however, was not detected in their study but was here found to similar levels in \cite{Fu_2022}. \\

For WASP-39\,b, we detect the spectral signatures of H$_2$O, CO, CO$_2$, Na, and K. The atmosphere is here consistent with superstellar metallicity---log(Z) $\sim$ 5x solar for the \textsc{FRECKLL} run and log(Z) $\sim$ 20x solar for the free run. This is mainly due to high water abundance (up to 2.3\% in the free run), which is interesting given the sub-solar O/H of the host star. We also infer a sub-solar C/O ratio, with C/O $\sim$ 0.35 for the \textsc{FRECKLL} retrieval, but as with HAT-P-18\,b, we caution against over-interpreting the C/O derived from free retrievals. Both Na and K seem to have significantly enhanced abundances (between 10x and 100x solar depending on the run) with clear spectral signature in the NIRISS spectrum, but this does not seem degenerate with the aerosol solution in this case. Those results are overall consistent with other studies of this dataset \citep{Fenstein_2023, Constantinou_2023, Arfaux_2024}. Using the \citetalias{Lee_2013_clouds} model, clouds are found at high altitude ($p < 0.1$\,mbar) and made from sub-micrometer sized particles. While the retrieved $T-p$ profile is on average hotter than for HAT-P-18\,b and does not cross the KCl and ZnS condensation lines (i.e., this could favor Na$_2$S as a possibility for the clouds), the phenomenological properties of the retrieved aerosols are similar to those of HAT-P-18\,b. A recent study \citep{Arfaux_2024} employing more complex haze and cloud microphysics has also suggested Na$_2$S as a good contender for the clouds on WASP-39\,b. However, their study also indicates that the formation of such clouds would deplete atomic Na in gas, which seems contradictory with the findings from our retrievals. Further iterations between first principle and data oriented approaches are needed here. Recently, a study \citep{Ma_2025} has also performed atmospheric retrievals on the full WASP-39\,b data, discussing the possibility of important Si chemistry to explain various spectral features seen at longer wavelengths. These disparate results highlight the need for observations covering a large wavelength range---necessary to break the chemical-cloud degeneracies for exoplanets in this regime---and iterations between free retrieval approaches and self-consistent modeling. We discuss the advantage of the wide wavelength coverage in the next section for WASP-107\,b. \\

In WASP-96\,b, we detect signatures of H$_2$O, CO, Na, and K. Both retrievals (free and \textsc{FRECKLL}) suggest a clear atmosphere. While this appear to contradict the final interpretations of \cite{Taylor_2023} with the same raw data (but not the same reduction), we note that they also find that such a clear solution can fit their data. In our work, the NIRISS spectrum of WASP-96\,b is the only one that can be fitted with a model that does not include clouds nor hazes. Note that we also performed a tholin retrieval to simulate high altitude hazes but this was not favored. Interestingly, a cloud-free atmosphere is in line with prior interpretations of the HST and VLT data \citep{Nicolov_2018, Yip_2021_W96}. We note that the spectra used in \cite{Taylor_2023} were reduced in \cite{Radica_2023}, but we could not find these resources to conduct further comparisons. In \cite{Holmberg_2023}---our original source for the WASP-96\,b NIRISS spectrum---they did not perform atmospheric retrievals to support their interpretation, but highlighted that assumptions on the stellar limb-darkening have a significant impact on the blue-ward slope (where most of the cloud constraints originate from). Alternatively, we also stress that past observations showing a clear atmosphere might not be fully relevant for the interpretation of JWST data since hot-Jupiters like WASP-96\,b are expected to be temporally variable: they could for instance have transient cloud coverage \citep{Skinner_2021_modons, Changeat_2024}. Either way, the spectrum of WASP-96\,b presents a blue-ward slope favoring a solution with high abundances of Na and K in our retrievals. The large wings of those atomic species could easily explain the increased absorption in the data from \cite{Holmberg_2023}, and this interpretation is independent from our chemical assumptions (free vs \textsc{FRECKLL}). In our runs, the metallicity is consistent with slightly sub-solar to solar, log(Z) $\sim$ 0.5x solar for the \textsc{FRECKLL} run and log(Z) $\sim$ 4x solar for the free case. The only C-bearing species to be detected is CO, potentially suggesting higher than solar abundances of this molecule---C/H $\sim$ 4x solar in the free case---which could explain the inferred super-solar C/O ratio for the free run. However, robust estimates of C/O ratios from free runs might require constraints on multiple carbon species, which we remind is not achieved here for any of our targets. The \textsc{FRECKLL} retrieval instead suggests a solar C/O ratio. As previously said, Na and K are used by the retrieval to explain the blue-ward slope of our retrievals, which translates into a retrieved super-solar abundance for those species, log(Na) $\sim$ 100x solar, and log(K) $\sim$ 25x solar (see values in Table \ref{tab:niriss_retrievals}). Comparing the retrieved $T-p$ profiles with cloud condensation lines (see Figure \ref{fig:niriss_spec}, we expected clouds to be detected as in HAT-P-18\,b and WASP-39\,b, and similarly to what is favored in \cite{Taylor_2023}. These different interpretations likely originate from model degeneracies between Na/K and cloud scattering slopes, and/or have contribution from instrumental origin \citep{Taylor_2023, Rotman_2025}.  \\

For aerosols, JWST/NIRISS gives access to scattering processes by probing the blue-ward absorption slope of scattering particles at $\lambda < 1\mu$m. However, in the cases investigated here, the wavelength coverage is not sufficient to identify the type of particles responsible for the enhanced absorption (see e.g., Figure \ref{fig:corner_hp18}). The inferred properties, however, seem robust to assumptions on the cloud type, potentially allowing us to safely utilize phenomenological models such as \citetalias{Lee_2013_clouds}. Interestingly, the case of WASP-96\,b---which visually shows an increased absorption at blue wavelengths---reveals some level of degeneracies with Na and K continuous wing absorption. For this case, aerosols are not needed to explain the spectrum. For all three targets our self-consistent chemical approach (i.e., \textsc{FRECKLL}) requires some form of dis-equilibrium processes marked by the retrieved vertical mixing (Kzz $> 10^7$ cm$^2$s$^{-1}$), despite relatively high equilibrium temperatures. Other processes such as photo-chemistry---not included in this study---could also play a significant role and bias our interpretations. Retrieval sensitivity and their modeling assumptions should therefore be carefully tested to ensure robust constraints with JWST.

\subsection{Silicate aerosol feature: WASP-107\,b observations}

\begin{figure*}[!h]
\centering
    \includegraphics[width = 0.98\textwidth]{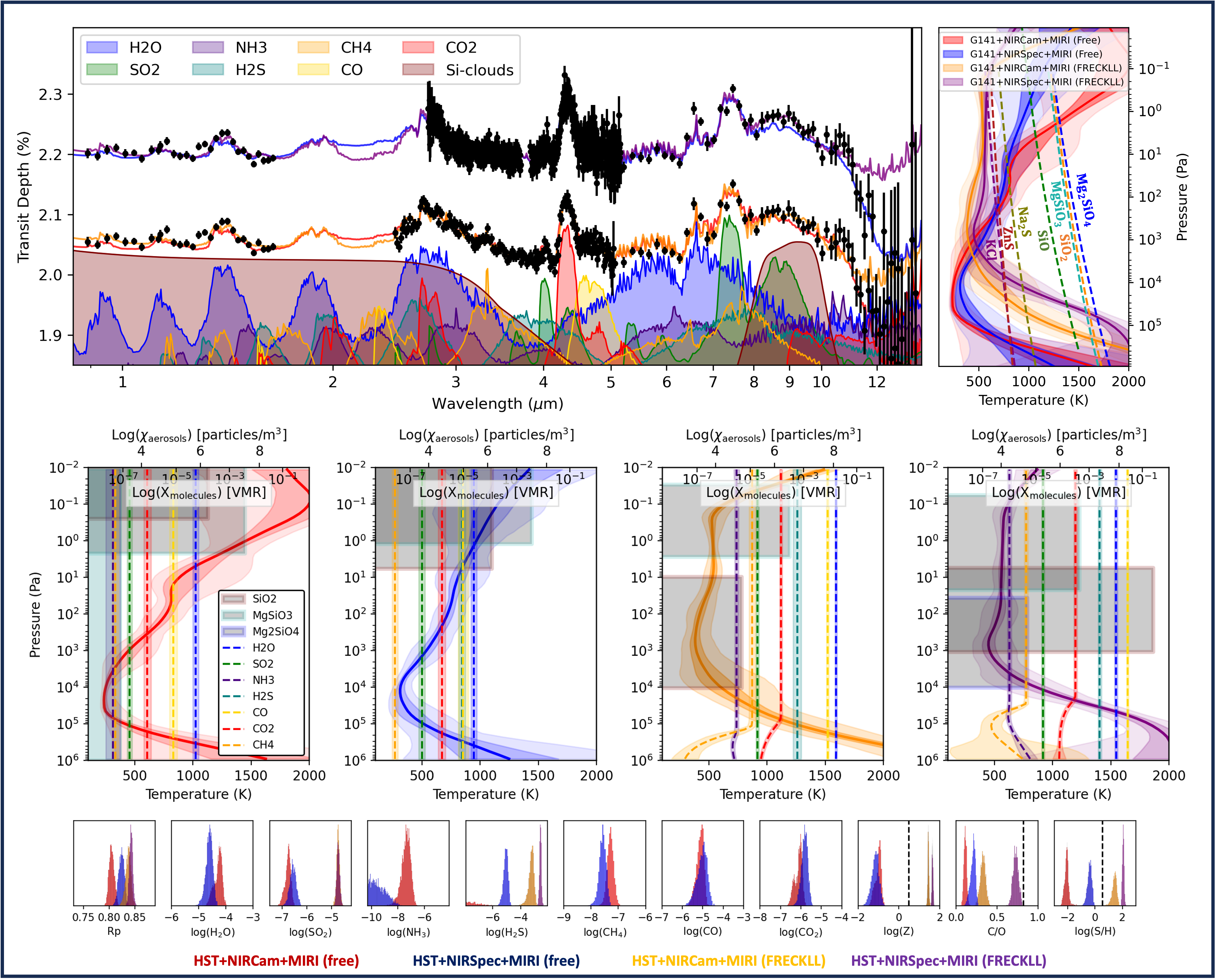}
    \caption{Summarized results of the retrievals on the WASP-107\,b HST and JWST data. The top left panel shows the observed spectra corrected for offsets for the \textsc{FRECKLL} retrievals (datapoints), the best-fits of free and \textsc{FRECKLL} retrievals (solid lines), and the contributions of the \textsc{FRECKLL} HST+NIRCam+MIRI retrieval (shaded areas). The top right panel shows the retrieved thermal structures including 1\,$\sigma$ and 3\,$\sigma$ confidence regions (shaded areas), as well as cloud condensation curves from \cite{Wetzel_2013, Gao_2021, Grant_2023}. The middle raw shows the chemistry and aerosol structure (SiO$_2$, MgSiO$_3$, and Mg$_2$SiO$_4$) for the four retrievals. The bottom row shows the retrieved probability density. The metallicity (Z) is directly retrieved in the \textsc{FRECKLL} retrievals, but estimated from the O/H (normalized by WASP-107 value) in the free retrievals. Overall the atmosphere of WASP-107\,b is consistent with high-altitude Si-clouds but the exact solution somewhat depend on the considered data (NIRSpec or NIRCam) and model assumptions. Figure \ref{fig:w107_zoom_sp}) shows incompatibility of the NIRSpec and NIRCam spectra, leading to slightly different solutions.}\label{fig:w107_spec}
\end{figure*}

For WASP-107\,b, recent works \citep{Dyrek_2023, Welbanks_2024, Sing_2024} have employed atmospheric retrievals and grid model fits to interpret the observations. \cite{Welbanks_2024} is the only work to date to have performed retrievals of the full data (including HST and MIRI data), so we compare more directly with their results\footnote{\cite{Welbanks_2024} tested two models: a self-consistent radiative-convective equilibrium with photochemistry using a two-stage grid interpolation fitting procedure; a free retrieval employing the six-parameter T--p profile from \cite{Madhu_retrieval_method}. Their approach is less flexible than our chosen free node $T-p$ approach \citep{changeat_2021_phasecurve2, Rowland_2023, Schleich_2024}. }. We note that they did not include cloud properties directly, and instead modeled the 10\,$\mu$m silicate Si-O stretch feature using a Gaussian-like extinction. Here, we aim to constrain the type of clouds present in WASP-107\,b's atmosphere, so we use \textsc{TauREx-PyMieScatt} to compute the extinction of various possible species: SiO, SiO$_2$, MgSiO$_3$, and Mg$_2$SiO$_4$. This is a significant departure from their study. The results of our retrievals, including spectra, $T-p$ structures, and probability distributions for the free parameters, are shown in Figure \ref{fig:w107_spec}. In Table \ref{tab:miri_retrievals}, we provide the quantiles from our different retrievals. Many of our results are consistent within the models and datasets we consider: detection of H$_2$O, SO$_2$, CH$_4$, CO, and CO$_2$, and presence of high-altitude ($p < 10^4$\,Pa) clouds composed of sub-micrometer SiO$_2$ particles. Other results, however, are model or dataset dependent, which is explained below:. \\

In the retrievals assuming free chemistry, the metallicity is sub-solar: log(H$_2$O)$\sim -4$\,dex. On the contrary, the \textsc{FRECKLL} chemistry retrievals favors a super-solar metallicity: Z$\in [20, 50]$x solar. Note again that the \textsc{FRECKLL} retrievals do not include S-species and photochemistry (use of reduced scheme), so SO$_2$ production described by \cite{Polman_2023, Tsai_2023, Dyrek_2023} is not included. The differences between free and self-consistent chemistry for this planet are likely driven by the red end of the MIRI data, which cannot be easily fitted using the less flexible \textsc{FRECKLL} retrievals. This is either because of the added chemical constraints from the model, or our assumption of one-dimensional aerosols. Regarding the latter point, our free chemistry runs allow for scattered clouds to exist---we find a cloud fraction of F$_a \in [0.4,0.5]$---while the \textsc{FRECKLL} runs do not. Retrieving the type of particles is possible ({\it but challenging}) due to the clearly visible 10\,$\mu$m Si feature, and as illustrated by Figure \ref{fig:w107_cloud_contrib}. In all our retrievals, SiO$_2$ clouds composed of sub-micrometer particles ($\mu < 0.7\,\mu$m) are detected. This is because the Si-O feature starts from 8$\mu$m, which can only be matched by the SiO$_2$ clouds in our model. Except for the (\textsc{FRECKLL}) retrieval with NIRSpec, these SiO$_2$ particles are also responsible for the grey opacity at lower wavelengths (i.e., in the HST data), however, the broader Si-O feature extending up to 11\,$\mu$m requires an additional component. In these retrievals, small ($\mu < 0.1\,\mu$m) MgSiO$_3$ particles located above the SiO$_2$ clouds can play this role. Processes that would lead to the formation of such cloud structures at the terminator of hot-Jupiters remains to be modeled self-consistently and are outside the scope of this work. In the NIRSpec (\textsc{FRECKLL}) run, the solution is different: the retrieval finds that a larger quantity of much smaller SiO$_2$ particles ($\mu \sim 0.02\,\mu$m) mixed with $\sim 0.5\,\mu$m Mg$_2$SiO$_4$ particles could also work well. In this case, the visible-light continuum is best explained by the Mg$_2$SiO$_4$ particles with SiO$_2$ contributing only to the enhanced absorption at $\sim 8\,\mu$m as seen in Figure \ref{fig:w107_cloud_contrib}. These degeneracies highlight the difficulties in extracting robust constraints for exoplanet aerosols, especially as many possible particle shape and types can contribute to the shape of the 10\,$\mu$m Si feature (see discussion). For instance, a study focusing on the HST+MIRI data only \citep{Dyrek_2023} explained the Si-O feature using SiO clouds only (see their \textsc{ARCiS} retrievals), albeit they also present results where SiO$_2$ clouds are favored (see their \textsc{petitRADTRANS} retrievals). Independently, Si-based condensates , likely including SiO$_2$ particles, are found by all the retrievals run in our study \cite[and also other works:][]{Dyrek_2023, Welbanks_2024, Sing_2024}. The retrievals suggest that these Si-clouds are located above their respective condensation points (see Figure \ref{fig:w107_spec}), which remain difficult to explain without invoking significant particle transport. In particular, the free retrievals show high altitude clouds (i.e, $p < 10\,$Pa) while the retrieved thermal profiles are inverted, crossing the condensation curves of silicates again (see Figure \ref{fig:w107_spec}). In this situation, sublimation of the cloud particles should efficiently remove clouds, which appear inconsistent with our findings. In the \textsc{FRECKLL} retrievals, the thermal inversion is not present but a hotter internal temperature seems needed, which was also suggested by the previous works of \cite{Welbanks_2024, Sing_2024}. In this case, the silicate condensation curves are crossed around $p \sim 0.1-1$\,bar, and significant mixing is still needed to lift the particles up to the region probed by the observations. This could be consistent with the retrieved vertical mixing of Kzz $\sim 10^9$\,cm$^2$s$^{-1}$, which is constrained by the dis-equilibrium chemical profiles. Other aspects not modeled here, including a more complex 3D structure at the terminator could play a significant role and further bias our inference \citep{Caldas_2019, Changeat_2020_phasecurve1, MacDonald_2020, Lacy_2020}. Note that our free chemistry results contrast with the interpretations from \cite{Dyrek_2023, Welbanks_2024}, where we find a lower abundance for most of the molecules, while our \textsc{FRECKLL} retrievals are consistent. \\

Another interesting finding of our study is the difference in solutions extracted from the NIRCam and NIRSpec datasets. The retrieved molecular abundances of the detected species in both dataset combinations can vary by up to $\pm 1$\,dex. For NH$_3$, however, the molecule is only detected when NIRCam is included. For H$_2$S, it is only detected when NIRSpec is included. These differences arise from incompatible spectral shape between the NIRCam-F322W2 and the NIRSpec-G395H NRS1 spectra (see Figure \ref{fig:w107_zoom_sp}), with the NIRCam data being steeper than the NIRSpec data at $\lambda < 3.7\,\mu$m. This could be explained by temporal variability of the planet's atmosphere (e.g., variability in the aerosol layers), but it is most likely due to systematic differences in the observing conditions \cite[i.e., light-source effects: ][]{Sing_2024} and/or the reduction of the data. Differences such as {\it offsets} have also been identified in recently analyzed observations of WASP-39\,b \cite[see e.g.,][]{Lueber_2024, Carter_2024}, while other studies have highlighted significant and poorly understood, wavelength dependent, long-term trends in NRS1 light-curves \citep{Espinoza_2023}. Recent works \citep{Moran_2023, Edwards_2024} have also suggested that systematic offsets could occur between the NRS1 and NRS2 detectors (here, we do not fit systematics between the two detectors). Presence of those instrumental systematic signals  can introduce biases, at minimum in the form of offsets, raising caution in interpreting combined datasets \cite[see:][]{Edwards_2024_K11}. For reference, our retrievals find offsets between HST, NIRSpec, NIRCam, and MIRI data of up to 250 ppm. Our results suggest that a large wavelength coverage is necessary to break the degeneracies between chemical and aerosol signatures, but also provides clues about the difficulties of combining JWST datasets. \\

Our retrieval exploration of the WASP-107\,b data suggests that interpretations of current JWST data can be subject to various biases. Some of those biases were clearly highlighted in this paper: choice of input data (cross-sections, aerosol optical properties), observations (i.e., wavelength coverage, data reduction), and importantly model assumptions (free vs self-consistent chemistry, aerosol modeling). We explore some of these aspects further in the next section.

\section{Discussion}\label{sec:disc}

\begin{figure*}
\centering
    \includegraphics[width = 0.98\textwidth]{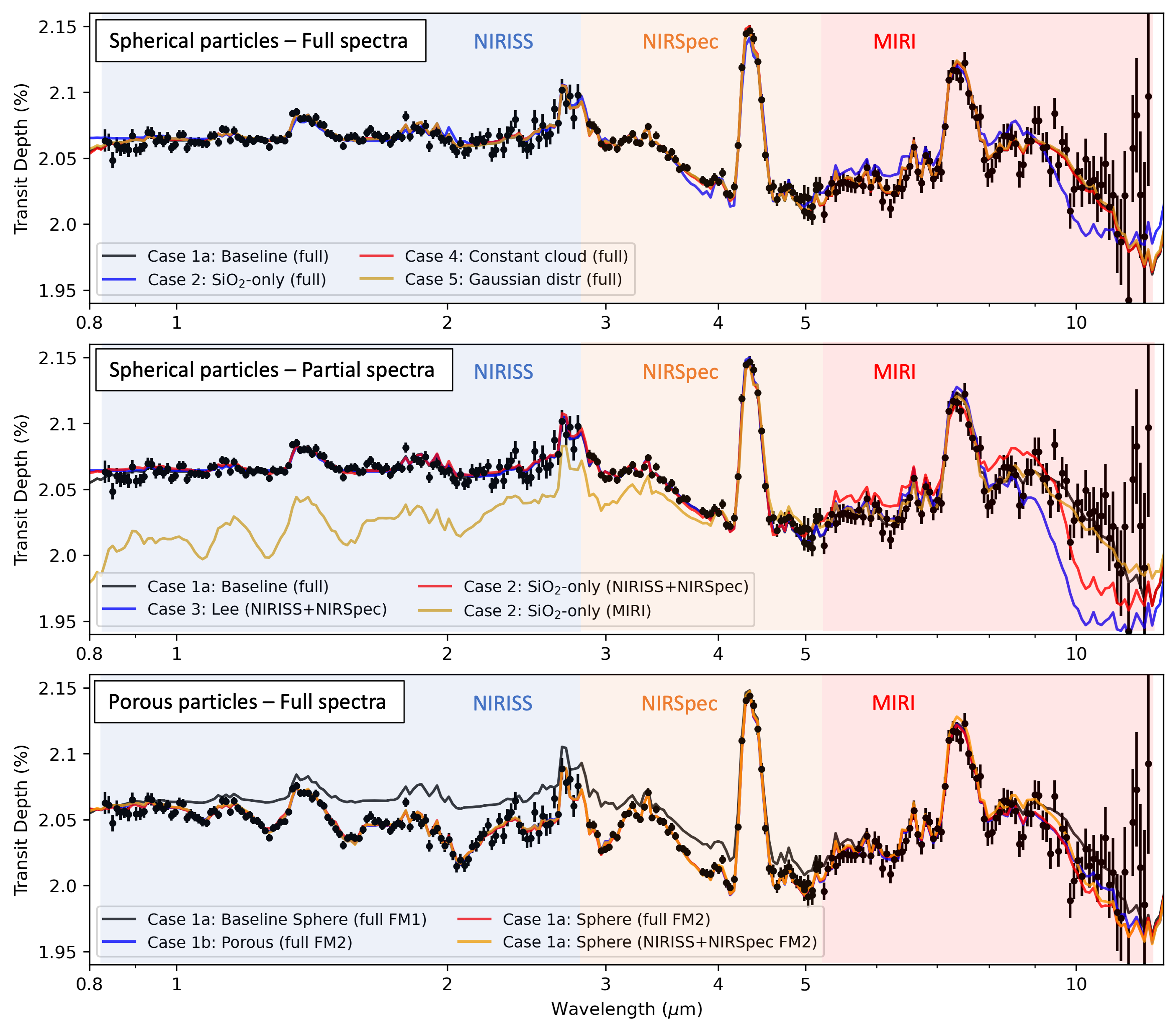}
    \caption{Simulated observations and retrieval best-fit models of our controlled experiments. The top panel shows retrievals on the full simulated spectrum with JWST/NIRISS, JWST/NIRSpec-G395H, and JWST/MIRI. The middle panel shows retrievals on subsets of the simulated data. Both panels use the forward models FM1, where aerosol particles are spherical. The bottom panel shows retrievals on the forward model FM2, where aerosol particles are 50\% porous. This ensemble of simulations (and corresponding posterior distributions in Figure \ref{fig:corner_jwst_simu} and Figure \ref{fig:corner_jwst_simu_porous}) illustrates the need for a wide wavelength coverage to constrain aerosol properties accurately. Retrievals restricted to a narrow wavelength coverage (NIRISS+NIRSpec) or (MIRI) are at risk of biases, especially when using incorrect refractive indexes (properties that are poorly known a-priori). Porosity is difficult to infer from JWST data (even in the case of full data), as it is degenerate with other parameters (i.e., models using spherical particles achieve good fits)}. 
    \label{fig:spec_jwst_simu}
\end{figure*}

We provide specific examples of cloud and haze retrievals with real JWST data. However, to build a more general intuition of the telescope's information content, additional controlled simulations are required. We utilize a synthetic hot Jupiter scenario based on the WASP-107\,b parameters. The atmosphere is assumed to be at chemical equilibrium \citep{Woitke_2018_GG} with a 20x solar atmosphere to which we add 10\,ppm of SO$_2$. This is motivated by recent JWST measurements for this planet \citep{Dyrek_2023, Welbanks_2024, Sing_2024}. A 10\,Pa aerosol layer of MgSiO$_3$ is added using an {\it exponentially decaying} profile with $\chi_\mathrm{max} = 10^5$ particles/m$^3$ and the \cite{Budaj_2015} distribution with $\mu_r = 0.5\,\mu$m. In Forward Model 1 (FM1), the particles are assumed spherical. In FM2, the particles have a porosity of 0.5 (see Appendix A). Following the methodology in \cite{Changeat_2019_2l}, the \textsc{TauREx} high-resolution spectrum from FM1 and FM2 are convolved with JWST instrument noise models \citep{Batalha_2017} for NIRISS-SOSS, NIRSpec-G395H, and MIRI-LRS using the instrument recommended setups (respectively using 3, 15, and 41 groups). Five retrieval cases are then investigated: \\ 
{\it -- Case 1a: Baseline with spheres}. It includes MgSiO$_3$ and SiO$_2$ clouds using spherical particles to test the feasibility of distinguishing aerosol species (i.e., it is a self-retrieval for FM1 but it was also attempted on FM2 to test the importance of the spherical particle assumption). \\
{\it -- Case 1b: Baseline with porous particles}. It includes MgSiO$_3$ and SiO$_2$ clouds using porous particles to test the feasibility of distinguishing aerosol species and their porosity with the FM2 spectrum (i.e., it is a self-retrieval for FM2). \\
{\it -- Case 2: SiO$_2$-only}. Same as Case 1a but without MgSiO$_3$ clouds. This case evaluates biases from using incorrect refractive indexes. \\
{\it -- Case 3: Lee clouds}. Same as Case 1a but using \citetalias{Lee_2013_clouds} prescription. \\
{\it -- Case 4: constant $\chi$}. Same as Case 1a but with a constant-with-altitude particle number density instead of the exponentially decaying profile. \\
{\it -- Case 5: log-normal particle size}. Same as Case 1a but assumes a Log-Gaussian particle distribution $n(r)$. \\
Note that only Case 1b uses porous particles, all the other cases assume spherical particles. Those cases are specifically designed to test the sensitivity of JWST to various important cloud properties. Since the full set of JWST observations (i.e., from $\lambda \in [0.8, 12]\,\mu$m) are not always available, or because temporal variability could render the datasets incompatible \citep{Cho_1996, Cho_2003, Skinner_2022_cyclogenesis, Changeat_2024}, we also produce Case 1a, Case 2, and Case 3 for the NIRISS+NirSpec data only and the MIRI data only. Figure \ref{fig:spec_jwst_simu} shows the simulated data and retrieved spectra, while the corner plots for relevant cases are available in Figure \ref{fig:corner_jwst_simu}. 

\subsection{Biases from assuming incorrect aerosol types}

Without prior knowledge from theoretical modeling, assuming incorrect cloud type could lead to retrieval biases. Comparing the baseline (Case 1a) and SiO$_2$-only (Case 2) cases, we conclude that incorrect refractive indexes or missing cloud species in retrievals generally lead to wrong inference of the chemistry, thermal structure, and cloud properties. Importantly, this issue (i.e., using the wrong refractive indexes) is only visible in the fit when the full wavelength coverage is available. When using the NIRISS+NIRSpec data only or the MIRI data only, the recovered parameters are incorrect---for instance Z is about 10$\sigma$ off in the NIRISS+NIRSpec case, while log(SO$_2$) is 4$\sigma$ off in the MIRI-only case---but the best-fit spectra in the region of the observations remain very similar. Due to this behaviour, narrow wavelength coverage (i.e., when simultaneous visible and infrared data is not available) should be interpreted with caution when prior knowledge on the cloud composition is not available. However, we note that a more heuristic approach (e.g., Case 3 using \citetalias{Lee_2013_clouds}) can be adapted when strong cloud absorption features are not present in the data (e.g., in the JWST/NIRISS cases studied in Section 4), leading to robust, unbiased results. For instance, the Case 3 \citetalias{Lee_2013_clouds} retrieval on NIRISS+NIRSpec obtains unbiased chemical parameters and relevant cloud properties: Z $= 22 \pm 2$, C/O $= 0.50\pm 0.05$, log(SO$_2$) = $-4.9 \pm 0.1$, log($P$) = $1.00 \pm 0.06$, log($\mu_r$) = $-1.0 \pm 0.5$. This is in line with the findings of Section \ref{sec:jwst} on the NIRISS data.

\subsection{Aerosol particle shape assumption}

Recent studies suggested the potential importance of particle porosity in interpreting atmospheric observations of exoplanets \citep{Adams_2019, Ohno_2020, Samra_2020, Lodge_2024, Vahidinia_2024}. In our examples, we assumed for simplicity that the aerosol particles are filled and spherical. Previous works \citep{ Fabian_2001, Min_2007, Akimasa_2014, Kiefer_2024} have demonstrated significant changes in extinction when more complex particle shapes (e.g., porous spheres, distribution of particles) are assumed. To illustrate such difficulties, we have tested Case 1b, which is the same as Case 1a but with porous particles. Figure \ref{fig:spec_jwst_simu} and Figure \ref{fig:corner_jwst_simu_porous} shows how this assumption impacts the forward model spectra (compare FM1 and FM2). The retrievals assuming spherical particles are able to explain the FM2 spectrum (when porous particles are in the forward model) convincingly, but the retrieved parameters are biased: the particles size and number density is incorrect, which also impacts the retrieved metallicity (Z). Intrinsic degeneracies exist, making it difficult to extract particle porosity. Such issues are also likely to occur when it comes to the particle arrangements (e.g., amorphous, crystalline), potentially leading to significant challenges in our ability to interpret the 10$\mu$m feature of silicate clouds in hot Jupiters. 

\subsection{Cloud information content in JWST data}

Comparing Case 1a to cases 4 and 5, we can evaluate the sensitivity of JWST to the vertical aerosol profile and particle radius distribution. Case 4 (when $\chi$ is constant with altitude) does not display significant differences compared to our baseline scenario, highlighting the poor sensitivity of WASP-107\,b-like data to vertical aerosol profile. In Case 5, however, we attempt to recover the particle size distribution directly, assuming it is normally distributed. We find in Figure \ref{fig:corner_jwst_simu} a very strong inverse {\it linear} correlation between the posterior distributions of $\mu_\mathrm{c}$ and $\sigma_\mathrm{c}$ in log space. The correlation indicates that the radiative contributions of aerosols in transit geometry are dominated by the largest particles of the distribution (i.e., information on the full distribution and its shape is redundant). From our results, detailed estimates of the vertical aerosol distribution and the particle size distributions seem difficult with JWST, meaning that simplified parameterizations for those factors might be sufficient to capture most of the available information.

\subsection{On synergies between micro-physical cloud models and information retrievals}

In this study, we quantify the sensitivity of JWST to particular properties of aerosols. For instance, we unveil that the actual particle radius distribution and the vertical aerosol abundance are second order properties affecting the observed spectra, and could be simplified or parameterized when comparing with current observations. This is important information as for teams developing complex microphysical cloud models \cite[e.g., ][]{Helling+08,Ohno&Okuzumi18,Gao&Benneke18,Powell+18,Ohno+20_fluffy,Samra+22,Lee23_mini-cloud,Powell&Zhang24} this allows to {more easily identify} the properties that could affect observables. Similar efforts have already been conducted for chemical properties \citep{Changeat_2019_2l, al-refaie_2021_taurex3.1, Al-Refaie_2022_FRECKLL}, or planetary mass \citep{Changeat_2020_mass}, providing relevant guidelines for parameterization and modeling in atmospheric retrievals. In this context, the development of information oriented approaches (i.e., free retrievals) should idealistically be done in conjunction with self-consistent approaches, with iterations until both techniques agree (i.e., the model complexity is adapted, and our current knowledge of physical processes is correct). For aerosols, such studies are just starting, motivated by recent access to high quality data (i.e., JWST) and recent advancements in self-consistent cloud models for retrieval applications \citep{Ormel&Min19,Min_2020, Ma_2023}. For chemistry, differences between our free and \textsc{FRECKLL} retrievals, potentially highlight that such level of agreement has not yet been reached for chemical models relevant to JWST data. 

\section{Conclusion}\label{sec:con}

In this work, we parameterized cloud and haze properties to explore the information content of JWST exoplanet spectra. We first evaluate the strategy on a solar-system example of Titan, observed in occultation by the {\it Cassini} spacecraft. We then perform similar atmospheric retrieval tests on recent JWST exoplanet data from the JWST/NIRISS and the JWST/MIRI instruments. We explored various retrieval assumptions using the \textsc{TauREx3} retrieval code. Notably, we performed retrievals beyond the common chemical equilibrium assumption using the kinetic chemistry code \textsc{FRECKLL} on JWST data for the first time. Our findings, also supported by additional controlled experiments using simulated data, confirm that characterizing the physical properties of aerosols (i.e., understanding their nature, size distribution, and abundances) requires a wide wavelength coverage as also suggested in other studies \cite[see e.g., ][]{Lee_2014, Wakeford_2015, Pinhas_2017, Mai_2019, Kawashima_2019, Lacy_2020_b, Gao_2021}. Specifically, to minimally constrain aerosols, observations need to be sensitive to {\it both} the visible light scattering slope and longer wavelength resonance features (e.g., the 10$\mu$m Si-O stretch). Without the combined information from those spectral regions---and in absence of further priors---the aerosol solution can be degenerate and easily lead to incorrect conclusions. Breaking this degeneracy with JWST alone may be a complicated task since the telescope does not possess an instrument covering those wavelengths simultaneously. When attempting to combine observations from different epochs to construct such a wavelength coverage, we noted incompatibilities in the data (i.e., WASP-107\,b NIRSpec and NIRISS) that might prevent us from using this strategy efficiently. Possible sources for those discrepancies are either the time-dependence of the observing conditions (i.e., changing instrument systematics, stellar variability) and/or the time-dependence of the astrophysical signal itself (exoplanet weather). To make further progress in this area, we suggest that synergies with other instruments (HST, ground-based, Ariel) should be explored to provide the required simultaneous coverage and mitigate the biases from repeated visits. 
Our investigations also highlighted important retrieval challenges inherently linked to modeling assumptions (i.e., free chemistry versus kinetic chemistry, aerosol model assumptions) and the assumed input sources (cross-sections, aerosol optical data). 

\section*{Data Availability}

The atmospheric retrieval code \textsc{TauREx3} is open source and available at: \url{https://github.com/ucl-exoplanets/TauREx3}. The \textsc{TauREx3} plugins developed and used for this article are also publicly available on Github: \textsc{TauREx-PyMieScatt} at \url{https://github.com/groningen-exoatmospheres/taurex-pymiescatt}, \textsc{TauREx-MultiModel} at \url{https://github.com/groningen-exoatmospheres/taurex-multimodel}, and \textsc{TauREx-InstrumentSystematics} at \url{https://github.com/groningen-exoatmospheres/taurex-instrumentsystematics}. The opacity sources that were compiled for this article are compatible with the \textsc{TauREx3} framework and can be found in a Zenodo repository: \url{https://doi.org/10.5281/zenodo.15495830}.
This work is based upon observations with the NASA/ESA/CSA James Webb Space Telescope, obtained at the Space Telescope Science Institute (STScI) operated by AURA, Inc. The raw data used in this work are available as part of the Mikulski Archive for Space Telescopes. We are thankful to those who operate these telescopes and their corresponding archives, the public nature of which increases scientific productivity and accessibility \citep{Peek_2019}.

\begin{acknowledgements}

We thank the anonymous referee for their useful comments that significantly improved our manuscript. \\

This publication is part of the project "Interpreting exoplanet atmospheres with JWST" with file number 2024.034 of the research programme "Rekentijd nationale computersystemen" that is (partly) funded by the Netherlands Organisation for Scientific Research (NWO) under grant \url{https://doi.org/10.61686/QXVQT85756}. This work used the Dutch national e-infrastructure with the support of the SURF Cooperative using grant no. 2024.034. DB and OV acknowledge funding from Agence Nationale de la Recherche (ANR), project ``EXACT'' (ANR-21-CE49-0008-01). \\

We also acknowledge the availability and support from the High-Performance Computing platforms (HPC) from DIRAC, and OzSTAR, which provided the computing resources necessary to perform this work. This work utilised the Cambridge Service for Data-Driven Discovery (CSD3), part of which is operated by the University of Cambridge Research Computing on behalf of the STFC DiRAC HPC Facility (www.dirac.ac.uk). The DiRAC component of CSD3 was funded by BEIS capital funding via STFC capital grants ST/P002307/1 and ST/R002452/1 and STFC operations grant ST/R00689X/1. DiRAC is part of the National e-Infrastructure. This work utilised the OzSTAR national facility at Swinburne University of Technology. The OzSTAR program receives funding in part from the Astronomy National Collaborative Research Infrastructure Strategy (NCRIS) allocation provided by the Australian Government.
\end{acknowledgements}

\bibliographystyle{aasjournal}
\bibliography{main}   

\begin{thebibliography}{}
\expandafter\ifx\csname natexlab\endcsname\relax\def\natexlab#1{#1}\fi
\providecommand{\url}[1]{\href{#1}{#1}}
\providecommand{\dodoi}[1]{doi:~\href{http://doi.org/#1}{\nolinkurl{#1}}}
\providecommand{\doeprint}[1]{\href{http://ascl.net/#1}{\nolinkurl{http://ascl.net/#1}}}
\providecommand{\doarXiv}[1]{\href{https://arxiv.org/abs/#1}{\nolinkurl{https://arxiv.org/abs/#1}}}

\bibitem[{Abel {et~al.}(2011)Abel, Frommhold, Li, \& Hunt}]{abel_h2-h2}
Abel, M., Frommhold, L., Li, X., \& Hunt, K.~L. 2011, The Journal of Physical Chemistry A, 115, 6805

\bibitem[{Abel {et~al.}(2012)Abel, Frommhold, Li, \& Hunt}]{abel_h2-he}
---. 2012, The Journal of chemical physics, 136, 044319

\bibitem[{{{\'A}d{\'a}mkovics} {et~al.}(2016){{\'A}d{\'a}mkovics}, {Mitchell}, {Hayes}, {Rojo}, {Corlies}, {Barnes}, {Ivanov}, {Brown}, {Baines}, {Buratti}, {Clark}, {Nicholson}, \& {Sotin}}]{Adamkovics_2016}
{{\'A}d{\'a}mkovics}, M., {Mitchell}, J.~L., {Hayes}, A.~G., {et~al.} 2016, \icarus, 270, 376, \dodoi{10.1016/j.icarus.2015.05.023}

\bibitem[{{Adams} {et~al.}(2019){Adams}, {Gao}, {de Pater}, \& {Morley}}]{Adams_2019}
{Adams}, D., {Gao}, P., {de Pater}, I., \& {Morley}, C.~V. 2019, \apj, 874, 61, \dodoi{10.3847/1538-4357/ab074c}

\bibitem[{{Al Derzi} {et~al.}(2015){Al Derzi}, {Furtenbacher}, {Tennyson}, {Yurchenko}, \& {Cs{\'a}sz{\'a}r}}]{al-derzi_2015_nh3}
{Al Derzi}, A.~R., {Furtenbacher}, T., {Tennyson}, J., {Yurchenko}, S.~N., \& {Cs{\'a}sz{\'a}r}, A.~G. 2015, Journal of Quantitative Spectroscopy and Radiative Transfer, 161, 117, \dodoi{10.1016/j.jqsrt.2015.03.034}

\bibitem[{{Al-Refaie} {et~al.}(2022){Al-Refaie}, {Changeat}, {Venot}, {Waldmann}, \& {Tinetti}}]{al-refaie_2021_taurex3.1}
{Al-Refaie}, A.~F., {Changeat}, Q., {Venot}, O., {Waldmann}, I.~P., \& {Tinetti}, G. 2022, \apj, 932, 123, \dodoi{10.3847/1538-4357/ac6dcd}

\bibitem[{{Al-Refaie} {et~al.}(2021){Al-Refaie}, {Changeat}, {Waldmann}, \& {Tinetti}}]{2019_al-refaie_taurex3}
{Al-Refaie}, A.~F., {Changeat}, Q., {Waldmann}, I.~P., \& {Tinetti}, G. 2021, The Astrophysical Journal, 917, 37, \dodoi{10.3847/1538-4357/ac0252}

\bibitem[{{Al-Refaie} {et~al.}(2024){Al-Refaie}, {Venot}, {Changeat}, \& {Edwards}}]{Al-Refaie_2022_FRECKLL}
{Al-Refaie}, A.~F., {Venot}, O., {Changeat}, Q., \& {Edwards}, B. 2024, \apj, 967, 132, \dodoi{10.3847/1538-4357/ad3dee}

\bibitem[{{Allard} {et~al.}(2016){Allard}, {Spiegelman}, \& {Kielkopf}}]{Allard_2016_K}
{Allard}, N.~F., {Spiegelman}, F., \& {Kielkopf}, J.~F. 2016, \aap, 589, A21, \dodoi{10.1051/0004-6361/201628270}

\bibitem[{{Allard} {et~al.}(2019){Allard}, {Spiegelman}, {Leininger}, \& {Molliere}}]{Allard_2019_Na}
{Allard}, N.~F., {Spiegelman}, F., {Leininger}, T., \& {Molliere}, P. 2019, \aap, 628, A120, \dodoi{10.1051/0004-6361/201935593}

\bibitem[{{Anderson} {et~al.}(2018){Anderson}, {Samuelson}, \& {Nna-Mvondo}}]{anderson_2018_titan}
{Anderson}, C.~M., {Samuelson}, R.~E., \& {Nna-Mvondo}, D. 2018, \ssr, 214, 125, \dodoi{10.1007/s11214-018-0559-5}

\bibitem[{{Anisman} {et~al.}(2022){Anisman}, {Chubb}, {Elsey}, {Al-Refaie}, {Changeat}, {Yurchenko}, {Tennyson}, \& {Tinetti}}]{Anisman_2022}
{Anisman}, L.~O., {Chubb}, K.~L., {Elsey}, J., {et~al.} 2022, \jqsrt, 278, 108013, \dodoi{10.1016/j.jqsrt.2021.108013}

\bibitem[{{Arfaux} \& {Lavvas}(2024)}]{Arfaux_2024}
{Arfaux}, A., \& {Lavvas}, P. 2024, \mnras, 530, 482, \dodoi{10.1093/mnras/stae826}

\bibitem[{{Asplund} {et~al.}(2009){Asplund}, {Grevesse}, {Sauval}, \& {Scott}}]{Asplund_2009}
{Asplund}, M., {Grevesse}, N., {Sauval}, A.~J., \& {Scott}, P. 2009, \araa, 47, 481, \dodoi{10.1146/annurev.astro.46.060407.145222}

\bibitem[{{Atreya} \& {Wong}(2005)}]{Atreya_2005}
{Atreya}, S., \& {Wong}, A. 2005, Space Sci Rev, 116, 121, \dodoi{10.1007/s11214-005-1951-5}

\bibitem[{{Atreya} {et~al.}(2022){Atreya}, {Crida}, {Guillot}, {Li}, {Lunine}, {Madhusudhan}, {Mousis}, \& {Wong}}]{Atreya_2022}
{Atreya}, S.~K., {Crida}, A., {Guillot}, T., {et~al.} 2022, arXiv e-prints, arXiv:2205.06914, \dodoi{10.48550/arXiv.2205.06914}

\bibitem[{{Azzam} {et~al.}(2016){Azzam}, {Tennyson}, {Yurchenko}, \& {Naumenko}}]{Azzam_2016_h2s}
{Azzam}, A. A.~A., {Tennyson}, J., {Yurchenko}, S.~N., \& {Naumenko}, O.~V. 2016, Monthly Notices of the Royal Astronomical Society, 460, 4063, \dodoi{10.1093/mnras/stw1133}

\bibitem[{Barber {et~al.}(2013)Barber, Strange, Hill, Polyansky, Mellau, Yurchenko, \& Tennyson}]{Barber_2013_HCN}
Barber, R.~J., Strange, J.~K., Hill, C., {et~al.} 2013, Monthly Notices of the Royal Astronomical Society, 437, 1828–1835, \dodoi{10.1093/mnras/stt2011}

\bibitem[{{Barth}(2017)}]{Barth_2017}
{Barth}, E.~L. 2017, \planss, 137, 20, \dodoi{10.1016/j.pss.2017.01.003}

\bibitem[{Batalha \& Marley(2020)}]{natasha_batalha_2020_5179187}
Batalha, N., \& Marley, M. 2020, Refractive Indices For Virga Exoplanet Cloud Model, 1.2,  Zenodo, \dodoi{10.5281/zenodo.5179187}

\bibitem[{{Batalha} {et~al.}(2017){Batalha}, {Mandell}, {Pontoppidan}, {Stevenson}, {Lewis}, {Kalirai}, {Earl}, {Greene}, {Albert}, \& {Nielsen}}]{Batalha_2017}
{Batalha}, N.~E., {Mandell}, A., {Pontoppidan}, K., {et~al.} 2017, \pasp, 129, 064501, \dodoi{10.1088/1538-3873/aa65b0}

\bibitem[{{Bellucci} {et~al.}(2009){Bellucci}, {Sicardy}, {Drossart}, {Rannou}, {Nicholson}, {Hedman}, {Baines}, \& {Burrati}}]{Bellucci_2009}
{Bellucci}, A., {Sicardy}, B., {Drossart}, P., {et~al.} 2009, \icarus, 201, 198, \dodoi{10.1016/j.icarus.2008.12.024}

\bibitem[{Bohren \& Huffman(2008)}]{Bohren_2008}
Bohren, C.~F., \& Huffman, D.~R. 2008, Absorption and scattering of light by small particles (John Wiley \& Sons)

\bibitem[{{Borysow} \& {Frommhold}(1987)}]{Borysow_1987}
{Borysow}, A., \& {Frommhold}, L. 1987, \apj, 318, 940, \dodoi{10.1086/165426}

\bibitem[{{Borysow} \& {Tang}(1993)}]{Borysow_1993}
{Borysow}, A., \& {Tang}, C. 1993, \icarus, 105, 175, \dodoi{10.1006/icar.1993.1117}

\bibitem[{Bruggeman(1935)}]{Bruggeman_1935}
Bruggeman, V.~D. 1935, Annalen der physik, 416, 636

\bibitem[{{Budaj} {et~al.}(2015){Budaj}, {Kocifaj}, {Salmeron}, \& {Hubeny}}]{Budaj_2015}
{Budaj}, J., {Kocifaj}, M., {Salmeron}, R., \& {Hubeny}, I. 2015, \mnras, 454, 2, \dodoi{10.1093/mnras/stv1711}

\bibitem[{{Cabane} {et~al.}(1993){Cabane}, {Rannou}, {Chassefiere}, \& {Israel}}]{Cabane+93}
{Cabane}, M., {Rannou}, P., {Chassefiere}, E., \& {Israel}, G. 1993, \planss, 41, 257, \dodoi{10.1016/0032-0633(93)90021-S}

\bibitem[{Caldas {et~al.}(2019)Caldas, Leconte, Selsis, Waldmann, Bordé, Rocchetto, \& Charnay}]{Caldas_2019}
Caldas, A., Leconte, J., Selsis, F., {et~al.} 2019, Astronomy \& Astrophysics, 623, A161, \dodoi{10.1051/0004-6361/201834384}

\bibitem[{{Carter} {et~al.}(2024){Carter}, {May}, {Espinoza}, {Welbanks}, {Ahrer}, {Alderson}, {Brahm}, {Feinstein}, {Grant}, {Line}, {Morello}, {O'Steen}, {Radica}, {Rustamkulov}, {Stevenson}, {Turner}, {Alam}, {Anderson}, {Batalha}, {Battley}, {Bayliss}, {Bean}, {Benneke}, {Berta-Thompson}, {Brande}, {Bryant}, {Burleigh}, {Coulombe}, {Crossfield}, {Damiano}, {D{\'e}sert}, {Flagg}, {Gill}, {Inglis}, {Kirk}, {Knutson}, {Kreidberg}, {L{\'o}pez Morales}, {Mansfield}, {Moran}, {Murray}, {Nixon}, {Petit dit de la Roche}, {Rackham}, {Schlawin}, {Sing}, {Wakeford}, {Wallack}, {Wheatley}, {Zieba}, {Aggarwal}, {Barstow}, {Bell}, {Blecic}, {Caceres}, {Crouzet}, {Cubillos}, {Daylan}, {de Val-Borro}, {Decin}, {Fortney}, {Gibson}, {Heng}, {Hu}, {Kempton}, {Lagage}, {Lothringer}, {Lustig-Yaeger}, {Mancini}, {Mayne}, {Mayorga}, {Molaverdikhani}, {Nasedkin}, {Ohno}, {Parmentier}, {Powell}, {Redfield}, {Roy}, {Taylor}, \& {Zhang}}]{Carter_2024}
{Carter}, A.~L., {May}, E.~M., {Espinoza}, N., {et~al.} 2024, Nature Astronomy, 8, 1008, \dodoi{10.1038/s41550-024-02292-x}

\bibitem[{Changeat(2025)}]{changeat_2025_15495830}
Changeat, Q. 2025, Reference HDF5 cross-sections for TauREx3,  Zenodo, \dodoi{10.5281/zenodo.15495830}

\bibitem[{{Changeat} \& {Al-Refaie}(2020)}]{Changeat_2020_phasecurve1}
{Changeat}, Q., \& {Al-Refaie}, A. 2020, The Astrophysical Journal, 898, 155, \dodoi{10.3847/1538-4357/ab9b82}

\bibitem[{{Changeat} {et~al.}(2021){Changeat}, {Al-Refaie}, {Edwards}, {Waldmann}, \& {Tinetti}}]{changeat_2021_phasecurve2}
{Changeat}, Q., {Al-Refaie}, A.~F., {Edwards}, B., {Waldmann}, I.~P., \& {Tinetti}, G. 2021, The Astrophysical Journal, 913, 73, \dodoi{10.3847/1538-4357/abf2bb}

\bibitem[{{Changeat} {et~al.}(2020{\natexlab{a}}){Changeat}, {Edwards}, {Al-Refaie}, {Morvan}, {Tsiaras}, {Waldmann}, \& {Tinetti}}]{Changeat_2020_K11}
{Changeat}, Q., {Edwards}, B., {Al-Refaie}, A.~F., {et~al.} 2020{\natexlab{a}}, The Astronomical Journal, 160, 260, \dodoi{10.3847/1538-3881/abbe12}

\bibitem[{Changeat {et~al.}(2019)Changeat, Edwards, Waldmann, \& Tinetti}]{Changeat_2019_2l}
Changeat, Q., Edwards, B., Waldmann, I.~P., \& Tinetti, G. 2019, The Astrophysical Journal, 886, 39, \dodoi{10.3847/1538-4357/ab4a14}

\bibitem[{{Changeat} {et~al.}(2020{\natexlab{b}}){Changeat}, {Keyte}, {Waldmann}, \& {Tinetti}}]{Changeat_2020_mass}
{Changeat}, Q., {Keyte}, L., {Waldmann}, I.~P., \& {Tinetti}, G. 2020{\natexlab{b}}, The Astrophysical Journal, 896, 107, \dodoi{10.3847/1538-4357/ab8f8b}

\bibitem[{{Changeat} {et~al.}(2024){Changeat}, {Skinner}, {Cho}, {N{\"a}ttil{\"a}}, {Waldmann}, {Al-Refaie}, {Dyrek}, {Edwards}, {Mikal-Evans}, {Joshua}, {Morello}, {Skaf}, {Tsiaras}, {Venot}, \& {Yip}}]{Changeat_2024}
{Changeat}, Q., {Skinner}, J.~W., {Cho}, J.~Y.~K., {et~al.} 2024, \apjs, 270, 34, \dodoi{10.3847/1538-4365/ad1191}

\bibitem[{{Chistikov} {et~al.}(2019){Chistikov}, {Finenko}, {Lokshtanov}, {Petrov}, \& {Vigasin}}]{Chistikov_2019}
{Chistikov}, D.~N., {Finenko}, A.~A., {Lokshtanov}, S.~E., {Petrov}, S.~V., \& {Vigasin}, A.~A. 2019, \jcp, 151, 194106, \dodoi{10.1063/1.5125756}

\bibitem[{{Cho} \& {Polvani}(1996)}]{Cho_1996}
{Cho}, J. Y.~K., \& {Polvani}, L.~M. 1996, Physics of Fluids, 8, 1531, \dodoi{10.1063/1.868929}

\bibitem[{Cho {et~al.}(2003)Cho, Menou, Hansen, \& Seager}]{Cho_2003}
Cho, J.~{\relax Y-K}., Menou, K., Hansen, B. M.~S., \& Seager, S. 2003, The Astrophysical Journal, 587, L117–L120, \dodoi{10.1086/375016}

\bibitem[{{Chubb} {et~al.}(2020){Chubb}, {Tennyson}, \& {Yurchenko}}]{Chubb_2020_c2h2}
{Chubb}, K.~L., {Tennyson}, J., \& {Yurchenko}, S.~N. 2020, Monthly Notices of the Royal Astronomical Society, 493, 1531, \dodoi{10.1093/mnras/staa229}

\bibitem[{Chubb {et~al.}(2021)Chubb, Rocchetto, Yurchenko, Min, Waldmann, Barstow, Mollière, Al-Refaie, Phillips, \& Tennyson}]{Chubb_2021_exomol}
Chubb, K.~L., Rocchetto, M., Yurchenko, S.~N., {et~al.} 2021, Astronomy \& Astrophysics, 646, A21, \dodoi{10.1051/0004-6361/202038350}

\bibitem[{{Chubb} {et~al.}(2024){Chubb}, {Robert}, {Sousa-Silva}, {Yurchenko}, {Allard}, {Boudon}, {Buldyreva}, {Bultel}, {Coustenis}, {Foltynowicz}, {Gordon}, {Hargreaves}, {Helling}, {Hill}, {Hrodmarsson}, {Karman}, {Lecoq-Molinos}, {Migliorini}, {Rey}, {Richard}, {Sadiek}, {Schmidt}, {Sokolov}, {Stefani}, {Tennyson}, {Venot}, {Wright}, {Arenales-Lope}, {Barstow}, {Bocchieri}, {Carrasco}, {Dubey}, {Egorov}, {Mu{\~n}oz}, {Gharib-Nezhad}, {Gkouvelis}, {Gr{\"u}bel}, {Irwin}, {Kn{\'\i}{\v{z}}ek}, {Lewis}, {Lodge}, {Ma}, {Martins}, {Molaverdikhani}, {Morello}, {Nikitin}, {Panek}, {Rengel}, {Rinaldi}, {Skinner}, {Tinetti}, {van Kempen}, {Yang}, \& {Zingales}}]{Chubb_2024}
{Chubb}, K.~L., {Robert}, S., {Sousa-Silva}, C., {et~al.} 2024, RAS Techniques and Instruments, 3, 636, \dodoi{10.1093/rasti/rzae039}

\bibitem[{{Coles} {et~al.}(2019){Coles}, {Yurchenko}, \& {Tennyson}}]{Coles_2019_nh3}
{Coles}, P.~A., {Yurchenko}, S.~N., \& {Tennyson}, J. 2019, Monthly Notices of the Royal Astronomical Society, 490, 4638, \dodoi{10.1093/mnras/stz2778}

\bibitem[{{Col{\'o}n} {et~al.}(2020){Col{\'o}n}, {Kreidberg}, {Welbanks}, {Line}, {Madhusudhan}, {Beatty}, {Tamburo}, {Stevenson}, {Mandell}, {Rodriguez}, {Barclay}, {Lopez}, {Stassun}, {Angerhausen}, {Fortney}, {James}, {Pepper}, {Ahlers}, {Plavchan}, {Awiphan}, {Kotnik}, {McLeod}, {Murawski}, {Chotani}, {LeBrun}, {Matzko}, {Rea}, {Vidaurri}, {Webster}, {Williams}, {Cox}, {Tan}, \& {Gilbert}}]{Colon_2020}
{Col{\'o}n}, K.~D., {Kreidberg}, L., {Welbanks}, L., {et~al.} 2020, \aj, 160, 280, \dodoi{10.3847/1538-3881/abc1e9}

\bibitem[{{Constantinou} {et~al.}(2023){Constantinou}, {Madhusudhan}, \& {Gandhi}}]{Constantinou_2023}
{Constantinou}, S., {Madhusudhan}, N., \& {Gandhi}, S. 2023, \apjl, 943, L10, \dodoi{10.3847/2041-8213/acaead}

\bibitem[{{Cours} {et~al.}(2020){Cours}, {Cordier}, {Seignovert}, {Maltagliati}, \& {Biennier}}]{Cours_2020}
{Cours}, T., {Cordier}, D., {Seignovert}, B., {Maltagliati}, L., \& {Biennier}, L. 2020, \icarus, 339, 113571, \dodoi{10.1016/j.icarus.2019.113571}

\bibitem[{{Coustenis} {et~al.}(2003){Coustenis}, {Salama}, {Schulz}, {Ott}, {Lellouch}, {Encrenaz}, {Gautier}, \& {Feuchtgruber}}]{Coustenis_2003}
{Coustenis}, A., {Salama}, A., {Schulz}, B., {et~al.} 2003, \icarus, 161, 383, \dodoi{10.1016/S0019-1035(02)00028-3}

\bibitem[{{Coustenis} {et~al.}(2010){Coustenis}, {Jennings}, {Nixon}, {Achterberg}, {Lavvas}, {Vinatier}, {Teanby}, {Bjoraker}, {Carlson}, {Piani}, {Bampasidis}, {Flasar}, \& {Romani}}]{Coustenis_2010}
{Coustenis}, A., {Jennings}, D.~E., {Nixon}, C.~A., {et~al.} 2010, \icarus, 207, 461, \dodoi{10.1016/j.icarus.2009.11.027}

\bibitem[{{Coustenis} {et~al.}(2016){Coustenis}, {Jennings}, {Achterberg}, {Bampasidis}, {Lavvas}, {Nixon}, {Teanby}, {Anderson}, {Cottini}, \& {Flasar}}]{Coustenis_2016}
{Coustenis}, A., {Jennings}, D.~E., {Achterberg}, R.~K., {et~al.} 2016, \icarus, 270, 409, \dodoi{10.1016/j.icarus.2015.08.027}

\bibitem[{Cox(2015)}]{cox_allen_rayleigh}
Cox, A.~N. 2015, Allen’s astrophysical quantities (Springer)

\bibitem[{{Deirmendjian}(1964)}]{Deirmendjian_1964}
{Deirmendjian}, D. 1964, \ao, 3, 187, \dodoi{10.1364/AO.3.000187}

\bibitem[{{Di Maio} {et~al.}(2023){Di Maio}, {Changeat}, {Benatti}, \& {Micela}}]{DiMaio2023}
{Di Maio}, C., {Changeat}, Q., {Benatti}, S., \& {Micela}, G. 2023, \aap, 669, A150, \dodoi{10.1051/0004-6361/202244881}

\bibitem[{{Dyrek} {et~al.}(2024){Dyrek}, {Min}, {Decin}, {Bouwman}, {Crouzet}, {Molli{\`e}re}, {Lagage}, {Konings}, {Tremblin}, {G{\"u}del}, {Pye}, {Waters}, {Henning}, {Vandenbussche}, {Ardevol Martinez}, {Argyriou}, {Ducrot}, {Heinke}, {van Looveren}, {Absil}, {Barrado}, {Baudoz}, {Boccaletti}, {Cossou}, {Coulais}, {Edwards}, {Gastaud}, {Glasse}, {Glauser}, {Greene}, {Kendrew}, {Krause}, {Lahuis}, {Mueller}, {Olofsson}, {Patapis}, {Rouan}, {Royer}, {Scheithauer}, {Waldmann}, {Whiteford}, {Colina}, {van Dishoeck}, {{\"O}stlin}, {Ray}, \& {Wright}}]{Dyrek_2023}
{Dyrek}, A., {Min}, M., {Decin}, L., {et~al.} 2024, \nat, 625, 51, \dodoi{10.1038/s41586-023-06849-0}

\bibitem[{{Edwards} \& {Changeat}(2024)}]{Edwards_2024}
{Edwards}, B., \& {Changeat}, Q. 2024, \apjl, 962, L30, \dodoi{10.3847/2041-8213/ad2000}

\bibitem[{{Edwards} {et~al.}(2024){Edwards}, {Tsiaras}, {Changeat}, \& {Yip}}]{Edwards_2024_K11}
{Edwards}, B., {Tsiaras}, A., {Changeat}, Q., \& {Yip}, K.~H. 2024, RAS Techniques and Instruments, 3, 415, \dodoi{10.1093/rasti/rzae023}

\bibitem[{{Edwards} {et~al.}(2023){Edwards}, {Changeat}, {Tsiaras}, {Yip}, {Al-Refaie}, {Anisman}, {Bieger}, {Gressier}, {Shibata}, {Skaf}, {Bouwman}, {Cho}, {Ikoma}, {Venot}, {Waldmann}, {Lagage}, \& {Tinetti}}]{edwards_pop}
{Edwards}, B., {Changeat}, Q., {Tsiaras}, A., {et~al.} 2023, \apjs, 269, 31, \dodoi{10.3847/1538-4365/ac9f1a}

\bibitem[{{Espinoza} {et~al.}(2023){Espinoza}, {{\'U}beda}, {Birkmann}, {Ferruit}, {Valenti}, {Sing}, {Rustamkulov}, {Regan}, {Kendrew}, {Sabbi}, {Schlawin}, {Beatty}, {Albert}, {Greene}, {Nikolov}, {Karakla}, {Keyes}, {Alves de Oliveira}, {B{\"o}ker}, {Pena-Guerrero}, {Giardino}, {Kumari}, {Manjavacas}, {Proffitt}, \& {Rawle}}]{Espinoza_2023}
{Espinoza}, N., {{\'U}beda}, L., {Birkmann}, S.~M., {et~al.} 2023, \pasp, 135, 018002, \dodoi{10.1088/1538-3873/aca3d3}

\bibitem[{{Estrela} {et~al.}(2022){Estrela}, {Swain}, \& {Roudier}}]{Estrela_2022}
{Estrela}, R., {Swain}, M.~R., \& {Roudier}, G.~M. 2022, \apjl, 941, L5, \dodoi{10.3847/2041-8213/aca2aa}

\bibitem[{{Fabian} {et~al.}(2001){Fabian}, {Henning}, {J{\"a}ger}, {Mutschke}, {Dorschner}, \& {Wehrhan}}]{Fabian_2001}
{Fabian}, D., {Henning}, T., {J{\"a}ger}, C., {et~al.} 2001, \aap, 378, 228, \dodoi{10.1051/0004-6361:20011196}

\bibitem[{{Fairman} {et~al.}(2024){Fairman}, {Wakeford}, \& {MacDonald}}]{Fairman_2024}
{Fairman}, C., {Wakeford}, H.~R., \& {MacDonald}, R.~J. 2024, \aj, 167, 240, \dodoi{10.3847/1538-3881/ad3454}

\bibitem[{{Feinstein} {et~al.}(2023){Feinstein}, {Radica}, {Welbanks}, {Murray}, {Ohno}, {Coulombe}, {Espinoza}, {Bean}, {Teske}, {Benneke}, {Line}, {Rustamkulov}, {Saba}, {Tsiaras}, {Barstow}, {Fortney}, {Gao}, {Knutson}, {MacDonald}, {Mikal-Evans}, {Rackham}, {Taylor}, {Parmentier}, {Batalha}, {Berta-Thompson}, {Carter}, {Changeat}, {dos Santos}, {Gibson}, {Goyal}, {Kreidberg}, {L{\'o}pez-Morales}, {Lothringer}, {Miguel}, {Molaverdikhani}, {Moran}, {Morello}, {Mukherjee}, {Sing}, {Stevenson}, {Wakeford}, {Ahrer}, {Alam}, {Alderson}, {Allen}, {Batalha}, {Bell}, {Blecic}, {Brande}, {Caceres}, {Casewell}, {Chubb}, {Crossfield}, {Crouzet}, {Cubillos}, {Decin}, {D{\'e}sert}, {Harrington}, {Heng}, {Henning}, {Iro}, {Kempton}, {Kendrew}, {Kirk}, {Krick}, {Lagage}, {Lendl}, {Mancini}, {Mansfield}, {May}, {Mayne}, {Nikolov}, {Palle}, {Petit dit de la Roche}, {Piaulet}, {Powell}, {Redfield}, {Rogers}, {Roman}, {Roy}, {Nixon}, {Schlawin}, {Tan}, {Tremblin}, {Turner}, {Venot}, {Waalkes}, {Wheatley}, \&
  {Zhang}}]{Fenstein_2023}
{Feinstein}, A.~D., {Radica}, M., {Welbanks}, L., {et~al.} 2023, \nat, 614, 670, \dodoi{10.1038/s41586-022-05674-1}

\bibitem[{{Feng} {et~al.}(2016){Feng}, {Line}, {Fortney}, {Stevenson}, {Bean}, {Kreidberg}, \& {Parmentier}}]{Feng_2016_inomogeneou}
{Feng}, Y.~K., {Line}, M.~R., {Fortney}, J.~J., {et~al.} 2016, The Astrophysical Journal, 829, 52, \dodoi{10.3847/0004-637X/829/1/52}

\bibitem[{Fletcher {et~al.}(2018)Fletcher, Gustafsson, \& Orton}]{fletcher_h2-h2}
Fletcher, L.~N., Gustafsson, M., \& Orton, G.~S. 2018, The Astrophysical Journal Supplement Series, 235, 24

\bibitem[{{Fonte} {et~al.}(2023){Fonte}, {Turrini}, {Pacetti}, {Schisano}, {Molinari}, {Polychroni}, {Politi}, \& {Changeat}}]{Fonte_2023}
{Fonte}, S., {Turrini}, D., {Pacetti}, E., {et~al.} 2023, \mnras, 520, 4683, \dodoi{10.1093/mnras/stad245}

\bibitem[{{Fournier-Tondreau} {et~al.}(2024){Fournier-Tondreau}, {MacDonald}, {Radica}, {Lafreni{\`e}re}, {Welbanks}, {Piaulet}, {Coulombe}, {Allart}, {Morel}, {Artigau}, {Albert}, {Lim}, {Doyon}, {Benneke}, {Rowe}, {Darveau-Bernier}, {Cowan}, {Lewis}, {Cook}, {Flagg}, {Genest}, {Pelletier}, {Johnstone}, {Dang}, {Kaltenegger}, {Taylor}, \& {Turner}}]{Fournier_2024}
{Fournier-Tondreau}, M., {MacDonald}, R.~J., {Radica}, M., {et~al.} 2024, \mnras, 528, 3354, \dodoi{10.1093/mnras/stad3813}

\bibitem[{{Fu} {et~al.}(2022){Fu}, {Espinoza}, {Sing}, {Lothringer}, {Dos Santos}, {Rustamkulov}, {Deming}, {Kempton}, {Komacek}, {Knutson}, {Albert}, {Pontoppidan}, {Volk}, \& {Filippazzo}}]{Fu_2022}
{Fu}, G., {Espinoza}, N., {Sing}, D.~K., {et~al.} 2022, \apjl, 940, L35, \dodoi{10.3847/2041-8213/ac9977}

\bibitem[{{Fulchignoni} {et~al.}(2005){Fulchignoni}, {Ferri}, {Angrilli}, {Ball}, {Bar-Nun}, {Barucci}, {Bettanini}, {Bianchini}, {Borucki}, {Colombatti}, {Coradini}, {Coustenis}, {Debei}, {Falkner}, {Fanti}, {Flamini}, {Gaborit}, {Grard}, {Hamelin}, {Harri}, {Hathi}, {Jernej}, {Leese}, {Lehto}, {Lion Stoppato}, {L{\'o}pez-Moreno}, {M{\"a}kinen}, {McDonnell}, {McKay}, {Molina-Cuberos}, {Neubauer}, {Pirronello}, {Rodrigo}, {Saggin}, {Schwingenschuh}, {Seiff}, {Sim{\~o}es}, {Svedhem}, {Tokano}, {Towner}, {Trautner}, {Withers}, \& {Zarnecki}}]{Fulchignoni_2005}
{Fulchignoni}, M., {Ferri}, F., {Angrilli}, F., {et~al.} 2005, \nat, 438, 785, \dodoi{10.1038/nature04314}

\bibitem[{{Gao} \& {Benneke}(2018)}]{Gao&Benneke18}
{Gao}, P., \& {Benneke}, B. 2018, \apj, 863, 165, \dodoi{10.3847/1538-4357/aad461}

\bibitem[{{Gao} {et~al.}(2021){Gao}, {Wakeford}, {Moran}, \& {Parmentier}}]{Gao_2021}
{Gao}, P., {Wakeford}, H.~R., {Moran}, S.~E., \& {Parmentier}, V. 2021, Journal of Geophysical Research (Planets), 126, e06655, \dodoi{10.1029/2020JE006655}

\bibitem[{{Gharib-Nezhad} {et~al.}(2024){Gharib-Nezhad}, {Batalha}, {Chubb}, {Freedman}, {Gordon}, {Gamache}, {Hargreaves}, {Lewis}, {Tennyson}, \& {Yurchenko}}]{Garib_2024}
{Gharib-Nezhad}, E.~S., {Batalha}, N.~E., {Chubb}, K., {et~al.} 2024, RAS Techniques and Instruments, 3, 44, \dodoi{10.1093/rasti/rzad058}

\bibitem[{{Grant} {et~al.}(2023){Grant}, {Lewis}, {Wakeford}, {Batalha}, {Glidden}, {Goyal}, {Mullens}, {MacDonald}, {May}, {Seager}, {Stevenson}, {Valenti}, {Visscher}, {Alderson}, {Allen}, {Ca{\~n}as}, {Col{\'o}n}, {Clampin}, {Espinoza}, {Gressier}, {Huang}, {Lin}, {Long}, {Louie}, {Pe{\~n}a-Guerrero}, {Ranjan}, {Sotzen}, {Valentine}, {Anderson}, {Balmer}, {Bellini}, {Hoch}, {Kammerer}, {Libralato}, {Mountain}, {Perrin}, {Pueyo}, {Rickman}, {Rebollido}, {Sohn}, {van der Marel}, \& {Watkins}}]{Grant_2023}
{Grant}, D., {Lewis}, N.~K., {Wakeford}, H.~R., {et~al.} 2023, \apjl, 956, L29, \dodoi{10.3847/2041-8213/acfc3b}

\bibitem[{Harrison {et~al.}(2010)Harrison, Allen, \& Bernath}]{Harrison_2010_C2H6}
Harrison, J.~J., Allen, N. D.~C., \& Bernath, P.~F. 2010, Journal of Quantitative Spectroscopy and Radiative Transfer, 111, 357, \dodoi{10.1016/j.jqsrt.2009.09.010}

\bibitem[{Harrison \& Bernath(2010)}]{Harrison_2010_C3H8}
Harrison, J.~J., \& Bernath, P.~F. 2010, Journal of Quantitative Spectroscopy and Radiative Transfer, 111, 1282, \dodoi{10.1016/j.jqsrt.2009.11.027}

\bibitem[{{Hejazi} {et~al.}(2023){Hejazi}, {Crossfield}, {Nordlander}, {Mansfield}, {Souto}, {Marfil}, {Coria}, {Brande}, {Polanski}, {Hand}, \& {Wienke}}]{Hejazi_2023}
{Hejazi}, N., {Crossfield}, I. J.~M., {Nordlander}, T., {et~al.} 2023, \apj, 949, 79, \dodoi{10.3847/1538-4357/accb97}

\bibitem[{{Helling} {et~al.}(2008){Helling}, {Woitke}, \& {Thi}}]{Helling+08}
{Helling}, C., {Woitke}, P., \& {Thi}, W.~F. 2008, \aap, 485, 547, \dodoi{10.1051/0004-6361:20078220}

\bibitem[{{Hinkel} {et~al.}(2014){Hinkel}, {Timmes}, {Young}, {Pagano}, \& {Turnbull}}]{Hinkel_2014}
{Hinkel}, N.~R., {Timmes}, F.~X., {Young}, P.~A., {Pagano}, M.~D., \& {Turnbull}, M.~C. 2014, \aj, 148, 54, \dodoi{10.1088/0004-6256/148/3/54}

\bibitem[{{Holmberg} \& {Madhusudhan}(2023)}]{Holmberg_2023}
{Holmberg}, M., \& {Madhusudhan}, N. 2023, arXiv e-prints, arXiv:2306.04676, \dodoi{10.48550/arXiv.2306.04676}

\bibitem[{{H{\"o}rst}(2017)}]{Horst_2017}
{H{\"o}rst}, S.~M. 2017, Journal of Geophysical Research (Planets), 122, 432, \dodoi{10.1002/2016JE005240}

\bibitem[{{Kataoka} {et~al.}(2014){Kataoka}, {Okuzumi}, {Tanaka}, \& {Nomura}}]{Akimasa_2014}
{Kataoka}, A., {Okuzumi}, S., {Tanaka}, H., \& {Nomura}, H. 2014, \aap, 568, A42, \dodoi{10.1051/0004-6361/201323199}

\bibitem[{{Kawashima} \& {Ikoma}(2019)}]{Kawashima_2019}
{Kawashima}, Y., \& {Ikoma}, M. 2019, \apj, 877, 109, \dodoi{10.3847/1538-4357/ab1b1d}

\bibitem[{{Khachai} {et~al.}(2009){Khachai}, {Khenata}, {Bouhemadou}, {Haddou}, {Reshak}, {Amrani}, {Rached}, \& {Soudini}}]{Khachai_2009}
{Khachai}, H., {Khenata}, R., {Bouhemadou}, A., {et~al.} 2009, Journal of Physics Condensed Matter, 21, 095404, \dodoi{10.1088/0953-8984/21/9/095404}

\bibitem[{{Khare} {et~al.}(1984){Khare}, {Sagan}, {Arakawa}, {Suits}, {Callcott}, \& {Williams}}]{Khare_1984}
{Khare}, B.~N., {Sagan}, C., {Arakawa}, E.~T., {et~al.} 1984, \icarus, 60, 127, \dodoi{10.1016/0019-1035(84)90142-8}

\bibitem[{{Kiefer} {et~al.}(2024){Kiefer}, {Samra}, {Lewis}, {Schneider}, {Min}, {Carone}, {Decin}, \& {Helling}}]{Kiefer_2024}
{Kiefer}, S., {Samra}, D., {Lewis}, D.~A., {et~al.} 2024, \aap, 690, A244, \dodoi{10.1051/0004-6361/202450526}

\bibitem[{{Kitzmann} \& {Heng}(2018)}]{Kitzmann_2018}
{Kitzmann}, D., \& {Heng}, K. 2018, \mnras, 475, 94, \dodoi{10.1093/mnras/stx3141}

\bibitem[{{Kochanov} {et~al.}(2016){Kochanov}, {Gordon}, {Rothman}, {Wcis{\l}o}, {Hill}, \& {Wilzewski}}]{Kochanov_2016}
{Kochanov}, R.~V., {Gordon}, I.~E., {Rothman}, L.~S., {et~al.} 2016, \jqsrt, 177, 15, \dodoi{10.1016/j.jqsrt.2016.03.005}

\bibitem[{{Lacy} \& {Burrows}(2020{\natexlab{a}})}]{Lacy_2020}
{Lacy}, B.~I., \& {Burrows}, A. 2020{\natexlab{a}}, \apj, 905, 131, \dodoi{10.3847/1538-4357/abc01c}

\bibitem[{{Lacy} \& {Burrows}(2020{\natexlab{b}})}]{Lacy_2020_b}
---. 2020{\natexlab{b}}, \apj, 904, 25, \dodoi{10.3847/1538-4357/abbc6c}

\bibitem[{{Lavvas} {et~al.}(2010){Lavvas}, {Yelle}, \& {Griffith}}]{Lavvas_2010}
{Lavvas}, P., {Yelle}, R.~V., \& {Griffith}, C.~A. 2010, \icarus, 210, 832, \dodoi{10.1016/j.icarus.2010.07.025}

\bibitem[{{Lee}(2023)}]{Lee23_mini-cloud}
{Lee}, E. K.~H. 2023, \mnras, 524, 2918, \dodoi{10.1093/mnras/stad2037}

\bibitem[{{Lee} {et~al.}(2013){Lee}, {Heng}, \& {Irwin}}]{Lee_2013_clouds}
{Lee}, J.-M., {Heng}, K., \& {Irwin}, P. G.~J. 2013, The Astrophysical Journal, 778, 97, \dodoi{10.1088/0004-637X/778/2/97}

\bibitem[{{Lee} {et~al.}(2014){Lee}, {Irwin}, {Fletcher}, {Heng}, \& {Barstow}}]{Lee_2014}
{Lee}, J.-M., {Irwin}, P. G.~J., {Fletcher}, L.~N., {Heng}, K., \& {Barstow}, J.~K. 2014, \apj, 789, 14, \dodoi{10.1088/0004-637X/789/1/14}

\bibitem[{Li {et~al.}(2015)Li, Gordon, Rothman, Tan, Hu, Kassi, Campargue, \& Medvedev}]{li_co_2015}
Li, G., Gordon, I.~E., Rothman, L.~S., {et~al.} 2015, The Astrophysical Journal Supplement Series, 216, 15

\bibitem[{{Lodge} {et~al.}(2024){Lodge}, {Wakeford}, \& {Leinhardt}}]{Lodge_2024}
{Lodge}, M.~G., {Wakeford}, H.~R., \& {Leinhardt}, Z.~M. 2024, \mnras, 527, 11113, \dodoi{10.1093/mnras/stad3743}

\bibitem[{{Lueber} {et~al.}(2024){Lueber}, {Novais}, {Fisher}, \& {Heng}}]{Lueber_2024}
{Lueber}, A., {Novais}, A., {Fisher}, C., \& {Heng}, K. 2024, \aap, 687, A110, \dodoi{10.1051/0004-6361/202348802}

\bibitem[{{Ma} {et~al.}(2023){Ma}, {Ito}, {Al-Refaie}, {Changeat}, {Edwards}, \& {Tinetti}}]{Ma_2023}
{Ma}, S., {Ito}, Y., {Al-Refaie}, A.~F., {et~al.} 2023, \apj, 957, 104, \dodoi{10.3847/1538-4357/acf8ca}

\bibitem[{{Ma} {et~al.}(2025){Ma}, {Saba}, {Faris Al-Refaie}, {Tinetti}, {Yurchenko}, {Tennyson}, \& {Cecchi Pestellini}}]{Ma_2025}
{Ma}, S., {Saba}, A., {Faris Al-Refaie}, A., {et~al.} 2025, arXiv e-prints, arXiv:2504.07823, \dodoi{10.48550/arXiv.2504.07823}

\bibitem[{MacDonald {et~al.}(2020)MacDonald, Goyal, \& Lewis}]{MacDonald_2020}
MacDonald, R.~J., Goyal, J.~M., \& Lewis, N.~K. 2020, The Astrophysical Journal, 893, L43, \dodoi{10.3847/2041-8213/ab8238}

\bibitem[{{Madhusudhan} \& {Seager}(2009)}]{Madhu_retrieval_method}
{Madhusudhan}, N., \& {Seager}, S. 2009, The Astrophysical Journal, 707, 24, \dodoi{10.1088/0004-637X/707/1/24}

\bibitem[{{Mai} \& {Line}(2019)}]{Mai_2019}
{Mai}, C., \& {Line}, M.~R. 2019, \apj, 883, 144, \dodoi{10.3847/1538-4357/ab3e6d}

\bibitem[{Maltagliati {et~al.}(2015)Maltagliati, Bézard, Vinatier, Hedman, Lellouch, Nicholson, Sotin, {de Kok}, \& Sicardy}]{Maltagliati_2015}
Maltagliati, L., Bézard, B., Vinatier, S., {et~al.} 2015, Icarus, 248, 1, \dodoi{https://doi.org/10.1016/j.icarus.2014.10.004}

\bibitem[{Mant {et~al.}(2018)Mant, Yachmenev, Tennyson, \& Yurchenko}]{2018_Mant_C2H4}
Mant, B.~P., Yachmenev, A., Tennyson, J., \& Yurchenko, S.~N. 2018, Monthly Notices of the Royal Astronomical Society, 478, 3220–3232, \dodoi{10.1093/mnras/sty1239}

\bibitem[{{Martonchik} \& {Orton}(1994)}]{Martonchik_1994}
{Martonchik}, J.~V., \& {Orton}, G.~S. 1994, \ao, 33, 8306, \dodoi{10.1364/AO.33.008306}

\bibitem[{Maxwell-Garnett(1904)}]{Maxwell_1904}
Maxwell-Garnett, J.~C. 1904, Philosophical Transactions of the Royal Society of London. Series A, Containing Papers of a Mathematical or Physical Character, 203, 385

\bibitem[{{McKay} {et~al.}(2001){McKay}, {Coustenis}, {Samuelson}, {Lemmon}, {Lorenz}, {Cabane}, {Rannou}, \& {Drossart}}]{McKay_2001}
{McKay}, C.~P., {Coustenis}, A., {Samuelson}, R.~E., {et~al.} 2001, \planss, 49, 79, \dodoi{10.1016/S0032-0633(00)00051-9}

\bibitem[{{Miles} {et~al.}(2023){Miles}, {Biller}, {Patapis}, {Worthen}, {Rickman}, {Hoch}, {Skemer}, {Perrin}, {Whiteford}, {Chen}, {Sargent}, {Mukherjee}, {Morley}, {Moran}, {Bonnefoy}, {Petrus}, {Carter}, {Choquet}, {Hinkley}, {Ward-Duong}, {Leisenring}, {Millar-Blanchaer}, {Pueyo}, {Ray}, {Sallum}, {Stapelfeldt}, {Stone}, {Wang}, {Absil}, {Balmer}, {Boccaletti}, {Bonavita}, {Booth}, {Bowler}, {Chauvin}, {Christiaens}, {Currie}, {Danielski}, {Fortney}, {Girard}, {Grady}, {Greenbaum}, {Henning}, {Hines}, {Janson}, {Kalas}, {Kammerer}, {Kennedy}, {Kenworthy}, {Kervella}, {Lagage}, {Lew}, {Liu}, {Macintosh}, {Marino}, {Marley}, {Marois}, {Matthews}, {Matthews}, {Mawet}, {McElwain}, {Metchev}, {Meyer}, {Molliere}, {Pantin}, {Quirrenbach}, {Rebollido}, {Ren}, {Schneider}, {Vasist}, {Wyatt}, {Zhou}, {Briesemeister}, {Bryan}, {Calissendorff}, {Cantalloube}, {Cugno}, {De Furio}, {Dupuy}, {Factor}, {Faherty}, {Fitzgerald}, {Franson}, {Gonzales}, {Hood}, {Howe}, {Kraus}, {Kuzuhara}, {Lagrange}, {Lawson}, {Lazzoni},
  {Liu}, {Llop-Sayson}, {Lloyd}, {Martinez}, {Mazoyer}, {Quanz}, {Redai}, {Samland}, {Schlieder}, {Tamura}, {Tan}, {Uyama}, {Vigan}, {Vos}, {Wagner}, {Wolff}, {Ygouf}, {Zhang}, {Zhang}, \& {Zhang}}]{Miles_2023}
{Miles}, B.~E., {Biller}, B.~A., {Patapis}, P., {et~al.} 2023, \apjl, 946, L6, \dodoi{10.3847/2041-8213/acb04a}

\bibitem[{{Min} {et~al.}(2020){Min}, {Ormel}, {Chubb}, {Helling}, \& {Kawashima}}]{Min_2020}
{Min}, M., {Ormel}, C.~W., {Chubb}, K., {Helling}, C., \& {Kawashima}, Y. 2020, Astronomy and Astrophysics, 642, A28, \dodoi{10.1051/0004-6361/201937377}

\bibitem[{{Min} {et~al.}(2007){Min}, {Waters}, {de Koter}, {Hovenier}, {Keller}, \& {Markwick-Kemper}}]{Min_2007}
{Min}, M., {Waters}, L.~B.~F.~M., {de Koter}, A., {et~al.} 2007, \aap, 462, 667, \dodoi{10.1051/0004-6361:20065436}

\bibitem[{{Moran} {et~al.}(2023){Moran}, {Stevenson}, {Sing}, {MacDonald}, {Kirk}, {Lustig-Yaeger}, {Peacock}, {Mayorga}, {Bennett}, {L{\'o}pez-Morales}, {May}, {Rustamkulov}, {Valenti}, {Adams Redai}, {Alam}, {Batalha}, {Fu}, {Gonzalez-Quiles}, {Highland}, {Kruse}, {Lothringer}, {Ortiz Ceballos}, {Sotzen}, \& {Wakeford}}]{Moran_2023}
{Moran}, S.~E., {Stevenson}, K.~B., {Sing}, D.~K., {et~al.} 2023, \apjl, 948, L11, \dodoi{10.3847/2041-8213/accb9c}

\bibitem[{{Niemann} {et~al.}(2010){Niemann}, {Atreya}, {Demick}, {Gautier}, {Haberman}, {Harpold}, {Kasprzak}, {Lunine}, {Owen}, \& {Raulin}}]{Niemann_2010}
{Niemann}, H.~B., {Atreya}, S.~K., {Demick}, J.~E., {et~al.} 2010, Journal of Geophysical Research (Planets), 115, E12006, \dodoi{10.1029/2010JE003659}

\bibitem[{{Nikolov} {et~al.}(2018){Nikolov}, {Sing}, {Fortney}, {Goyal}, {Drummond}, {Evans}, {Gibson}, {De Mooij}, {Rustamkulov}, {Wakeford}, {Smalley}, {Burgasser}, {Hellier}, {Helling}, {Mayne}, {Madhusudhan}, {Kataria}, {Baines}, {Carter}, {Ballester}, {Barstow}, {McCleery}, \& {Spake}}]{Nicolov_2018}
{Nikolov}, N., {Sing}, D.~K., {Fortney}, J.~J., {et~al.} 2018, \nat, 557, 526, \dodoi{10.1038/s41586-018-0101-7}

\bibitem[{{Niraula} {et~al.}(2022){Niraula}, {de Wit}, {Gordon}, {Hargreaves}, {Sousa-Silva}, \& {Kochanov}}]{Niraula_2022}
{Niraula}, P., {de Wit}, J., {Gordon}, I.~E., {et~al.} 2022, Nature Astronomy, 6, 1287, \dodoi{10.1038/s41550-022-01773-1}

\bibitem[{{Nixon} \& {Madhusudhan}(2022)}]{Nixon_2022}
{Nixon}, M.~C., \& {Madhusudhan}, N. 2022, \apj, 935, 73, \dodoi{10.3847/1538-4357/ac7c09}

\bibitem[{{Ohno} \& {Okuzumi}(2018)}]{Ohno&Okuzumi18}
{Ohno}, K., \& {Okuzumi}, S. 2018, \apj, 859, 34, \dodoi{10.3847/1538-4357/aabee3}

\bibitem[{{Ohno} {et~al.}(2020{\natexlab{a}}){Ohno}, {Okuzumi}, \& {Tazaki}}]{Ohno_2020}
{Ohno}, K., {Okuzumi}, S., \& {Tazaki}, R. 2020{\natexlab{a}}, \apj, 891, 131, \dodoi{10.3847/1538-4357/ab44bd}

\bibitem[{{Ohno} {et~al.}(2020{\natexlab{b}}){Ohno}, {Okuzumi}, \& {Tazaki}}]{Ohno+20_fluffy}
---. 2020{\natexlab{b}}, \apj, 891, 131, \dodoi{10.3847/1538-4357/ab44bd}

\bibitem[{{Ormel} \& {Min}(2019)}]{Ormel&Min19}
{Ormel}, C.~W., \& {Min}, M. 2019, \aap, 622, A121, \dodoi{10.1051/0004-6361/201833678}

\bibitem[{{Palik}(1991)}]{Palik_1991}
{Palik}, E.~D. 1991, {Handbook of optical constants of solids II}

\bibitem[{{Peek} {et~al.}(2019){Peek}, {Desai}, {White}, {D'Abrusco}, {Mazzarella}, {Grant}, {Novacescu}, {Scire}, \& {Winkelman}}]{Peek_2019}
{Peek}, J., {Desai}, V., {White}, R.~L., {et~al.} 2019, in Bulletin of the American Astronomical Society, Vol.~51, 105.
\newblock \doarXiv{1907.06234}

\bibitem[{Perrin {et~al.}(2025)Perrin, Carrasco, Gautier, Ruscassier, Maillard, Afonso, \& Vettier}]{Perrin_2025}
Perrin, Z., Carrasco, N., Gautier, T., {et~al.} 2025, Icarus, 429, 116418, \dodoi{10.1016/j.icarus.2024.116418}

\bibitem[{{Pinhas} \& {Madhusudhan}(2017)}]{Pinhas_2017}
{Pinhas}, A., \& {Madhusudhan}, N. 2017, \mnras, 471, 4355, \dodoi{10.1093/mnras/stx1849}

\bibitem[{Pluriel {et~al.}(2020)Pluriel, Zingales, Leconte, \& Parmentier}]{Pluriel_2020}
Pluriel, W., Zingales, T., Leconte, J., \& Parmentier, V. 2020, Astronomy \& Astrophysics, 636, A66, \dodoi{10.1051/0004-6361/202037678}

\bibitem[{{Polman} {et~al.}(2023){Polman}, {Waters}, {Min}, {Miguel}, \& {Khorshid}}]{Polman_2023}
{Polman}, J., {Waters}, L.~B.~F.~M., {Min}, M., {Miguel}, Y., \& {Khorshid}, N. 2023, \aap, 670, A161, \dodoi{10.1051/0004-6361/202244647}

\bibitem[{Polyansky {et~al.}(2018)Polyansky, Kyuberis, Zobov, Tennyson, Yurchenko, \& Lodi}]{polyansky_h2o}
Polyansky, O.~L., Kyuberis, A.~A., Zobov, N.~F., {et~al.} 2018, Monthly Notices of the Royal Astronomical Society, 480, 2597

\bibitem[{{Powell} \& {Zhang}(2024)}]{Powell&Zhang24}
{Powell}, D., \& {Zhang}, X. 2024, \apj, 969, 5, \dodoi{10.3847/1538-4357/ad3de4}

\bibitem[{{Powell} {et~al.}(2018){Powell}, {Zhang}, {Gao}, \& {Parmentier}}]{Powell+18}
{Powell}, D., {Zhang}, X., {Gao}, P., \& {Parmentier}, V. 2018, \apj, 860, 18, \dodoi{10.3847/1538-4357/aac215}

\bibitem[{Querry {et~al.}(1987)Querry, Chemical~Research, United States. Army~Armament, Command, \& Service}]{Querry_1987}
Querry, M., Chemical~Research, D. . E. C.~U., United States. Army~Armament, M., Command, C., \& Service, U. S. N. T.~I. 1987, Optical Constants of Minerals and Other Materials from the Millimeter to the Ultraviolet, CRDC-CR (Chemical Research, Development \& Engineering Center, US Army Armament, Munitions, Chemical Command).
\newblock \url{https://books.google.com/books?id=FVeENwAACAAJ}

\bibitem[{{Radica} {et~al.}(2023){Radica}, {Welbanks}, {Espinoza}, {Taylor}, {Coulombe}, {Feinstein}, {Goyal}, {Scarsdale}, {Albert}, {Baghel}, {Bean}, {Blecic}, {Lafreni{\`e}re}, {MacDonald}, {Zamyatina}, {Allart1}, {Artigau}, {Batalha}, {Cook}, {Cowan}, {Dang}, {Doyon}, {Fournier-Tondreau}, {Johnstone}, {Line}, {Moran}, {Mukherjee}, {Pelletier}, {Roy}, {Talens}, {Filippazzo}, {Pontoppidan}, \& {Volk}}]{Radica_2023}
{Radica}, M., {Welbanks}, L., {Espinoza}, N., {et~al.} 2023, \mnras, 524, 835, \dodoi{10.1093/mnras/stad1762}

\bibitem[{{Rannou} {et~al.}(1995){Rannou}, {Cabane}, {Chassefiere}, {Botet}, {McKay}, \& {Courtin}}]{Rannou+95}
{Rannou}, P., {Cabane}, M., {Chassefiere}, E., {et~al.} 1995, \icarus, 118, 355, \dodoi{10.1006/icar.1995.1196}

\bibitem[{{Rannou} {et~al.}(2010){Rannou}, {Cours}, {Le Mou{\'e}lic}, {Rodriguez}, {Sotin}, {Drossart}, \& {Brown}}]{Rannou_2010}
{Rannou}, P., {Cours}, T., {Le Mou{\'e}lic}, S., {et~al.} 2010, \icarus, 208, 850, \dodoi{10.1016/j.icarus.2010.03.016}

\bibitem[{{Rannou} {et~al.}(2022){Rannou}, {Coutelier}, {Rey}, \& {Vinatier}}]{Rannou_2022}
{Rannou}, P., {Coutelier}, M., {Rey}, M., \& {Vinatier}, S. 2022, \aap, 666, A140, \dodoi{10.1051/0004-6361/202243045}

\bibitem[{{Robinson} {et~al.}(2014){Robinson}, {Maltagliati}, {Marley}, \& {Fortney}}]{Robinson_2014}
{Robinson}, T.~D., {Maltagliati}, L., {Marley}, M.~S., \& {Fortney}, J.~J. 2014, Proceedings of the National Academy of Science, 111, 9042, \dodoi{10.1073/pnas.1403473111}

\bibitem[{Rocchetto {et~al.}(2016)Rocchetto, Waldmann, Venot, Lagage, \& Tinetti}]{Rocchetto_2016}
Rocchetto, M., Waldmann, I.~P., Venot, O., Lagage, P.-O., \& Tinetti, G. 2016, The Astrophysical Journal, 833, 120, \dodoi{10.3847/1538-4357/833/1/120}

\bibitem[{{Rotman} {et~al.}(2025){Rotman}, {Welbanks}, {Line}, {McGill}, {Radica}, \& {Nixon}}]{Rotman_2025}
{Rotman}, Y., {Welbanks}, L., {Line}, M.~R., {et~al.} 2025, arXiv e-prints, arXiv:2503.21702, \dodoi{10.48550/arXiv.2503.21702}

\bibitem[{{Rowland} {et~al.}(2023){Rowland}, {Morley}, \& {Line}}]{Rowland_2023}
{Rowland}, M.~J., {Morley}, C.~V., \& {Line}, M.~R. 2023, \apj, 947, 6, \dodoi{10.3847/1538-4357/acbb07}

\bibitem[{{Rustamkulov} {et~al.}(2023){Rustamkulov}, {Sing}, {Mukherjee}, {May}, {Kirk}, {Schlawin}, {Line}, {Piaulet}, {Carter}, {Batalha}, {Goyal}, {L{\'o}pez-Morales}, {Lothringer}, {MacDonald}, {Moran}, {Stevenson}, {Wakeford}, {Espinoza}, {Bean}, {Batalha}, {Benneke}, {Berta-Thompson}, {Crossfield}, {Gao}, {Kreidberg}, {Powell}, {Cubillos}, {Gibson}, {Leconte}, {Molaverdikhani}, {Nikolov}, {Parmentier}, {Roy}, {Taylor}, {Turner}, {Wheatley}, {Aggarwal}, {Ahrer}, {Alam}, {Alderson}, {Allen}, {Banerjee}, {Barat}, {Barrado}, {Barstow}, {Bell}, {Blecic}, {Brande}, {Casewell}, {Changeat}, {Chubb}, {Crouzet}, {Daylan}, {Decin}, {D{\'e}sert}, {Mikal-Evans}, {Feinstein}, {Flagg}, {Fortney}, {Harrington}, {Heng}, {Hong}, {Hu}, {Iro}, {Kataria}, {Kempton}, {Krick}, {Lendl}, {Lillo-Box}, {Louca}, {Lustig-Yaeger}, {Mancini}, {Mansfield}, {Mayne}, {Miguel}, {Morello}, {Ohno}, {Palle}, {Petit dit de la Roche}, {Rackham}, {Radica}, {Ramos-Rosado}, {Redfield}, {Rogers}, {Shkolnik}, {Southworth}, {Teske}, {Tremblin},
  {Tucker}, {Venot}, {Waalkes}, {Welbanks}, {Zhang}, \& {Zieba}}]{Rustamkulov_2023}
{Rustamkulov}, Z., {Sing}, D.~K., {Mukherjee}, S., {et~al.} 2023, \nat, 614, 659, \dodoi{10.1038/s41586-022-05677-y}

\bibitem[{{Samra} {et~al.}(2022){Samra}, {Helling}, \& {Birnstiel}}]{Samra+22}
{Samra}, D., {Helling}, C., \& {Birnstiel}, T. 2022, \aap, 663, A47, \dodoi{10.1051/0004-6361/202142651}

\bibitem[{{Samra} {et~al.}(2020){Samra}, {Helling}, \& {Min}}]{Samra_2020}
{Samra}, D., {Helling}, C., \& {Min}, M. 2020, \aap, 639, A107, \dodoi{10.1051/0004-6361/202037553}

\bibitem[{{Schleich} {et~al.}(2024){Schleich}, {Boro Saikia}, {Changeat}, {G{\"u}del}, {Voigt}, \& {Waldmann}}]{Schleich_2024}
{Schleich}, S., {Boro Saikia}, S., {Changeat}, Q., {et~al.} 2024, arXiv e-prints, arXiv:2409.09127, \dodoi{10.48550/arXiv.2409.09127}

\bibitem[{{Scott} \& {Duley}(1996)}]{Scott_1996}
{Scott}, A., \& {Duley}, W.~W. 1996, \apjs, 105, 401, \dodoi{10.1086/192321}

\bibitem[{{Sing} {et~al.}(2016){Sing}, {Fortney}, {Nikolov}, {Wakeford}, {Kataria}, {Evans}, {Aigrain}, {Ballester}, {Burrows}, {Deming}, {D{\'e}sert}, {Gibson}, {Henry}, {Huitson}, {Knutson}, {Lecavelier Des Etangs}, {Pont}, {Showman}, {Vidal-Madjar}, {Williamson}, \& {Wilson}}]{sing_pop}
{Sing}, D.~K., {Fortney}, J.~J., {Nikolov}, N., {et~al.} 2016, \nat, 529, 59, \dodoi{10.1038/nature16068}

\bibitem[{{Sing} {et~al.}(2024){Sing}, {Rustamkulov}, {Thorngren}, {Barstow}, {Tremblin}, {Alves de Oliveira}, {Beck}, {Birkmann}, {Challener}, {Crouzet}, {Espinoza}, {Ferruit}, {Giardino}, {Gressier}, {Lee}, {Lewis}, {Maiolino}, {Manjavacas}, {Rauscher}, {Sirianni}, \& {Valenti}}]{Sing_2024}
{Sing}, D.~K., {Rustamkulov}, Z., {Thorngren}, D.~P., {et~al.} 2024, \nat, 630, 831, \dodoi{10.1038/s41586-024-07395-z}

\bibitem[{Skinner \& Cho(2022)}]{Skinner_2021_modons}
Skinner, J.~W., \& Cho, J. Y.-K. 2022, Monthly Notices of the Royal Astronomical Society, 511, 3584, \dodoi{10.1093/mnras/stab2809}

\bibitem[{{Skinner} {et~al.}(2022){Skinner}, {Cho}, \& N\"attil\"a}]{Skinner_2022_cyclogenesis}
{Skinner}, J.~W., {Cho}, J.~Y.~K., \& N\"attil\"a, J. 2022, Submitted

\bibitem[{{Sumlin} {et~al.}(2018){Sumlin}, {Heinson}, \& {Chakrabarty}}]{Sumlin_2018}
{Sumlin}, B.~J., {Heinson}, W.~R., \& {Chakrabarty}, R.~K. 2018, \jqsrt, 205, 127, \dodoi{10.1016/j.jqsrt.2017.10.012}

\bibitem[{{Sylvestre} {et~al.}(2018){Sylvestre}, {Teanby}, {Vinatier}, {Lebonnois}, \& {Irwin}}]{Sylvestre_2018}
{Sylvestre}, M., {Teanby}, N.~A., {Vinatier}, S., {Lebonnois}, S., \& {Irwin}, P.~G.~J. 2018, \aap, 609, A64, \dodoi{10.1051/0004-6361/201630255}

\bibitem[{{Taylor} {et~al.}(2020){Taylor}, {Parmentier}, {Irwin}, {Aigrain}, {Lee}, \& {Krissansen-Totton}}]{Taylor_2020}
{Taylor}, J., {Parmentier}, V., {Irwin}, P. G.~J., {et~al.} 2020, Monthly Notices of the Royal Astronomical Society, 493, 4342, \dodoi{10.1093/mnras/staa552}

\bibitem[{{Taylor} {et~al.}(2023){Taylor}, {Radica}, {Welbanks}, {MacDonald}, {Blecic}, {Zamyatina}, {Roth}, {Bean}, {Parmentier}, {Coulombe}, {Feinstein}, {Espinoza}, {Benneke}, {Lafreni{\`e}re}, {Doyon}, \& {Ahrer}}]{Taylor_2023}
{Taylor}, J., {Radica}, M., {Welbanks}, L., {et~al.} 2023, \mnras, \dodoi{10.1093/mnras/stad1547}

\bibitem[{Tennyson {et~al.}(2016)Tennyson, Yurchenko, Al-Refaie, Barton, Chubb, Coles, Diamantopoulou, Gorman, Hill, Lam, Lodi, McKemmish, Na, Owens, Polyansky, Rivlin, Sousa-Silva, Underwood, Yachmenev, \& Zak}]{Tennyson_exomol}
Tennyson, J., Yurchenko, S.~N., Al-Refaie, A.~F., {et~al.} 2016, Journal of Molecular Spectroscopy, 327, 73 , \dodoi{https://doi.org/10.1016/j.jms.2016.05.002}

\bibitem[{{Tennyson} {et~al.}(2024){Tennyson}, {Yurchenko}, {Zhang}, {Bowesman}, {Brady}, {Buldyreva}, {Chubb}, {Gamache}, {Gorman}, {Guest}, {Hill}, {Kefala}, {Lynas-Gray}, {Mellor}, {McKemmish}, {Mitev}, {Mizus}, {Owens}, {Peng}, {Perri}, {Pezzella}, {Polyansky}, {Qu}, {Semenov}, {Smola}, {Solokov}, {Somogyi}, {Upadhyay}, {Wright}, \& {Zobov}}]{Tennyson_2024}
{Tennyson}, J., {Yurchenko}, S.~N., {Zhang}, J., {et~al.} 2024, \jqsrt, 326, 109083, \dodoi{10.1016/j.jqsrt.2024.109083}

\bibitem[{{Tomasko} {et~al.}(2008){Tomasko}, {Doose}, {Engel}, {Dafoe}, {West}, {Lemmon}, {Karkoschka}, \& {See}}]{Tomasko_2008}
{Tomasko}, M.~G., {Doose}, L., {Engel}, S., {et~al.} 2008, \planss, 56, 669, \dodoi{10.1016/j.pss.2007.11.019}

\bibitem[{{Tsai} {et~al.}(2023){Tsai}, {Lee}, {Powell}, {Gao}, {Zhang}, {Moses}, {H{\'e}brard}, {Venot}, {Parmentier}, {Jordan}, {Hu}, {Alam}, {Alderson}, {Batalha}, {Bean}, {Benneke}, {Bierson}, {Brady}, {Carone}, {Carter}, {Chubb}, {Inglis}, {Leconte}, {Line}, {L{\'o}pez-Morales}, {Miguel}, {Molaverdikhani}, {Rustamkulov}, {Sing}, {Stevenson}, {Wakeford}, {Yang}, {Aggarwal}, {Baeyens}, {Barat}, {de Val-Borro}, {Daylan}, {Fortney}, {France}, {Goyal}, {Grant}, {Kirk}, {Kreidberg}, {Louca}, {Moran}, {Mukherjee}, {Nasedkin}, {Ohno}, {Rackham}, {Redfield}, {Taylor}, {Tremblin}, {Visscher}, {Wallack}, {Welbanks}, {Youngblood}, {Ahrer}, {Batalha}, {Behr}, {Berta-Thompson}, {Blecic}, {Casewell}, {Crossfield}, {Crouzet}, {Cubillos}, {Decin}, {D{\'e}sert}, {Feinstein}, {Gibson}, {Harrington}, {Heng}, {Henning}, {Kempton}, {Krick}, {Lagage}, {Lendl}, {Lothringer}, {Mansfield}, {Mayne}, {Mikal-Evans}, {Palle}, {Schlawin}, {Shorttle}, {Wheatley}, \& {Yurchenko}}]{Tsai_2023}
{Tsai}, S.-M., {Lee}, E. K.~H., {Powell}, D., {et~al.} 2023, \nat, 617, 483, \dodoi{10.1038/s41586-023-05902-2}

\bibitem[{{Tsiaras} {et~al.}(2018){Tsiaras}, {Waldmann}, {Zingales}, {Rocchetto}, {Morello}, {Damiano}, {Karpouzas}, {Tinetti}, {McKemmish}, {Tennyson}, \& {Yurchenko}}]{tsiaras_30planets}
{Tsiaras}, A., {Waldmann}, I.~P., {Zingales}, T., {et~al.} 2018, The Astronomical Journal, 155, 156, \dodoi{10.3847/1538-3881/aaaf75}

\bibitem[{{Underwood} {et~al.}(2016){Underwood}, {Tennyson}, {Yurchenko}, {Huang}, {Schwenke}, {Lee}, {Clausen}, \& {Fateev}}]{Underwood_2016}
{Underwood}, D.~S., {Tennyson}, J., {Yurchenko}, S.~N., {et~al.} 2016, \mnras, 459, 3890, \dodoi{10.1093/mnras/stw849}

\bibitem[{{Vahidinia} {et~al.}(2024){Vahidinia}, {Moran}, {Marley}, \& {Cuzzi}}]{Vahidinia_2024}
{Vahidinia}, S., {Moran}, S.~E., {Marley}, M.~S., \& {Cuzzi}, J.~N. 2024, \pasp, 136, 084404, \dodoi{10.1088/1538-3873/ad6cf2}

\bibitem[{{Venot} {et~al.}(2020){Venot}, {Cavali{\'e}}, {Bounaceur}, {Tremblin}, {Brouillard}, \& {Lhoussaine Ben Brahim}}]{Venot_2020_new}
{Venot}, O., {Cavali{\'e}}, T., {Bounaceur}, R., {et~al.} 2020, \aap, 634, A78, \dodoi{10.1051/0004-6361/201936697}

\bibitem[{{Voyer} {et~al.}(2025){Voyer}, {Changeat}, {Lagage}, {Tremblin}, {Waters}, {G{\"u}del}, {Henning}, {Absil}, {Barrado}, {Boccaletti}, {Bouwman}, {Coulais}, {Decin}, {Glauser}, {Pye}, {Glasse}, {Gastaud}, {Kendrew}, {Patapis}, {Rouan}, {van Dishoeck}, {{\"O}stlin}, {Ray}, \& {Wright}}]{Voyer_2024}
{Voyer}, M., {Changeat}, Q., {Lagage}, P.-O., {et~al.} 2025, \apjl, 982, L38, \dodoi{10.3847/2041-8213/adbd46}

\bibitem[{{Wakeford} \& {Sing}(2015)}]{Wakeford_2015}
{Wakeford}, H.~R., \& {Sing}, D.~K. 2015, \aap, 573, A122, \dodoi{10.1051/0004-6361/201424207}

\bibitem[{{Welbanks} {et~al.}(2024){Welbanks}, {Bell}, {Beatty}, {Line}, {Ohno}, {Fortney}, {Schlawin}, {Greene}, {Rauscher}, {McGill}, {Murphy}, {Parmentier}, {Tang}, {Edelman}, {Mukherjee}, {Wiser}, {Lagage}, {Dyrek}, \& {Arnold}}]{Welbanks_2024}
{Welbanks}, L., {Bell}, T.~J., {Beatty}, T.~G., {et~al.} 2024, \nat, 630, 836, \dodoi{10.1038/s41586-024-07514-w}

\bibitem[{{West} \& {Smith}(1991)}]{West&Smith91}
{West}, R.~A., \& {Smith}, P.~H. 1991, \icarus, 90, 330, \dodoi{10.1016/0019-1035(91)90113-8}

\bibitem[{{Wetzel} {et~al.}(2013){Wetzel}, {Klevenz}, {Gail}, {Pucci}, \& {Trieloff}}]{Wetzel_2013}
{Wetzel}, S., {Klevenz}, M., {Gail}, H.~P., {Pucci}, A., \& {Trieloff}, M. 2013, \aap, 553, A92, \dodoi{10.1051/0004-6361/201220803}

\bibitem[{Whitten {et~al.}(2008)Whitten, Borucki, O'brien, \& Tripathi}]{Whitten_2008}
Whitten, R.~C., Borucki, W.~J., O'brien, K., \& Tripathi, S.~N. 2008, Journal of Geophysical Research, 113.
\newblock \url{https://api.semanticscholar.org/CorpusID:53446710}

\bibitem[{{Woitke} {et~al.}(2018){Woitke}, {Helling}, {Hunter}, {Millard}, {Turner}, {Worters}, {Blecic}, \& {Stock}}]{Woitke_2018_GG}
{Woitke}, P., {Helling}, C., {Hunter}, G.~H., {et~al.} 2018, Astronomy \& Astrophysics, 614, A1, \dodoi{10.1051/0004-6361/201732193}

\bibitem[{{Yip} {et~al.}(2021){Yip}, {Changeat}, {Edwards}, {Morvan}, {Chubb}, {Tsiaras}, {Waldmann}, \& {Tinetti}}]{Yip_2021_W96}
{Yip}, K.~H., {Changeat}, Q., {Edwards}, B., {et~al.} 2021, The Astronomical Journal, 161, 4, \dodoi{10.3847/1538-3881/abc179}

\bibitem[{Yurchenko {et~al.}(2017)Yurchenko, Amundsen, Tennyson, \& Waldmann}]{ExoMol_CH4_new}
Yurchenko, S.~N., Amundsen, D.~S., Tennyson, J., \& Waldmann, I.~P. 2017, A\&A, 605, A95, \dodoi{10.1051/0004-6361/201731026}

\bibitem[{Yurchenko {et~al.}(2020)Yurchenko, Mellor, Freedman, \& Tennyson}]{Yurchenko_2020}
Yurchenko, S.~N., Mellor, T.~M., Freedman, R.~S., \& Tennyson, J. 2020, Monthly Notices of the Royal Astronomical Society, 496, 5282–5291, \dodoi{10.1093/mnras/staa1874}

\end{thebibliography}

\clearpage

\begin{appendix}

\renewcommand\thesection{\Alph{section}}
\renewcommand\thesubsection{\thesection.\arabic{subsection}}
\value{section} = 0

\section{Plugin additions to \textsc{TauREx3}}

\renewcommand\thesubsection{A\thesection.\arabic{subsection}}

\setcounter{figure}{0}
\renewcommand{\thefigure}{A\arabic{figure}}
\renewcommand{\theHfigure}{A\arabic{figure}}

In this appendix, we describe the three new plugins \textsc{TauREx-PyMieScatt}, \textsc{TauREx-MultiModel} and \textsc{TauREx-InstrumentSystematics}.

\subsection{Aerosols modeling with \textsc{TauREx-PyMieScatt}}

For this work, a novel \textsc{TauREx3} plugin, \textsc{TauREx-PyMieScatt} is developed. The code is designed to model and retrieve parameterized aerosol properties, for application to exoplanet data. For each aerosol species, the plugin uses the python library \textsc{PyMieScatt} \citep{Sumlin_2018} to estimate the Mie Extinction Efficiency ($Q_\mathrm{ext}$), and cross-sections, of spherical particles from their complex refractive index\footnote{The complex refractive indexes for the modeled aerosols are required in the form of tables.}: $m = x+iy$. Since $Q_\mathrm{ext}$ is computed for a single homogeneous sphere, the contribution from poly-dispersed particles is estimated by weighting sampled points of the particle size distribution. Following \cite{Pinhas_2017}, the available particle size distributions ($n$) in \textsc{TauREx-PyMieScatt} are: \\

1) Log-Normal distribution:
\begin{equation}
    n(r) = \frac{1}{\sqrt{2 \pi} \sigma_r r} \exp \left\{ -0.5 \left(\frac{\ln(r) - \mu_r}{\sigma_r}\right)^2 \right\},
\end{equation}
where $r$ is the sampled particle radius, $\mu_r$ is the mean particle radius, and $\sigma_r$ is the scale of the particle size distribution. \\

2) Modified Gamma distribution \citep{Deirmendjian_1964}: 
\begin{equation}
    n(r) = a r^b \exp \left\{ -c r^d \right\},
\end{equation}
where $a$, $b$, $c$, and $d$ are {\it positive} coefficients. \\

3) A particular case of 2) from \cite{Budaj_2015}:
\begin{equation}
    n(r) = \left(\frac{r}{\mu_r} \right)^6 \exp \left\{ -6 \frac{r}{\mu_r} \right\},
\end{equation}
where here, $\mu_r$ represents the critical radius for which the function is maximum. As fitting for the full form of the particle size distribution cannot be done with current instruments, the latter distribution provides a convenient single parameter description of $n$.

Then, the cross-section $\sigma$ (in $m^2$), is given by:
\begin{equation}
    \sigma = \frac{1}{N} \int n(r) Q_\mathrm{ext}(r) \pi r^2 dr,
\end{equation}
where $N = \int n(r) dr$. \\

The contribution to the optical depth is obtained by multiplying with the particle number density $\chi$ (in $m^{-3}$), which is a function of the altitude $z$. In exoplanet studies, aerosol layers are often parameterized by a boxcar function, with $\chi(z) = 0$ outside reference bottom and top pressures and $\chi(z) = \chi$ inside of the reference pressures (i.e., $\chi$ is constant in the cloud layer). In \textsc{TauREx-PyMieScatt}, this boxcar function for a species S is defined by its mean pressure ($P_\mathrm{S}$) and its range ($\Delta P_\mathrm{S}$, always in log space). Other studies have also suggested an exponential decay with pressure \cite[e.g.,][]{Atreya_2005, Whitten_2008}, so this option is also offered in \textsc{TauREx-PyMieScatt}, with $\chi$ defining the number density at the bottom of the layer and $\alpha$ describing the exponential decay factor. In this case, $\chi(z)$ is given by:
\begin{equation}
    \chi(z) = \chi \left( \dfrac{p}{p_{ref}} \right)^\alpha,
\end{equation}
where $p_{ref}$ is the reference pressure at the bottom of the cloud layer.
In \textsc{TauREx}, contributions can be added dynamically. \textsc{TauREx-PyMieScatt} makes use of this feature, allowing multiple cloud layers and species to be added simultaneously. \\

Since precise calculations of components are difficult, we use Effective Medium Theory \cite[see][]{Bohren_2008, Akimasa_2014} to calculate the refractive index of porous particles and aggregates. In \textsc{TauREx-PyMieScatt}, the modified refractive indexes can be computed using the Bruggeman mixing rule \citep{Bruggeman_1935} or the Maxwell-Garnet mixing rule \citep{Maxwell_1904}. In this work, we focus on spherical particles, but include a retrieval example with porous particles in Section \ref{sec:disc}. \\

\textsc{TauREx-PyMieScatt} also provides an implementation of the phenomenological aerosol model from Equation A1 of \cite{Lee_2013_clouds}: it uses the same equations as already available in the base \textsc{TauREx} code but reformulates the parameterization to be consistent with the \textsc{PyMieScatt} implementation. In this model, $Q_\mathrm{ext}$ is inferred from an empirical fitting function rather than Mie theory The formulation does not use complex refractive indexes from aerosol species, so it cannot account for resonance (i.e, spectral feature), but it provides a simpler and more generic alternative for featureless aerosol absorption (see Section 2.3.2 in the original paper). The free parameters in this model are the similar to the \textsc{PyMieScatt} model: particle size ($\mu_r$), the extinction reference radius ($Q_0$), the particle number density ($\chi$), and the location of the cloud layer in pressure (labeled $P_\mathrm{lee}$ for the middle and $\Delta P_\mathrm{lee}$ for the extent of the cloud layer).

\begin{figure*}
\centering
    \includegraphics[width = 0.83\textwidth]{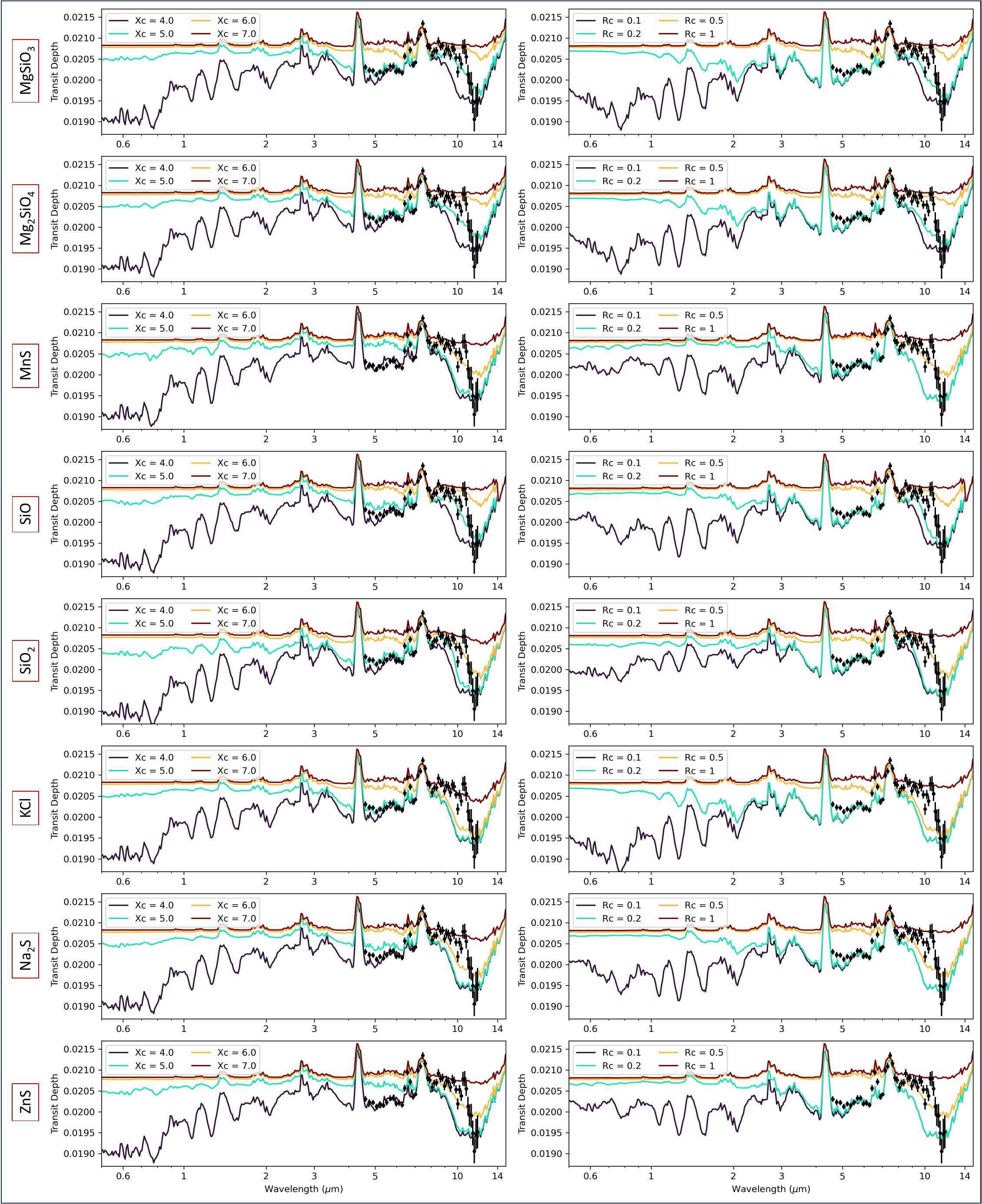}
    \caption{Example forward models of WASP-107\,b's atmosphere using various cloud species and compared to MIRI data (black datapoints) of \cite{Dyrek_2023}. Left column: the number density of the cloud particles, Xc in log part/m$^3$, is varied. Right column: the particle radius, Rc in $\mu$m, is varied. The atmosphere is modeled using the T--p profile from Fig. 3 of \cite{Dyrek_2023} and equilibrium chemistry assuming a solar C/O and a 20x solar metallicity. We manually force a constant SO$_2$ mixing ratio of 10ppm to capture the enhanced 8$\mu$m feature. Clouds are parameterized using the exponentially decaying layer, and spanning one log pressure scale centered at 10 Pa. It is difficult to distinguish which cloud species is present from observing a narrow wavelength range.}
    \label{fig:cloud_W107_exploration}
\end{figure*}

In Figure \ref{fig:cloud_W107_exploration}, forward models demonstrate the impact of particle types, size and number density on observed spectra.

\subsection{In-homogeneous atmospheres with \textsc{TauREx-MultiModel}}

Recent works have shown the importance of modeling spatial in-homogeneities when interpreting exoplanet atmospheres \citep{Feng_2016_inomogeneou, Caldas_2019, Changeat_2020_phasecurve1, Taylor_2020, MacDonald_2020, Pluriel_2020, Nixon_2022}. We here introduce a novel \textsc{TauREx3} plugin, \textsc{TauREx-MultiModel}, which is a mixin \cite[see:][]{al-refaie_2021_taurex3.1} enabling multiple forward models (or $regions$) to contribute to the final flux. It is compatible with all the available \textsc{TauREx3} forward models. The contribution of each region, noted $i$, is controlled by a contribution factor (F$_i$) between 0 and 1 (the sum of all region contributions is forced to equal 1). In this work, we use \textsc{TauREx-MultiModel} to add in-homogeneous cloud coverage at the limbs, but we highlight that the plugin is not limited to cloud parameters: all the forward model properties can be retrieved individually (i.e., for each region) or coupled flexibly. 
The \textsc{TauREx-MultiModel} plugin also adds the ability to post-process the forward model with an instrument function. Here, the instrument response has a Gaussian profile of width modeled by a polynomial function of $\lambda$ (the coefficients of the polynomial can be retrieved). This is useful when dealing with the JWST data at pixel resolution since the line spread functions of the JWST instruments are designed to span multiple pixels. This feature is not used in this work since the literature data is provided at much lower resolution than the instrument is capable.

\subsection{Treatment of instrument systematics with \textsc{TauREx-InstrumentSystematics}}

\textsc{TauREx-InstrumentSystematics} is a small \textsc{TauREx3} plugin to handle retrievals of combined datasets. Previous works have shown that combining datasets can lead to inconsistent results when instrument systematics remain present. This was particularly true in the HST era \citep{Yip_2021_W96}, when assumptions in the reduction pipelines could lead to significant (i.e., $\sim$500\,ppm) vertical offsets---with the spectral shape in the data remaining relatively conserved---and therefore widely different interpretations \cite[see a striking example of this for KELT-11\,b:][]{ Changeat_2020_K11, Colon_2020, Edwards_2024_K11}. Recent works on the ERS data of WASP-39\,b have also similar issues for JWST \citep{Constantinou_2023}. With \textsc{TauREx-InstrumentSystematics}, we provide retrieval capabilities for the most basic instrument systematics: vertical offsets and linear slopes. More complex systematics could eventually be handled in the future if warranted by the data. Additionally, with \textsc{TauREx-InstrumentSystematics}, each observation can be associated with an instrument response file. The response file is used to describe the shape of the Gaussian profile of the instrument response function (similar treatment to \textsc{TauREx-MultiModel}), which is then convolved with the atmospheric forward model.

\onecolumn

\clearpage

\vfill
\vfill
\clearpage

\section{Supplementary data for the Titan's retrievals}

\renewcommand\thesubsection{B\thesection.\arabic{subsection}}
\setcounter{figure}{0}
\renewcommand{\thefigure}{B\arabic{figure}}
\renewcommand{\theHfigure}{B\arabic{figure}}

\setcounter{table}{0}
\renewcommand{\thetable}{B\arabic{table}}
\renewcommand{\theHtable}{B\arabic{table}}

\begin{figure}[H]
\centering
    \includegraphics[width = 0.91\textwidth]{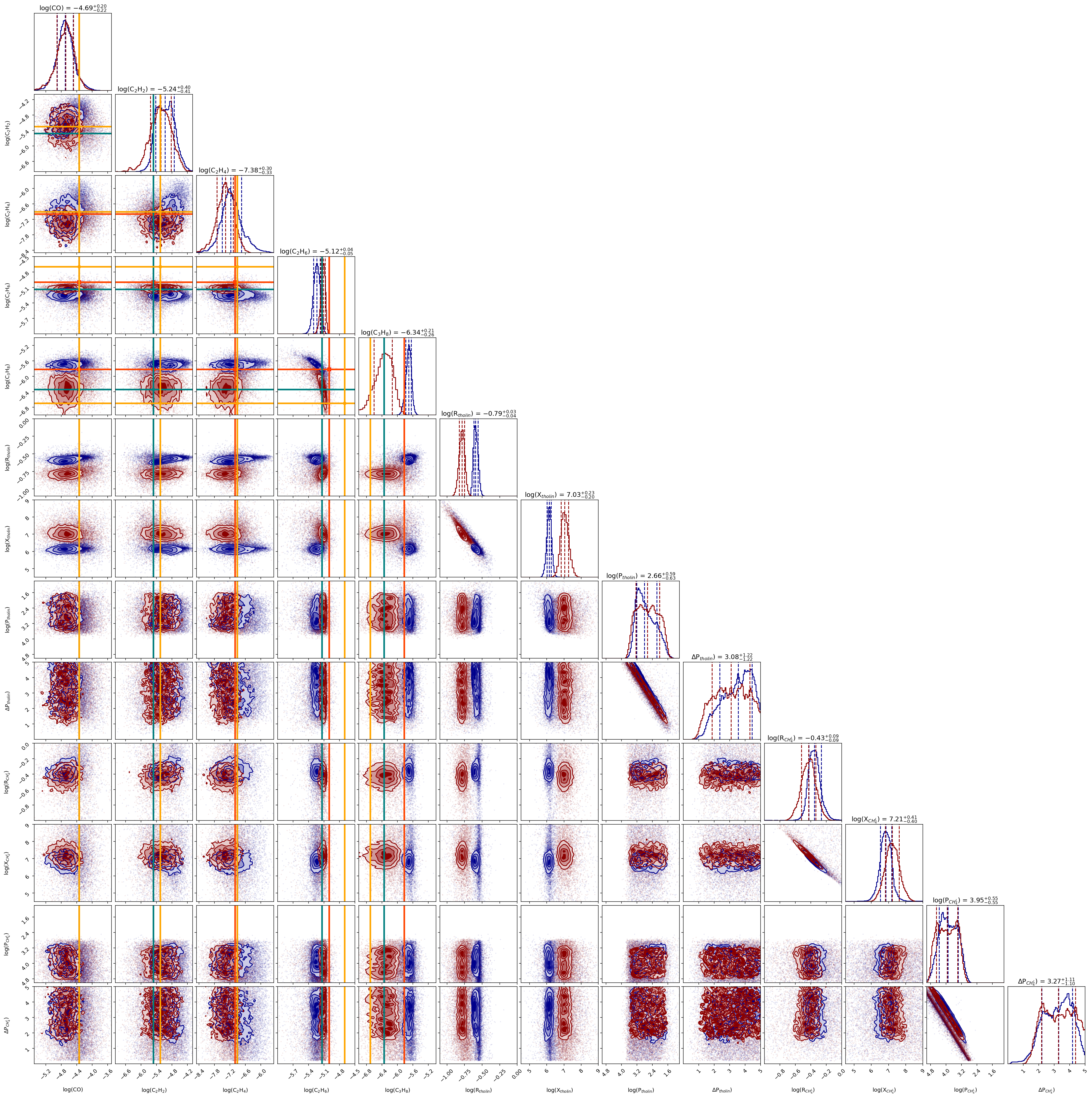}
    \caption{Posterior distribution for the {\it Cassini}/VIMS retrieval of Titan using the HITRAN cross-sections (red) and the ExoMol cross-sections (blue). The stratospheric hydrocarbon abundances, inferred from other studies, are also shown in orange for ISO/Hershel \citep{Coustenis_2003}, teal for {\it Cassini}/CIRS \cite[nominal mission 2004--2010, ][]{Coustenis_2010}, and red for {\it Cassini}/CIRS \cite[extended mission 2014,][]{Coustenis_2016, Sylvestre_2018}. The literature value for CO is from \cite{Maltagliati_2015}. While slightly different chemical compositions are quoted in the literature---depending on the probed altitudes and observations---overall, the retrieval agrees well with the prior knowledge of Titan's stratosphere. The atmosphere is also consistent with the presence of small ($\mu_{r}$ $\sim$ 0.2\,$\mu$m) haze particles extending as high as 0.1\,mbar, with a deeper cloud cover (here modeled with CH$_4$ condensates) composed of larger particles ($\mu_r$ $\sim$ 0.4\,$\mu$m). }
    \label{fig:titan_corner}
\end{figure}

Table \ref{tab:hazesRef} contains the complex refractive index for Tholin used in this study. Figure \ref{fig:titan_corner} shows the posterior distributions from \textsc{TauREx} retrievals of {\it Cassini} Titan's data. 

\begin{table}[H]
\centering
\caption{Refractive indexes for tholins used in this study, computed by interpolating \cite{Khare_1984, Rannou_2010}. See Data Availability section to access the full data table.}
\resizebox{0.5\textwidth}{!}{%
\begin{tabular}{ |c|c|c|c| } 
 \hline
 Wavenumber (cm$^-1$) & Wavelength ($\mu$m) & Real & Imaginary \\ \hline
 2.50e+04 & 4.00e-01 & 1.6745e+00 & 1.1762e-01 \\
 2.38e+04 & 4.20e-01 & 1.6968e+00 & 1.0197e-01 \\
 ... & ... & ... & ... \\
 \hline
\end{tabular}%
}\label{tab:hazesRef}
\end{table}

\vfill
\vfill
\clearpage

\vfill
\begin{figure}[H]
\centering
    \includegraphics[width = 0.80\textwidth]{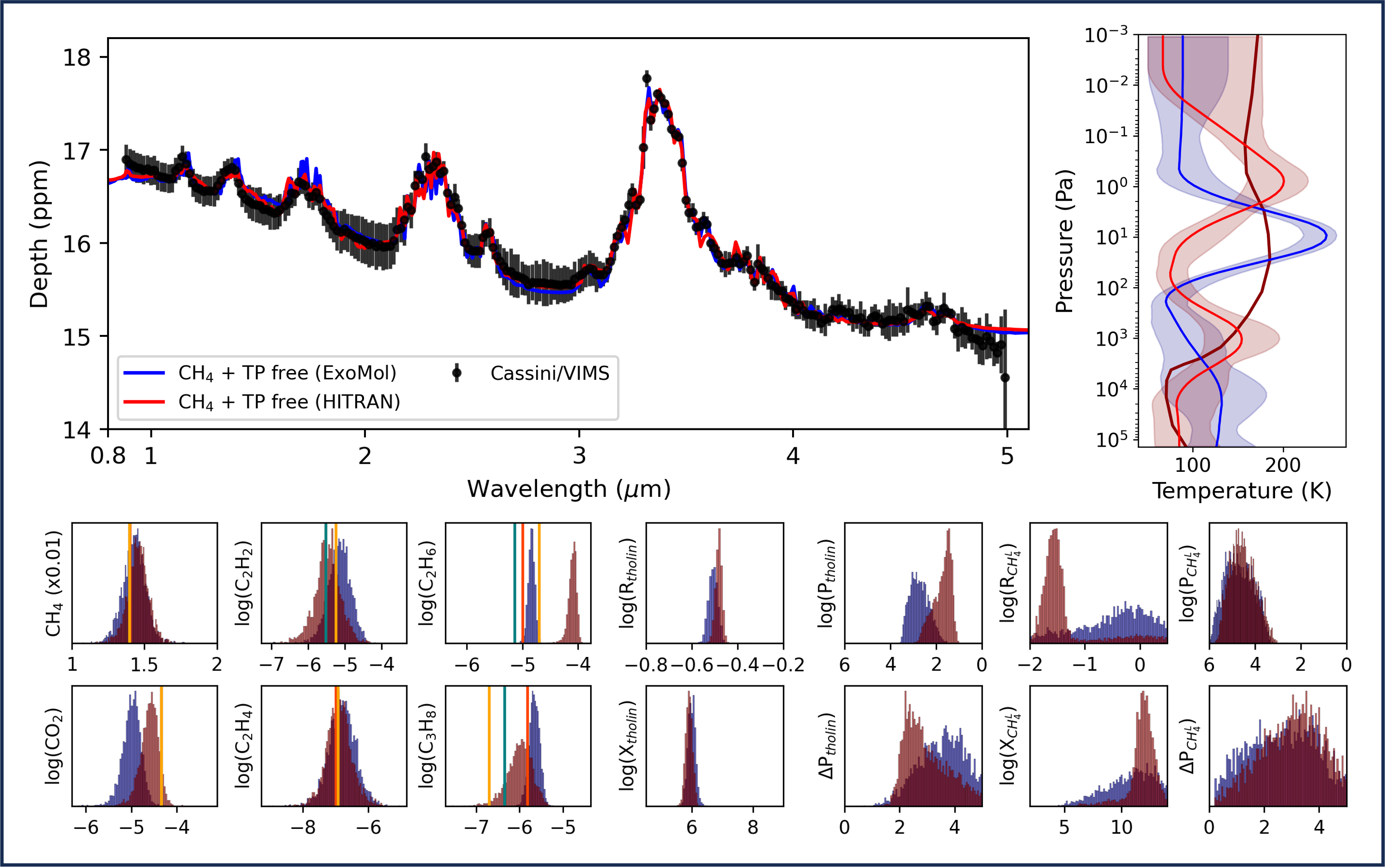}
    \caption{Retrieval results of {\it Cassini}/VIMS occultation of Titan when the temperature profile and methane is also retrieved. The reference values in the posteriors are from the same source as in Figure \ref{fig:titan_corner}. This setup would be more representative of an exoplanet approach where priors on the thermal structure and chemistry is unknown. The spectral fit and retrieved chemistry match the results from the main text. However, for the aerosol solution and the $T-p$ profiles, we inconsistent results. For the $T-p$ profile, in particular, the retrieved solution dos not match the known truth for Titan's atmosphere. Such degeneracies are expected for exoplanet transit retrievals where obtaining robust $T-p$ structure remains difficult \citep{Rocchetto_2016, Schleich_2024}. This is especially true for secondary and cloudy atmosphere cases \cite[see e.g.,][]{Changeat_2020_mass, DiMaio2023}.}
    \label{fig:titan_freeTP}
\end{figure}
\vfill
\clearpage

\section{Supplementary information on JWST retrievals}

\renewcommand\thesubsection{C\thesection.\arabic{subsection}}

\setcounter{figure}{0}
\renewcommand{\thefigure}{C\arabic{figure}}
\renewcommand{\theHfigure}{C\arabic{figure}}

\setcounter{table}{0}
\renewcommand{\thetable}{C\arabic{table}}
\renewcommand{\theHtable}{C\arabic{table}}

\begin{figure}[H]
\centering
    \includegraphics[width = 0.95\textwidth]{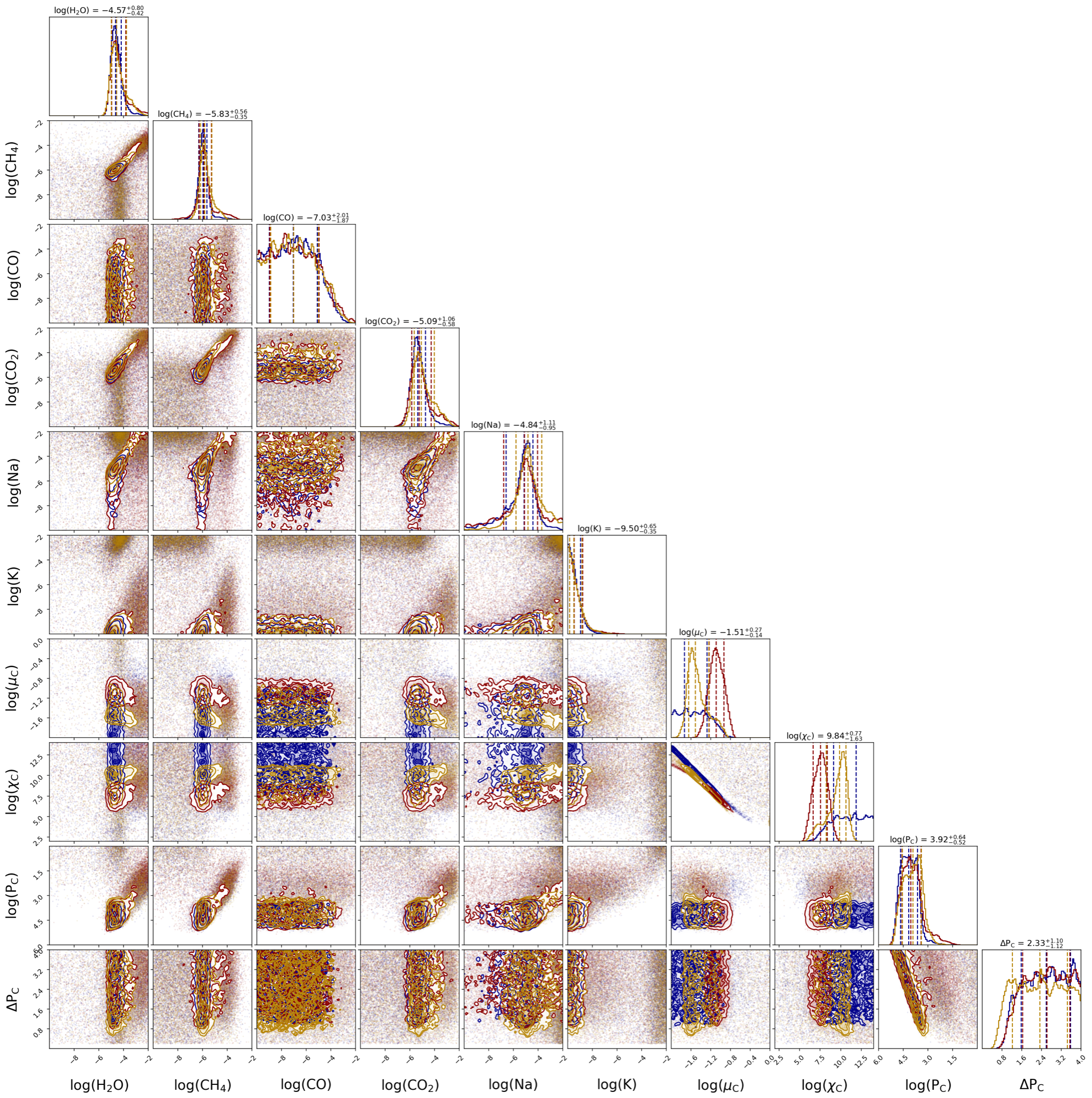}
    \caption{Posterior distribution for the JWST/NIRISS free retrievals of HAT-P-18\,b using various clouds species. Blue: KCl clouds; red: Na$_2$S; gold: ZnS. The abundances of chemical species is unchanged by the different types of clouds, likely because JWST/NIRISS probes the scattering slope at visible to near-infrared wavelenghts and does not cover cloud spectral features. Similar conclusions were inferred from WASP-39\,b and WASP-96\,b. }
    \label{fig:corner_hp18}
\end{figure}

\begin{figure}[H]
\centering
    \includegraphics[width = 0.95\textwidth]{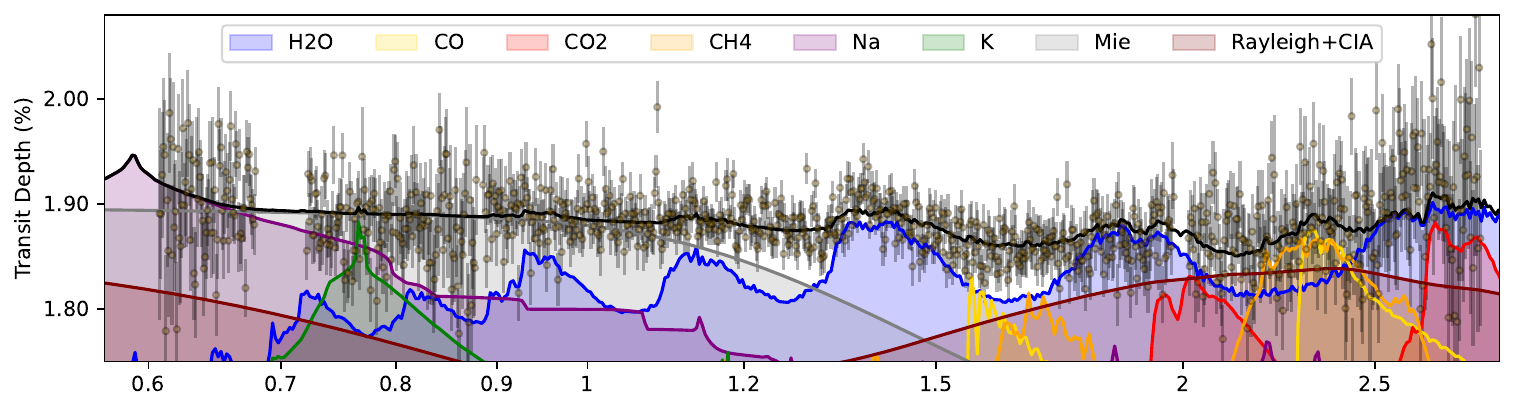}
    \includegraphics[width = 0.95\textwidth]{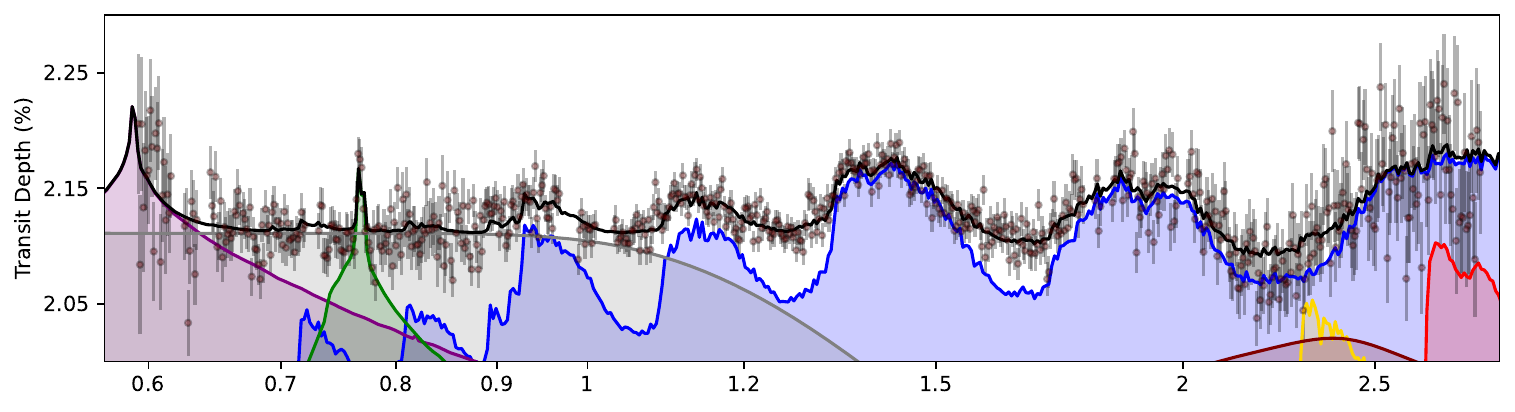}
    \includegraphics[width = 0.95\textwidth]{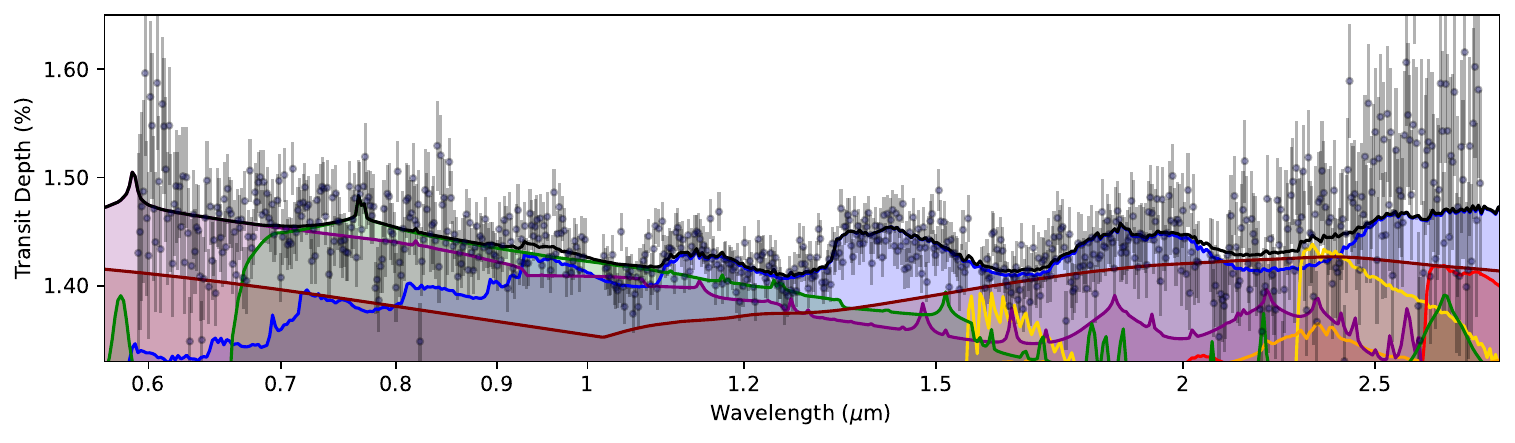}
    \caption{Contribution of different processes to the transmission spectra of HAT-P-18\,b (top), WASP-39\,b (middle), and WASP-96\,b (bottom) obtained from the \textsc{FRECKLL} retrievals. The contribution from water dominates most of the spectra after 1.2\,$\mu$m, while short wavelengths are dominated by the broadened lines of Na and K (WASP-96\,b), or by aerosols (HAT-P-18\,b and WASP-39\,b). In HAT-P-18\,b, H$_2$O, CH$_4$, CO$_2$ and Na are detected. In WASP-39\,b, H$_2$O, CO, CO$_2$, Na, and K are detected. In WASP-96\,b, H$_2$O, CO, Na, and K are detected. }
    \label{fig:niriss_contrib}
\end{figure}

\begin{figure}[H]
\centering
    \includegraphics[width = 0.95\textwidth]{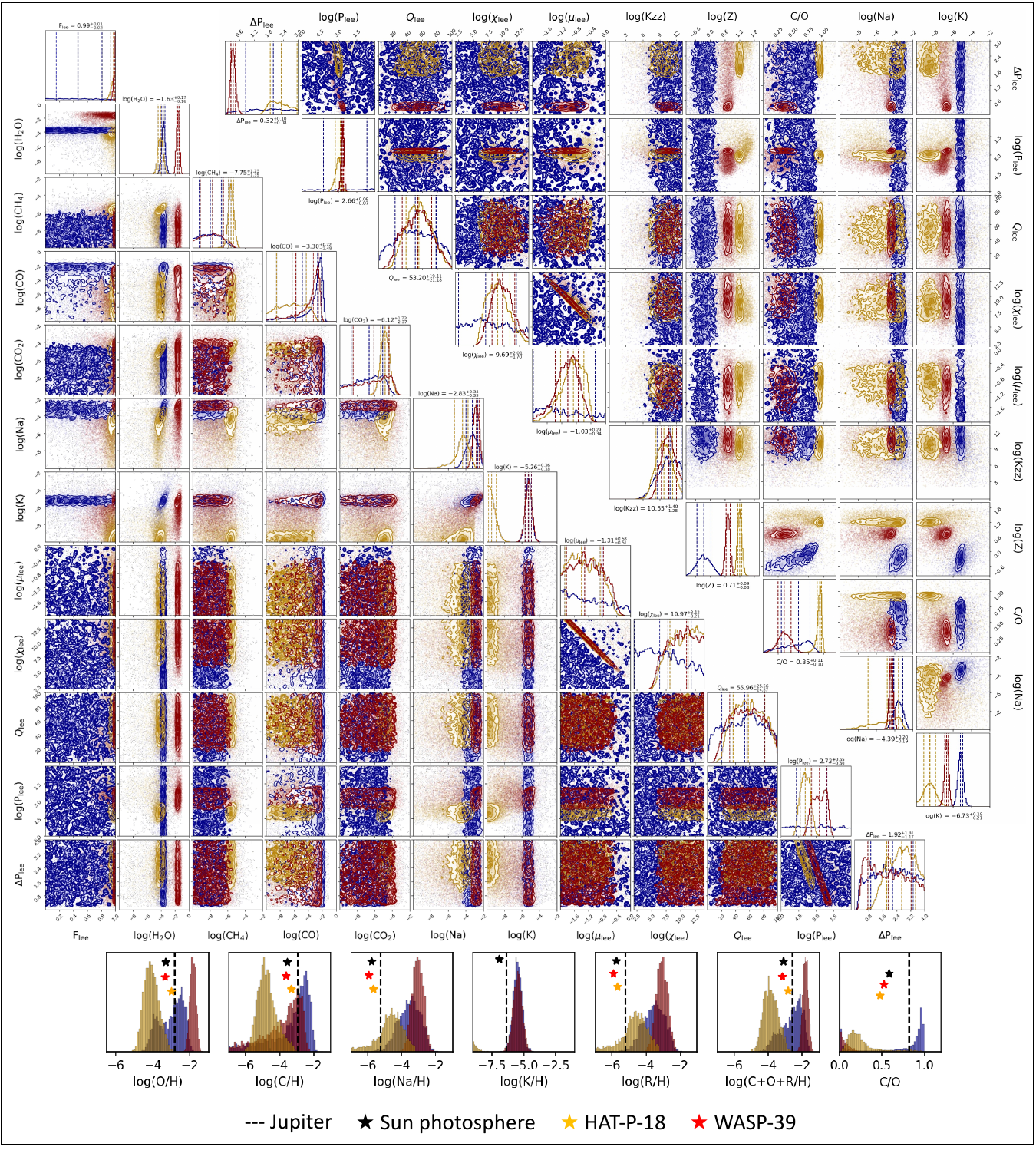}
    \caption{Posterior distribution for the JWST/NIRISS retrievals of HAT-P-18\,b (gold), WASP-39\,b (red), and WASP-96\,b (blue) using the \cite{Lee_2013_clouds} cloud model. The bottom left corner plot shows retrieved posterior distributions for the free chemistry retrievals. The top right inverted corner plots is showing the \textsc{FRECKLL} retrievals. To enable comparison, we also compute (bottom) the elemental ratio probability densities from post-processing the free chemistry runs. For those, we indicate reference solar-system values. The values for Jupiter are computed from \cite{Atreya_2022}, the values for the Sun are from \cite{Asplund_2009}, and the values for the stars HAT-P-18 and WASP-39 are from the Hypatia catalog \citep{Hinkel_2014}. Note that reference abundances for Na and K in Jupiter are not available due to the condensation of those species. We here estimate those values by using S/H as a proxy for Na and K enrichment (i.e., we assume S is a good tracer of refractory). Stellar elemental ratios for WASP-96 are not available.}
    \label{fig:niriss_corner}
\end{figure}

\begin{figure}[H]
\centering
    \includegraphics[width = 0.95\textwidth]{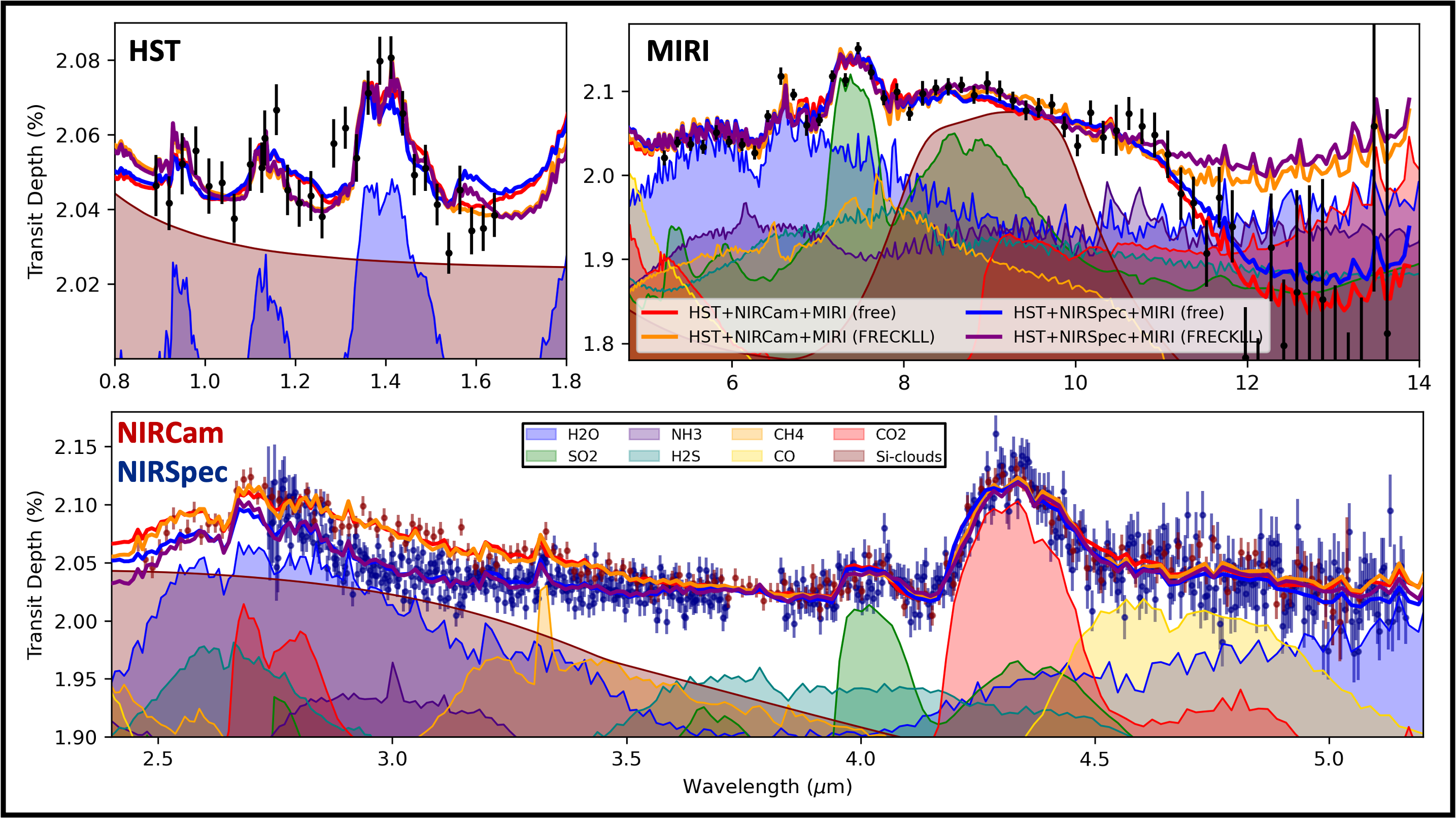}
    \caption{Spectra and best-fit models for the WASP-107\,b study. Red: case where the NIRCam data is considered; blue: case where the NIRSpec data is considered. We also show the contribution of the different components in the model to the best-fit spectrum for the NIRCam (\textsc{FRECKLL}) case. While the retrieval results are similar, systematic differences exist between the NIRCam and NIRSpec data for $\lambda < 3.7\,\mu$m. This difference could be due to remaining instrumental systematics or a time varying cloud cover. Note that the observations are plotted with correction of the vertical offsets. }
    \label{fig:w107_zoom_sp}
\end{figure}

This section contains supplementary information for the JWST retrievals. Figure \ref{fig:corner_hp18} shows the posterior distributions for the HAT-P-18\,b NIRISS retrievals using refractive index of KCl, ZnS and Na$_2$S. Figure \ref{fig:niriss_contrib} shows a breakdown of the relevant opacity contributions to the \textsc{FRECKLL} retrievals of HAT-P-18\,b, WASP-39\,b, and WASP-96\,b. Figure \ref{fig:niriss_corner} shows the posterior distributions of the NIRISS retrievals. Figure \ref{fig:w107_zoom_sp} shows a zoomed-in version of the data and best fit models for the WASP-107\,b retrievals. Figure \ref{fig:w107_corner_full} shows the posterior distributions of the WASP-107\,b retrievals. Finally, Table \ref{tab:niriss_retrievals} and Table \ref{tab:miri_retrievals} provide the retrieved properties from our JWST retrievals.

\vfill
\vfill

\clearpage

\vfill
\vfill

\clearpage

\vfill
\vfill

\clearpage

\vfill
\begin{figure}[H]
\centering
    \includegraphics[width = 0.95\textwidth]{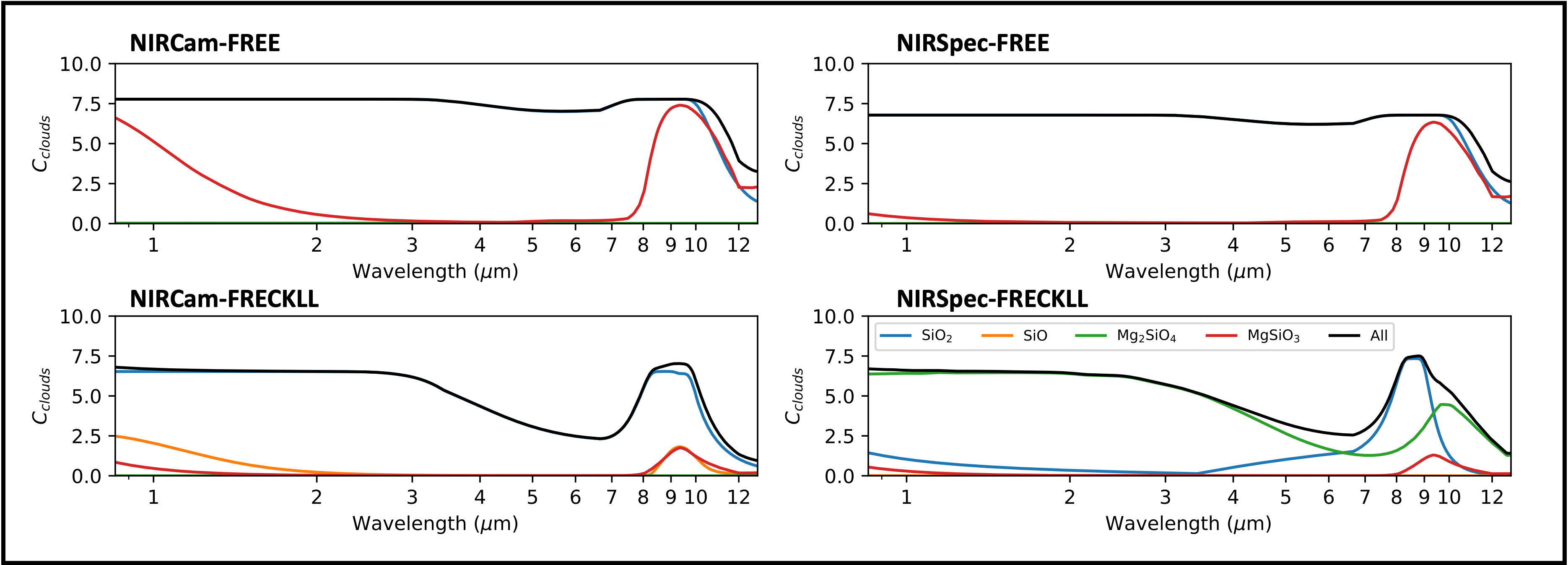}
    \caption{Contribution (C$_{clouds}$) of individual cloud species in the best-fit of the WASP-107\,b data normalized by the atmospheric scale height ($H = 1030$\,km), showing the contribution of the individual components. In all our retrievals, SiO$_2$ clouds (blue) seem necessary to explain the sharp opacity rise at $\lambda > 8\,\mu$m. However, they do not explain 100\% of the feature requiring secondary contributions from either MgSiO$_3$ or Mg$_2$SiO$_4$ particles.}
    \label{fig:w107_cloud_contrib}
\end{figure}
\vfill

\vfill
\begin{figure}[H]
\centering
    \includegraphics[width = 0.99\textwidth]{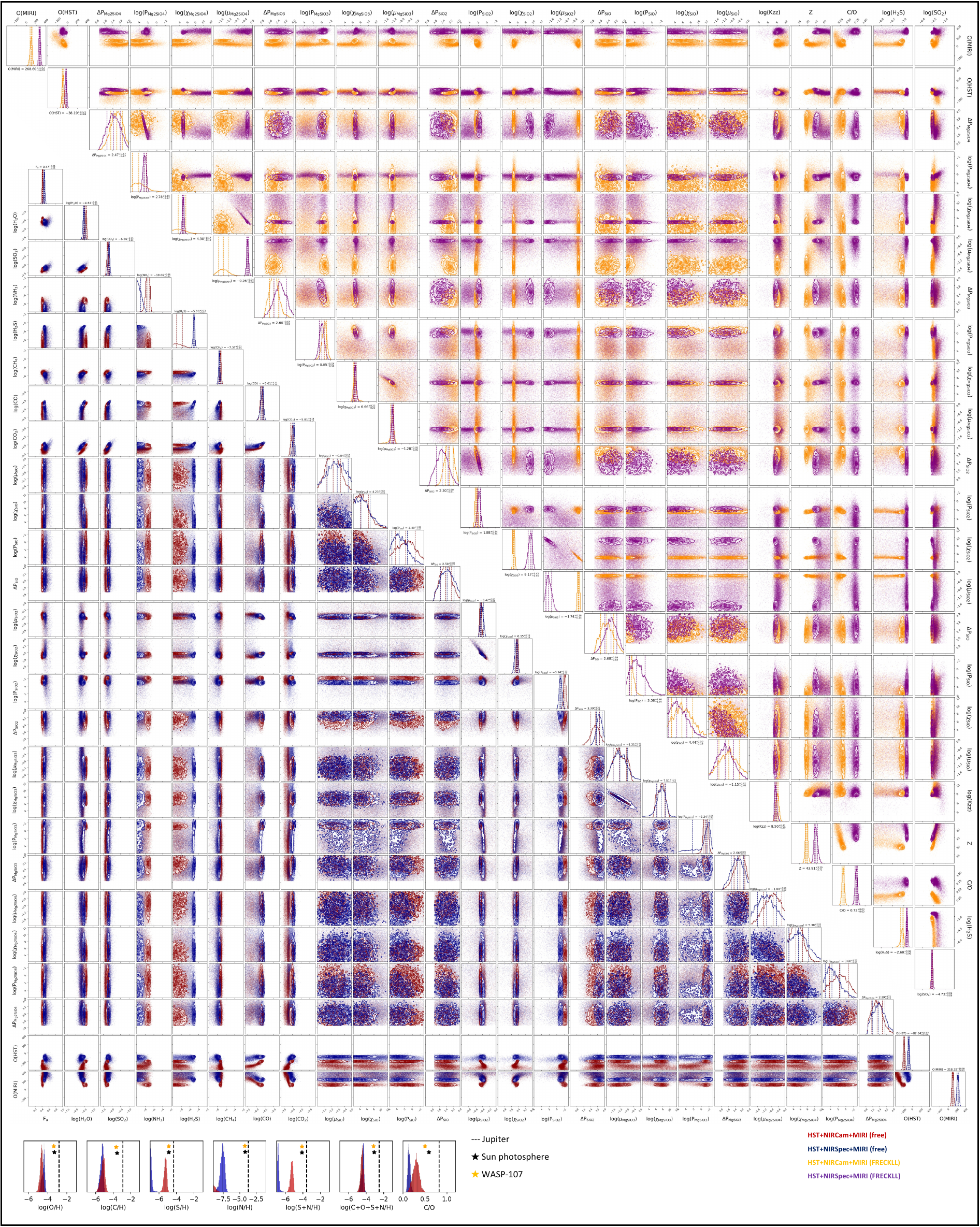}
    \caption{Posterior distributions of the retrievals on the WASP-107\,b data. The bottom left corner plot shows retrieved posterior distributions for the free chemistry retrievals. The top right inverted corner plot is showing the \textsc{FRECKLL} retrievals. To enable comparison, we also compute (bottom) the elemental ratio probability density by post-processing the free chemistry runs. For those, we indicate reference solar-system values. The values for Jupiter and the Sun are the same as in Figure \ref{fig:corner_hp18}. The value for WASP-107 are from \cite{Hejazi_2023}.  }
    \label{fig:w107_corner_full}
\end{figure}
\vfill

\clearpage

\clearpage

\begin{table}[]
\caption{Retrieved properties for the NIRISS observations of HAT-P-18\,b, WASP-39\,b, and WASP-96\,b with the free and \textsc{FRECKLL} retrievals.}
\centering
\begin{tabular}{@{}l|cc|cc|cc|@{}}
\cmidrule(l){2-7}
                              & \multicolumn{2}{c|}{HAT-P-18b} & \multicolumn{2}{c|}{WASP-39b} & \multicolumn{2}{c|}{WASP-96b} \\ \cmidrule(l){2-7} 
                              & Free         & FRECKLL         & Free         & FRECKLL        & Free         & FRECKLL        \\ \midrule
\multicolumn{1}{|l|}{log(H$_2$O)}    & $-4.11^{+0.35}_{-0.34}$ &                 & $-1.63^{+0.17}_{-0.16}$ &                & $-3.7^{+0.23}_{-0.23}$ &                \\
\multicolumn{1}{|l|}{log(CH$_4$)}    & $-5.64^{+0.3}_{-0.28}$ &                 & $-7.75^{+1.25}_{-1.39}$ &                & $-8.0^{+1.29}_{-1.26}$ &                \\ 
\multicolumn{1}{|l|}{log(CO)}        & $-6.32^{+2.04}_{-2.22}$ &                 & $-3.3^{+0.72}_{-2.48}$ &                & $-2.69^{+0.56}_{-2.19}$ &                \\ 
\multicolumn{1}{|l|}{log(CO$_2$)}    & $-4.84^{+0.49}_{-0.5}$ &                 & $-6.12^{+1.72}_{-2.37}$ &                & $-6.66^{+1.53}_{-2.05}$ &                \\ 
\multicolumn{1}{|l|}{log(Na)}        & $-4.41^{+0.65}_{-0.91}$ & $-4.84^{+1.23}_{-2.07}$ & $-2.83^{+0.34}_{-0.33}$ & $-4.39^{+0.2}_{-0.19}$ & $-3.29^{+0.63}_{-0.71}$ & $-3.6^{+0.51}_{-0.5}$ \\ 
\multicolumn{1}{|l|}{log(K)}         & $-9.48^{+0.54}_{-0.36}$ & $-8.57^{+0.63}_{-0.65}$ & $-5.26^{+0.36}_{-0.38}$ & $-6.73^{+0.19}_{-0.21}$ & $-5.28^{+0.35}_{-0.33}$ & $-5.21^{+0.26}_{-0.27}$ \\
\multicolumn{1}{|l|}{log(Z)}       & $-0.77^{+0.47}_{-0.4}$ & $1.19^{+0.08}_{-0.07}$ & $1.31^{+0.17}_{-0.16}$ & $0.71^{+0.09}_{-0.08}$ & $0.6^{+0.5}_{-0.87}$     & $-0.29^{+0.3}_{-0.28}$ \\
\multicolumn{1}{|l|}{C/O}           & $0.21^{+0.23}_{-0.1}$  & $0.93^{+0.02}_{-0.04}$ & $0.02^{+0.09}_{-0.02}$  & $0.35^{+0.11}_{-0.1}$ & $0.91^{+0.07}_{-0.8}$    & $0.58^{+0.19}_{-0.28}$ \\
\multicolumn{1}{|l|}{log(O/H)}       & $-0.8^{+0.43}_{-0.37}$ &                 & $1.47^{+0.16}_{-0.16}$  &                & $0.43^{+0.54}_{-0.95}$    &                \\
\multicolumn{1}{|l|}{log(C/H)}       & $-1.26^{+0.64}_{-0.5}$ &                 & $0.02^{+0.71}_{-1.71}$ &                & $0.62^{+0.56}_{-1.95}$     &                \\
\multicolumn{1}{|l|}{log(Na/H)}      & $1.14^{+0.65}_{-0.92}$ &                 & $2.71^{+0.34}_{-0.33}$ &                & $2.25^{+0.63}_{-0.72}$   &                \\
\multicolumn{1}{|l|}{log(K/H)}       & $-2.79^{+0.55}_{-0.36}$ &                 & $1.44^{+0.36}_{-0.38}$ &                & $1.42^{+0.35}_{-0.33}$    &                \\
\multicolumn{1}{|l|}{log(Kzz)}       &              & $9.71^{+1.55}_{-1.41}$ &                  & $10.55^{+1.41}_{-1.28}$ &              & $10.77^{+2.0}_{-2.2}$ \\
\multicolumn{1}{|l|}{log($\mu_\mathrm{lee}$)}            & $-1.41^{+0.53}_{-0.47}$ & $-0.89^{+0.29}_{-0.35}$ & $-1.31^{+0.53}_{-0.51}$ & $-1.03^{+0.29}_{-0.34}$ & $-2.25^{+1.43}_{-1.1}$ & $-1.98^{+1.7}_{-1.28}$ \\
\multicolumn{1}{|l|}{log($\chi_\mathrm{lee}$)}           & $11.27^{+2.84}_{-3.13}$ & $8.88^{+2.0}_{-1.76}$ & $10.97^{+3.11}_{-3.21}$ & $9.69^{+2.01}_{-1.7}$ & $6.37^{+5.38}_{-4.33}$ & $6.89^{+5.05}_{-4.39}$ \\
\multicolumn{1}{|l|}{$Q_\mathrm{lee}$}                  & $57.81^{+23.89}_{-21.65}$ & $54.59^{+17.78}_{-17.44}$ & $55.95^{+25.56}_{-24.87}$ & $53.2^{+19.1}_{-21.19}$ & $52.65^{+27.96}_{-33.15}$ & $49.6^{+28.7}_{-27.25}$ \\
\multicolumn{1}{|l|}{log(P$_\mathrm{lee}$)}              & $3.95^{+0.42}_{-0.43}$ & $3.03^{+0.25}_{-0.17}$ & $2.73^{+0.65}_{-0.6}$ & $2.66^{+0.09}_{-0.07}$ & $2.1^{+2.64}_{-3.01}$ & $0.67^{+3.55}_{-2.37}$ \\
\multicolumn{1}{|l|}{$\Delta\mathrm{P}_\mathrm{lee}$}   & $2.67^{+0.79}_{-0.85}$ & $2.29^{+0.66}_{-0.45}$ & $1.92^{+1.31}_{-1.17}$ & $0.32^{+0.1}_{-0.08}$ & $2.14^{+1.19}_{-1.23}$ & $1.98^{+1.13}_{-1.14}$ \\
\bottomrule
\end{tabular}\tablefoot{For the free retrievals, elemental ratios are calculated using the approach from \citetalias{Lee_2013_clouds}, with Z defined here as (C+O+R)/H. Contrary to Figure \ref{fig:niriss_corner}, the elemental ratios quoted here are normalized to the solar values.} \label{tab:niriss_retrievals}
\end{table}

\begin{table}[]
\caption{Retrieved properties for the MIRI data of WASP-107\,b with our different observational setups and models.}
\centering
\begin{tabular}{@{}l|cc|cc|@{}}
\cmidrule(l){2-5}
                                                        & \multicolumn{2}{c|}{HST+NIRCam+MIRI} & \multicolumn{2}{c|}{HST+NIRSpec+MIRI} \\ \cmidrule(l){2-5} 
                                                        & Free         & FRECKLL         & Free         & FRECKLL  \\ \midrule
\multicolumn{1}{|l|}{log(H$_2$O)}                        & $-4.26^{+0.11}_{-0.2}$     &                         &   $-4.61^{+0.11}_{-0.14}$  & \\
\multicolumn{1}{|l|}{log(SO$_2$)}                        & $-6.75^{+0.09}_{-0.14}$    & $-4.73^{+0.06}_{-0.06}$ &   $-6.56^{+0.13}_{-0.13}$  & $-4.73^{+0.06}_{-0.06}$  \\
\multicolumn{1}{|l|}{log(NH$_3$)}                        & $-7.37^{+0.29}_{-0.36}$    &                         &   $-10.02^{+0.96}_{-0.97}$ &   \\
\multicolumn{1}{|l|}{log(H$_2$S)}                        & $-9.14^{+1.4}_{-1.32}$     & $-3.23^{+0.15}_{-0.2}$  &   $-5.05^{+0.15}_{-0.17}$  & $-2.6^{+0.06}_{-0.06}$ \\
\multicolumn{1}{|l|}{log(CO)}                            &$-5.11^{+0.14}_{-0.21}$     &                         &   $-5.01^{+0.21}_{-0.3}$   &  \\
\multicolumn{1}{|l|}{log(CO$_2$)}                        & $-6.08^{+0.13}_{-0.26}$    &                         &   $-5.81^{+0.14}_{-0.17}$  &  \\
\multicolumn{1}{|l|}{log(CH$_4$)}                        & $-7.29^{+0.12}_{-0.12}$    &                         &   $-7.57^{+0.12}_{-0.13}$  &  \\
\multicolumn{1}{|l|}{log(Z)}                             & $-1.29^{+0.1}_{-0.21}$     & $1.44^{+0.03}_{-0.03}$ &   $-1.38^{+0.14}_{-0.17}$  & $1.64^{+0.03}_{-0.03}$\\
\multicolumn{1}{|l|}{C/O}                                & $0.14^{+0.03}_{-0.03}$     & $0.33^{+0.03}_{-0.03}$  &   $0.3^{+0.07}_{-0.08}$    & $0.73^{+0.03}_{-0.03}$\\
\multicolumn{1}{|l|}{log(O/H)}                           & $-1.11^{+0.1}_{-0.21}$     & &   $-1.34^{+0.13}_{-0.16}$  & \\
\multicolumn{1}{|l|}{log(C/H)}                           & $-1.73^{+0.14}_{-0.21}$    & &   $-1.6^{+0.2}_{-0.28}$    & \\
\multicolumn{1}{|l|}{log(S/H)}                           & $-2.08^{+0.12}_{-0.16}$    & &   $-0.39^{+0.15}_{-0.17}$  & \\
\multicolumn{1}{|l|}{log(N/H)}                           & $-3.43^{+0.3}_{-0.36}$     & &   $-6.08^{+0.96}_{-0.97}$  & \\
\multicolumn{1}{|l|}{log($\mu_\mathrm{SiO}$)}         & $-1.08^{+0.43}_{-0.4}$        &  $-0.97^{+0.53}_{-0.5}$  & $-0.84^{+0.64}_{-0.59}$ &   $-1.15^{+0.42}_{-0.4}$ \\
\multicolumn{1}{|l|}{log($\chi_\mathrm{SiO}$)}        & $4.09^{+2.3}_{-2.19}$         &  $5.71^{+2.69}_{-2.61}$  & $4.23^{+2.68}_{-2.44}$ &   $4.44^{+2.5}_{-2.18}$ \\
\multicolumn{1}{|l|}{log(P$_\mathrm{SiO}$)}           & $1.76^{+2.7}_{-2.44}$         &  $5.54^{+0.69}_{-0.78}$  & $3.49^{+1.9}_{-2.17}$ &   $3.58^{+1.6}_{-1.94}$  \\
\multicolumn{1}{|l|}{$\Delta\mathrm{P}_\mathrm{SiO}$} & $2.34^{+0.66}_{-0.66}$        &  $2.25^{+0.7}_{-0.63}$  & $2.5^{+0.8}_{-0.77}$ &    $2.68^{+0.58}_{-0.63}$ \\
\multicolumn{1}{|l|}{log($\mu_\mathrm{SiO2}$)}           & $-0.45^{+0.02}_{-0.02}$    & $-0.21^{+0.02}_{-0.03}$ &   $-0.42^{+0.05}_{-0.04}$ & $-1.74^{+0.16}_{-0.13}$ \\
\multicolumn{1}{|l|}{log($\chi_\mathrm{SiO2}$)}          & $6.27^{+0.08}_{-0.07}$     & $4.9^{+0.15}_{-0.12}$ &   $6.15^{+0.15}_{-0.18}$  & $9.17^{+0.43}_{-0.55}$ \\
\multicolumn{1}{|l|}{log(P$_\mathrm{SiO2}$)}             & $-2.07^{+0.29}_{-0.27}$    & $2.51^{+0.21}_{-0.21}$ &   $-0.94^{+0.33}_{-0.37}$ & $1.88^{+0.34}_{-0.34}$ \\
\multicolumn{1}{|l|}{$\Delta\mathrm{P}_\mathrm{SiO2}$}   & $2.93^{+0.49}_{-0.51}$     & $3.04^{+0.4}_{-0.43}$ &   $3.39^{+0.31}_{-0.33}$  & $2.3^{+0.6}_{-0.61}$ \\
\multicolumn{1}{|l|}{log($\mu_\mathrm{MgSiO3}$)}         & $-1.24^{+0.42}_{-0.35}$    & $-1.28^{+0.06}_{-0.09}$ &   $-1.21^{+0.45}_{-0.39}$ & $-1.28^{+0.05}_{-0.05}$ \\
\multicolumn{1}{|l|}{log($\chi_\mathrm{MgSiO3}$)}        & $7.52^{+1.07}_{-1.27}$     & $6.52^{+0.41}_{-0.31}$ &   $7.51^{+1.21}_{-1.43}$  & $6.66^{+0.23}_{-0.22}$ \\
\multicolumn{1}{|l|}{log(P$_\mathrm{MgSiO3}$)}           & $-1.13^{+0.48}_{-0.52}$    & $-0.55^{+0.57}_{-0.48}$ &   $-1.24^{+3.64}_{-0.89}$ & $0.05^{+0.54}_{-0.54}$  \\
\multicolumn{1}{|l|}{$\Delta\mathrm{P}_\mathrm{MgSiO3}$} & $2.96^{+0.55}_{-0.7}$      & $1.97^{+0.47}_{-0.45}$ &   $2.66^{+0.69}_{-0.73}$  & $2.6^{+0.64}_{-0.61}$ \\
\multicolumn{1}{|l|}{log($\mu_\mathrm{Mg2SiO4}$)}         & $-0.69^{+0.5}_{-0.55}$    & $-1.49^{+0.26}_{-0.25}$ &   $-1.08^{+0.58}_{-0.52}$ & $-0.26^{+0.04}_{-0.04}$ \\
\multicolumn{1}{|l|}{log($\chi_\mathrm{Mg2SiO4}$)}        & $4.62^{+1.98}_{-2.01}$     & $3.62^{+1.9}_{-1.73}$ &   $5.38^{+2.93}_{-2.8}$  & $4.9^{+0.17}_{-0.19}$ \\
\multicolumn{1}{|l|}{log(P$_\mathrm{Mg2SiO4}$)}           & $1.71^{+2.53}_{-2.35}$    & $4.13^{+1.48}_{-1.7}$ &   $3.68^{+1.73}_{-2.14}$ & $2.78^{+0.34}_{-0.37}$  \\
\multicolumn{1}{|l|}{$\Delta\mathrm{P}_\mathrm{Mg2SiO4}$} & $2.08^{+0.64}_{-0.61}$      & $2.8^{+0.56}_{-0.63}$ &   $2.29^{+0.72}_{-0.72}$  & $2.47^{+0.68}_{-0.67}$ \\
\multicolumn{1}{|l|}{log(Kzz)}                           &                            & $9.11^{+0.47}_{-0.45}$      &                           & $8.5^{+0.41}_{-0.3}$\\
\multicolumn{1}{|l|}{F$_\mathrm{clouds}$}                & $0.42^{+0.01}_{-0.01}$     & & $0.47^{+0.01}_{-0.01}$    & \\
\multicolumn{1}{|l|}{Offset HST}                            & $-194^{+15}_{-16}$      & $-96^{+16}_{-1.6}$  & $-88^{+14}_{-16}$        & $-38^{+13}_{-14}$ \\
\multicolumn{1}{|l|}{Offset MIRI}                            & $99^{+18}_{-17}$       & $96^{+17}_{-22}$& $218^{+20}_{-21}$         & $269^{+14}_{-14}$ \\
\bottomrule
\end{tabular}\tablefoot{Contrary to Figure \ref{fig:w107_corner_full}, the ratios are normalized to the solar values.} \label{tab:miri_retrievals}
\end{table}

\clearpage

\section{Supplementary information for the discussion section}

\renewcommand\thesubsection{D\thesection.\arabic{subsection}}

\setcounter{figure}{0}
\renewcommand{\thefigure}{D\arabic{figure}}
\renewcommand{\theHfigure}{D\arabic{figure}}

\begin{figure}[H]
\centering
    \includegraphics[width = 0.95\textwidth]{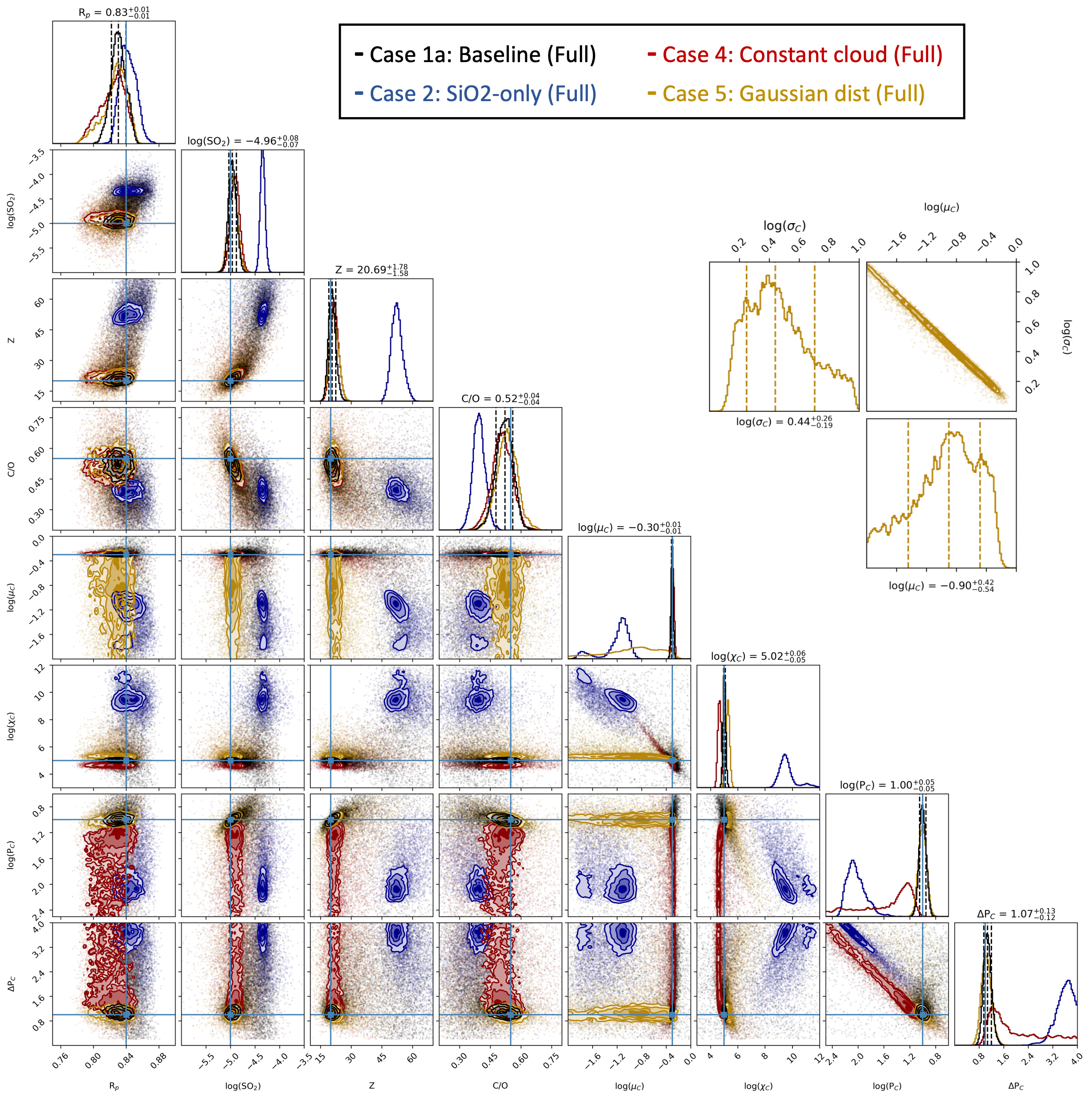}
    \caption{Posterior distributions of the retrievals from the top panel of Figure \ref{fig:spec_jwst_simu} (inset is for non shared parameters). The colors are conserved between the figures -- back: baseline retrieval, blue: SiO$_2$-only retrieval, red: constant cloud retrieval, and gold: Gaussian particle distribution retrieval. The SiO$_2$-only retrieval shows that a large wavelength coverage allows to distinguish aerosol properties. The other retrievals indicate that this dataset is poorly sensitive to vertical aerosol abundance distribution, but that the size of the largest particles in the distribution can be recovered (see negative correlation between mean and variance of the Gaussian distribution). }
    \label{fig:corner_jwst_simu}
\end{figure}

\begin{figure}[H]
\centering
    \includegraphics[width = 0.95\textwidth]{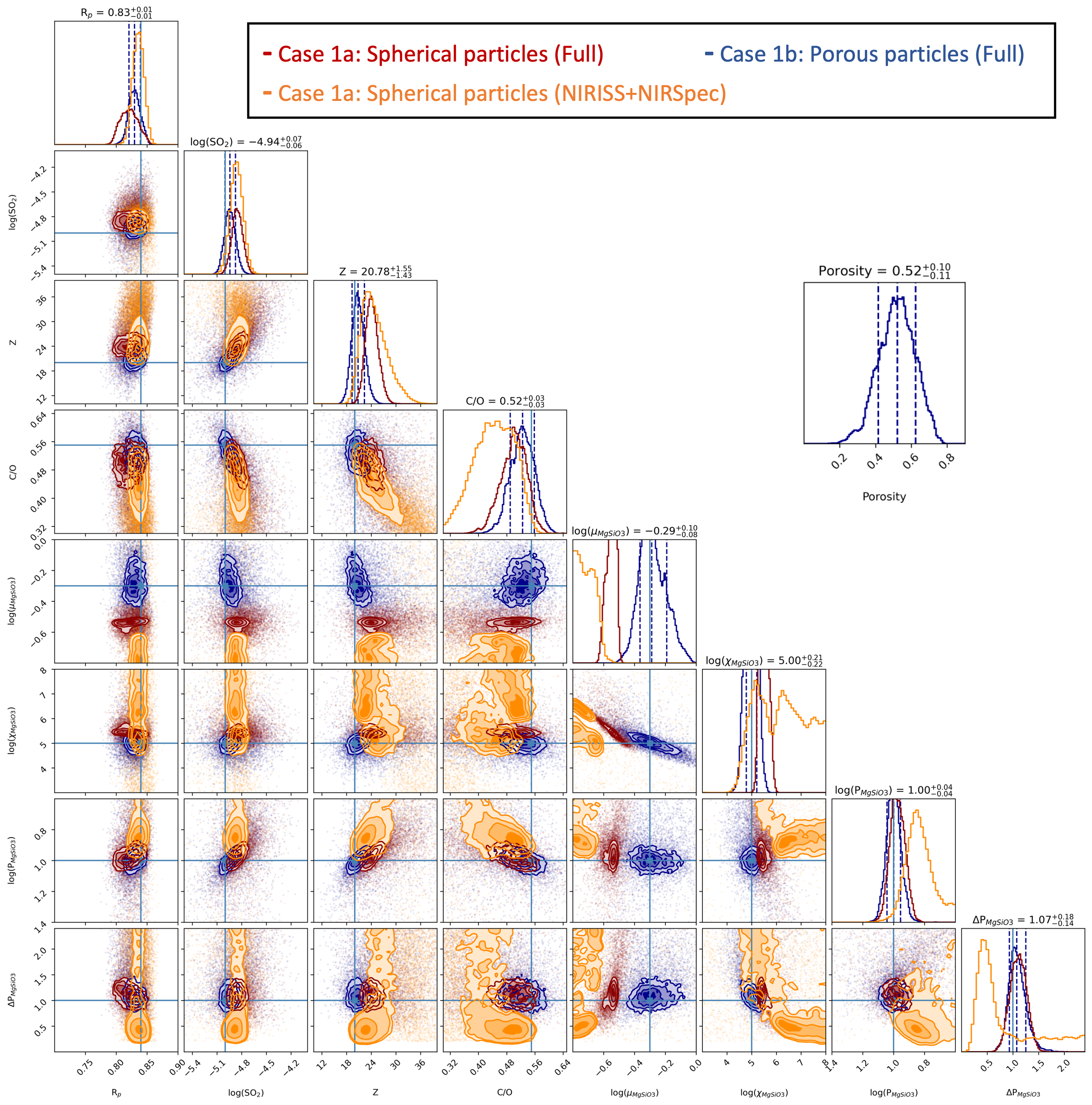}
    \caption{Posterior distributions of the retrievals from the bottom panel of Figure \ref{fig:spec_jwst_simu} (inset is for non shared parameters). The colors are conserved between the figures---red: spherical particles (Full), orange: spherical particles (NIRISS+NIRSpec only), and blue: porous particles (Full). Retrieving the porosity of aerosol particles is difficult with JWST data, as shown by the fit obtained by the spherical particle retrieval on the data of Figure \ref{fig:spec_jwst_simu}, bottom panel. The model compensates by introducing a bias in the particle size and abundance, but the chemistry remains similar. }
    \label{fig:corner_jwst_simu_porous}
\end{figure}

This appendix contains supplementary information regarding the simulations of Section \ref{sec:disc}. Figure \ref{fig:corner_jwst_simu} shows the corner plots of the full retrieval cases with spherical particles (forward model FM1). Figure \ref{fig:corner_jwst_simu_porous} shows the corner plots of the full retrieval cases when the forward model contains porous particles (forward model FM2).

\vfill
\vfill
\clearpage

\vfill
\vfill
\clearpage

\end{appendix}
\end{document}